\documentclass[11pt]{article}
\pdfoutput=1

\usepackage{cancel,slashed}
\usepackage{bbm}
\usepackage{bm}
\usepackage{amsmath,amsfonts,amssymb,mathtools,nicefrac,nccmath,cases}
\usepackage{graphicx}
\usepackage{cite}
\usepackage{skak}

\usepackage{dsfont}
\usepackage{latexsym}
\usepackage{color}
\usepackage[hyperfootnotes=false,linktocpage]{hyperref}
\usepackage{transparent}
\usepackage[hang,flushmargin]{footmisc}
\usepackage{blkarray}
\usepackage{multirow}
\usepackage{arydshln}
\usepackage{empheq}
\usepackage{comment}
\usepackage{wasysym}
\usepackage{tikzsymbols}

\usepackage{caption}
\usepackage{subcaption}
\usepackage[normalem]{ulem}

\newcommand{\bmat}{\left(\begin{array}}
\newcommand{\emat}{\end{array}\right)}

\def\R{\mathbbm{R}}

\def\a {\alpha}
\def\b {\beta}
\def\l{\lambda}
\def\m {\mu}
\def\n {\nu}

\def\K{\mathbf{K}}

\def\1{{\bf 1}}
\def\2{{\bf 2}}
\def\3{{\bf 3}}
\def\4{{\bf 4}}
\def\6{{\bf 6}}

\def\targ#1#2{\genfrac{[}{]}{0pt}{}{#1}{#2}}
\def\targ2#1#2{\genfrac{}{}{0pt}{}{#1}{#2}}

\definecolor{mygr}{rgb}{0,0.6,0}
\definecolor{mygrey}{rgb}{0,0.1,0.2}
\definecolor{myblue}{rgb}{0,0.5,0.9}
\definecolor{myblue2}{rgb}{0,0.5,0.5}
\definecolor{myblue3}{rgb}{0,0.7,0.9}
\definecolor{myblue4}{rgb}{0,0.6,0.6}
\definecolor{myorange}{rgb}{1,0.5,0}
\definecolor{mypurple}{rgb}{0.6,0,1}
\definecolor{mygolden}{rgb}{1,0.8,0.2}
\definecolor{mycyan}{rgb}{0,1,1}
\definecolor{mymagenta}{rgb}{1,0,1}
\definecolor{mykiwi}{rgb}{0.8,1,0.5}
\definecolor{mybrown}{cmyk}{0.14, 0.42, 0.56, 0.2}
\definecolor{myturq}{cmyk}{0.99, 0, 0.2, 0.4}
\definecolor{myaubergine2}{cmyk}{0.4, 0.5, 0, 0.1}
\definecolor{myaubergine}{cmyk}{0.6,0.85,0,0}
\definecolor{CycleGreen}{cmyk}{0.52,0,1,0}
\definecolor{CycleBrown}{cmyk}{0, 0.4, 0.9, 0.2}

\DeclareFontFamily{U}{rcjhbltx}{}
\DeclareFontShape{U}{rcjhbltx}{m}{n}{<->rcjhbltx}{}
\DeclareSymbolFont{hebrewletters}{U}{rcjhbltx}{m}{n}

\DeclareMathSymbol{\lamed}{\mathord}{hebrewletters}{108}
\DeclareMathSymbol{\mem}{\mathord}{hebrewletters}{109}
\DeclareMathSymbol{\ayin}{\mathord}{hebrewletters}{96}
\DeclareMathSymbol{\tsadi}{\mathord}{hebrewletters}{118}
\DeclareMathSymbol{\qof}{\mathord}{hebrewletters}{113}
\DeclareMathSymbol{\resh}{\mathord}{hebrewletters}{114}
\DeclareMathSymbol{\pe}{\mathord}{hebrewletters}{112}
\DeclareMathSymbol{\pesofit}{\mathord}{hebrewletters}{80}
\DeclareMathSymbol{\samekh}{\mathord}{hebrewletters}{115}
\DeclareMathSymbol{\tav}{\mathord}{hebrewletters}{116}
\DeclareMathSymbol{\vav}{\mathord}{hebrewletters}{119}
\DeclareMathSymbol{\het}{\mathord}{hebrewletters}{120}
\DeclareMathSymbol{\yod}{\mathord}{hebrewletters}{121}
\DeclareMathSymbol{\zayin}{\mathord}{hebrewletters}{122}
\DeclareMathSymbol{\alephdot}{\mathord}{hebrewletters}{128}
\DeclareMathSymbol{\tsadisofit}{\mathord}{hebrewletters}{90}
\DeclareMathSymbol{\shin}{\mathord}{hebrewletters}{152}

\def\CN {{\cal N}}

\def\CK {{\cal K}}
\def\CV {{\cal V}}

\def\sig{{\sigma}}
\def\del{{\delta}}

\def\be{\begin{equation}}
\def\ee{\end{equation}}
\def\bea{\begin{eqnarray}}
\def\eea{\end{eqnarray}}
\def\bes{\begin{subequations}}
\def\ees{\end{subequations}}

\def\oh{\frac{1}{2}}

\def\om{\omega}

\usepackage{multicol}
\usepackage{float}
\usepackage{caption}
\def\p {{\partial}}

\def\g {{\gamma}}

\def\s {{\sigma}}

\newcommand{\cF}{\mathcal{F}}

\newcommand{\cK}{\mathcal{K}}

\newcommand{\cN}{\mathcal{N}}
\newcommand{\cO}{\mathcal{O}}

\newcommand{\cI}{\mathcal{I}}

\newenvironment{eqn}{\begin{equation}\begin{aligned}}{\end{aligned}\end{equation}\noindent}
\newenvironment{eqn*}{\begin{equation*}\begin{aligned}}{\end{aligned}\end{equation*}\noindent}

\makeatletter
\newsavebox\myboxA
\newsavebox\myboxB
\newlength\mylenA

\newcommand*\xoverline[2][0.75]{%
\sbox{\myboxA}{$\m@th#2$}%
\setbox\myboxB\null
\ht\myboxB=\ht\myboxA%
\dp\myboxB=\dp\myboxA%
\wd\myboxB=#1\wd\myboxA
\sbox\myboxB{$\m@th\overline{\copy\myboxB}$}
\setlength\mylenA{\the\wd\myboxA}
\addtolength\mylenA{-\the\wd\myboxB}%
\ifdim\wd\myboxB<\wd\myboxA%
   \rlap{\hskip 0.5\mylenA\usebox\myboxB}{\usebox\myboxA}%
\else
    \hskip -0.5\mylenA\rlap{\usebox\myboxA}{\hskip 0.5\mylenA\usebox\myboxB}%
\fi}
\makeatother

\begin{document}
\pagestyle{plain}

\makeatletter
\@addtoreset{equation}{section}
\makeatother
\renewcommand{\theequation}{\thesection.\arabic{equation}}

\pagestyle{empty}
\rightline{IFT-UAM/CSIC-24-124}
\vspace{0.5cm}
\begin{center}
\Huge{{Asymptotic curvature divergences  \\ and non-gravitational theories} 
\\[10mm]}
\Large{Fernando Marchesano,$^{1}$ Luca Melotti\,$^{1,2}$ and Max Wiesner\,$^{3}$}\\[12mm]
\small{
${}^{1}$ Instituto de F\'{\i}sica Te\'orica UAM-CSIC, c/ Nicol\'as Cabrera 13-15, 28049 Madrid, Spain \\[2mm] 
${}^{2}$ Departamento de F\'{\i}sica Te\'orica, Universidad Aut\'onoma de Madrid, 28049 Madrid, Spain \\[2mm] 
${}^{3}$ Jefferson Physical Laboratory, Harvard University, 17 Oxford Street, Cambridge, MA 02138, USA
\\[10mm]} 
\small{\bf Abstract} \\[5mm]
\end{center}
\begin{center}
\begin{minipage}[h]{15.0cm}

We analyse divergences of the scalar curvature $R$ of the vector multiplet moduli space of type IIA string theory compactified on a Calabi--Yau $X$, along infinite-distance large volume limits. Extending previous results, we classify the origin of the divergence along trajectories which implement decompactifications to F-theory on $X$ and/or emergent heterotic string limits. In all cases, the curvature divergence can be traced back to a 4d rigid field theory that decouples from gravity along the limit. This can be quantified via the asymptotic relation $R \sim (\Lambda_{\rm WGC}/\Lambda_{\rm sp})^{2\nu}$, with $\Lambda_{\rm WGC} \equiv g_{\rm rigid} M_{\rm P}$ and $\Lambda_{\rm sp}$ the species scale. In the UV, the 4d rigid field theory becomes a higher-dimensional, strongly-coupled rigid theory that also decouples from gravity. The nature of this UV theory is encoded in the exponent $\nu$, and it either corresponds to a 5d SCFT, 6d SCFT or a Little String Theory.

\end{minipage}
\end{center}
\newpage
\setcounter{page}{1}
\pagestyle{plain}
\renewcommand{\thefootnote}{\arabic{footnote}}
\setcounter{footnote}{0}


\tableofcontents


\section{Introduction}
\label{s:intro}

A major motivation behind the Swampland Programme \cite{Vafa:2005ui,Brennan:2017rbf,Palti:2019pca,vanBeest:2021lhn,Grana:2021zvf,Agmon:2022thq} is the expectation that by understanding which effective field theories (EFTs) have a gravitational UV completion we learn significant lessons about our universe, which eventually lead us to predictions. In consequence, most of the conjectures that drive the programme are formulated from the viewpoint of the said EFT, in such a way that they become trivial as gravity is decoupled from the EFT.

In this context, the Distance Conjecture (SDC) \cite{Ooguri:2006in} is a remarkable proposal. It posits that along infinite distance geodesics in field space there is always an infinite tower of states with mass scale $m_*$, that becomes light exponentially fast in Planck units. If the EFT has a cut-off $\Lambda_{\rm EFT}$ fixed in units of the Planck scale $M_{\rm P}$, as it is typically the case, then the conjecture implies that at some point $m_*<\Lambda_{\rm EFT}$ and the EFT breaks down, removing the infinite distance. Alternatively, if one instead considers an EFT with $\Lambda_{\rm EFT}$ fixed in units of $m_*$ the EFT does not break down, but as $M_{\rm P}/\Lambda_{\rm EFT} \to \infty$ one obtains a non-gravitational theory that decouples from gravity along the limit. This EFT should neither contain gravity nor the field controlling the gravity-decoupling limit. In other words, it should only correspond to a subsector of the initial gravitational theory in which the infinite distance limit was formulated. 

In the context of the 4d ${\cal N}=2$ EFTs obtained from type II string theory compactified on Calabi--Yau (CY) manifolds this pattern was indeed observed in~\cite{Marchesano:2023thx}. The infinite-distance trajectories of the vector multiplet moduli space of such theories had been previously classified in \cite{Grimm:2018ohb,Grimm:2018cpv,Corvilain:2018lgw,Lee:2019wij}, both from the viewpoint of the geometry and the nature of the towers that realise the SDC. What \cite{Marchesano:2023thx} observed is that, along certain limits, a subsector of the vector multiplets remain dynamical below the SDC cut-off $m_*$, in the sense that its gauge interactions remain finite and its kinetic terms do not blow up in units of $m_*$. This subsector contains a 4d ${\cal N}=2$ rigid field theory (RFT), that decouples from gravity as one proceeds along the limit. Importantly, this RFT does not include the field describing the initial infinite distance trajectory. Considering the limit in the opposite direction, one realises the embedding of such a 4d RFT into a larger, UV complete gravitational theory. 

\begin{figure}[!t]
  \centering
\includegraphics[width=.6\linewidth]{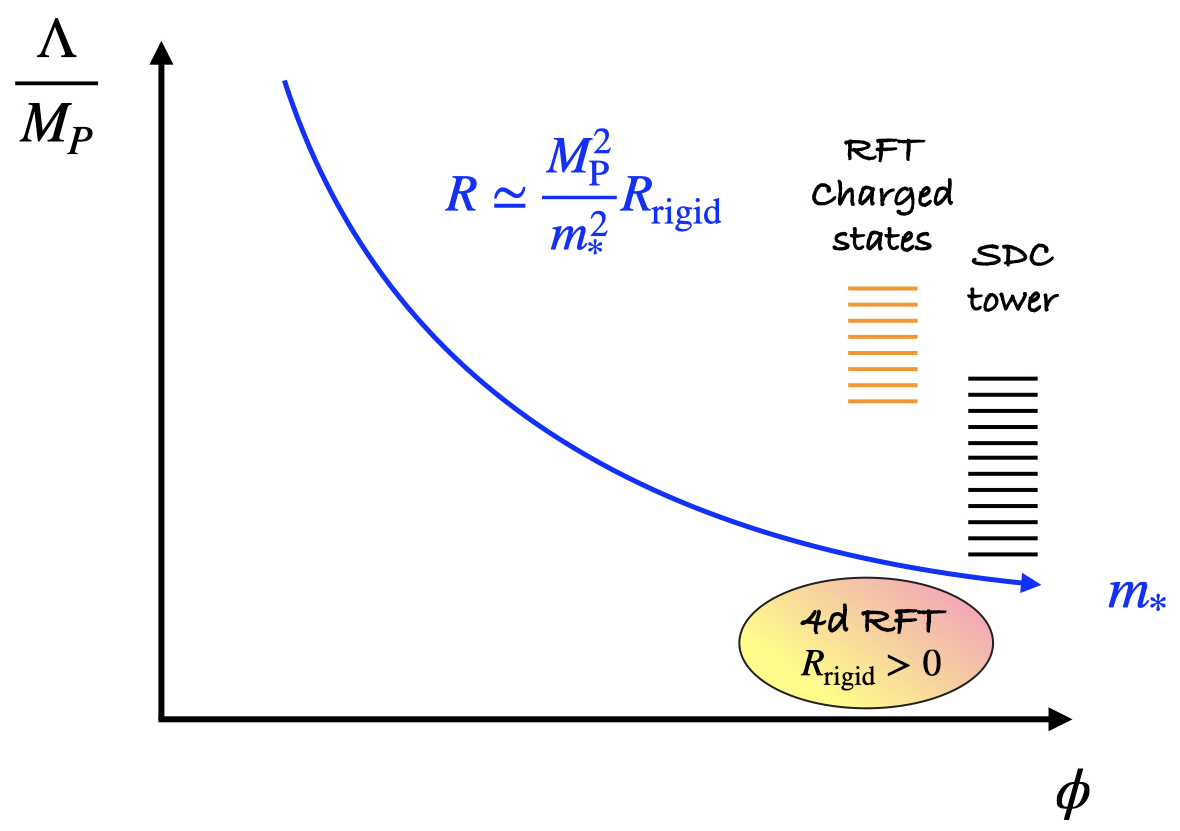}
\caption{Asymptotic curvature divergence sourced by a 4d RFT with a non-vanishing rigid curvature.}
\label{fig:introscales}
\end{figure}

Several interesting features were pointed out in \cite{Marchesano:2023thx} regarding this set of limits. In the context of large-volume limits of type IIA compactified on a Calabi--Yau $X$, it was observed that a 4d RFT gauge coupling that remains constant along the trajectory corresponds to a four-cycle divisor ${\cal D} \subset X$  whose volume remains constant in string units. This implies the existence of a whole sector of massive states charged under the 4d RFT. These states are made up from branes wrapping ${\cal D}$ and two-cycles contained in ${\cal D}$, and their mass scales like $m_* \simeq m_{\rm D0}$. These states are, however, unrelated to the tower of states predicted by the SDC, see figure \ref{fig:introscales}, even though their masses scale in the same way. This feature is particularly manifest in the subset of limits that implement decompactifications to a 5d EFT described by M-theory on $X$, dubbed $w=3$ limits in \cite{Marchesano:2023thx}. From the 5d M-theory viewpoint, ${\cal D}$ is a divisor that contracts to a point along the limit, and hosts a strongly-coupled 5d SCFT \cite{Witten:1996qb}. The 5d SCFT content is precisely the massive sector associated to ${\cal D}$, which suggests that the 4d RFT flows to this non-gravitational theory in the UV. 

The motivation underlying the analysis  in \cite{Marchesano:2023thx} was not the Distance Conjecture itself, but instead a related proposal of \cite{Ooguri:2006in} constraining the asymptotic behaviour of the field space scalar curvature.\footnote{See \cite{Raman:2024fcv} for a recent reformulation of such a conjecture on the curvature in moduli spaces.} In this regard, the lesson learnt was that any infinite-distance limit realising a 4d RFT with non-vanishing curvature $R_{\rm rigid}$ sources a positive divergence in the scalar curvature in Planck units as $R \simeq (M_{\rm P}/m_*)^2 R_{\rm rigid}$. This result motivated the  Curvature Criterion (CC)  \cite{Marchesano:2023thx}, which proposes that for any curvature divergence to occur, there must be a rigid theory that decouples from gravity along the limit and that its gauge interactions become parametrically stronger than gravity. One thus finds two separate effects related to the presence of a non-trivial 4d RFT below $m_*$, namely {\it i)} a source of curvature divergence and {\it ii)} a tower of charged states that signal the appearance of a strongly-coupled higher-dimensional theory that the 4d RFT flows to. Since both features have a common origin, it is natural to wonder if they are directly connected to each other. 

In this work we improve our understanding of infinite-distance limits with 4d RFTs, by extending the analysis of \cite{Marchesano:2023thx} to more general large-volume limits, and addressing some of the questions left open by this initial analysis. One of the key improvements that we implement is a quantitative realisation of the Curvature Criterion, by comparing the scale $\Lambda_{\rm WGC} \equiv g_{\rm rigid} M_{\rm P}$ with the species scale $\Lambda_{\rm sp}$. Following \cite{FierroCota:2023bsp}, one can interpret the relative value of these two scales as measuring the strength of the 4d RFT gauge interactions versus the gravitational ones, to conclude that the rigid theory decouples from gravity whenever $\Lambda_{\rm WGC}/\Lambda_{\rm sp} \to \infty$. As in the Curvature Criterion, this is only a necessary condition for a curvature divergence to be present. Still, in case such a  divergence occurs it is useful to express it as
\be
 R  \simeq \frac{M_{\rm P}^2}{m_*^2} R_{\rm rigid} \lesssim \left(\frac{\Lambda_{\rm WGC}}{\Lambda_{\rm sp}} \right)^{2\nu}\, ,
  \label{Rasintro}
\ee
for some integer $\nu$, that in our setup takes the values $\nu = 1,2,3$. As it turns out, this integer encodes the nature of the higher-dimensional rigid theory that the 4d RFT flows to in the UV. 

To establish this link, we analyse a wide set of large-volume infinite-distance limits in the vector multiplet moduli space of type IIA CY compactifications. Besides the  M-theory $w=3$ limits already considered in detail in \cite{Marchesano:2023thx}, we also look at those limits dubbed $w=2$ and $w=1$ in the nomenclature of \cite{Lanza:2020qmt,Lanza:2021udy,Lanza:2022zyg} regarding single-field EFT string limits, but also consider more general multi-field limits based on growth sectors \cite{Grimm:2018cpv}. In general, $w=2$ limits correspond to decompactification limits to a 6d EFT described by F-theory on $X$, which is an genus-one fibered CY. We extend the analysis of \cite{Marchesano:2023thx} to study both smooth and non-smooth fibrations, distinguishing the different kind of divisors that can give rise to a 4d RFT, their effect on the asymptotic curvature and the UV rigid theory that they host. We carry out a similar analysis for $w=1$ limits, focusing on the case where $X$ is a K3 fibration over $\mathbb{P}^1$, which correspond to emergent heterotic string limits \cite{Lee:2019wij}. In all of these cases, a tower of states with arbitrarily large charges under the 4d RFT appears at a certain scale $\Lambda_{\rm ch}$, which lies between $m_*$ and the species scale. The 4d RFT flows all the way up to this scale, where it becomes a strongly-coupled UV rigid theory whose nature is determined by the index $\nu$ in \eqref{Rasintro}.
 The relations that we find between curvature divergences and these UV rigid theories is captured in the tables of section \ref{s:summary} which, essentially, throw the following simple pattern:\footnote{Exceptions to this rule are  limits with no 6d EFT regime, in the sense that $\Lambda_{\rm sp} = m_{{\rm KK},6}$. There $\nu=1$ corresponds to a 5d LST and $\nu =2,3$ both describe a 5d SCFT. That is, we have a circle reduction of the previous dictionary.} 
\begin{center}
    $\nu =3$  \qquad 5d SCFT \\
    $\nu =2$  \qquad 6d SCFT \\
     $\nu =1$  \qquad \ \ 6d LST
\end{center}
where LST stands for a little string theory. We thus find a neat, direct connection between the asymptotic scalar curvature and the UV origin of the rigid theory that sources its divergence. 

The rest of the paper is organised as follows. Section \ref{s:typeIIA} describes the type IIA large volume limits that we analyse. Section \ref{s:decoupling} explains why we quantify asymptotic curvature divergences via \eqref{Rasintro}, and how we define the UV rigid theory for each limit. Section \ref{s:Ftheory} analyses the curvature divergences and the UV rigid theories for smooth elliptic fibrations, while section \ref{s:nonsmooth} extends this analysis to non-smooth ones. Section \ref{s:typesw1} classifies the different emergent string limits, and section \ref{s:LSTs} analyses the asymptotic curvature and UV rigid theories for each of them. Finally, section \ref{s:summary} summarises our findings and in section \ref{s:conclu} we draw our conclusions.


\section{Type IIA large volume limits and the moduli space curvature}
\label{s:typeIIA}

Let us review the infinite-distance limits analysed in  \cite{Marchesano:2023thx}. One first considers type IIA string theory compactified on a Calabi--Yau three-fold $X$ 
\be
ds^2 = ds_{\R^{1,3}}^2 + ds_{X}^2 \, ,
\label{comp}
\ee
whose Ricci-flat metric is determined by the periods of its holomorphic three-form $\Omega$ and its K\"ahler form $J$. The latter is  complexified by the internal B-field components. Together they are expanded as
\be
J_c = B + i J = T^a \om_a \, , \qquad a =1, \dots, h^{1,1}(X) ,
\ee
where $T^a = b^a + i t^a$ and $\ell_s^{-2} \omega_a$ is a basis of integral harmonic two-forms Poincar\'e dual to a basis of Nef divisors, with $\ell_s = 2\pi \sqrt{\a'}$ the string length. The $T^a$ parametrise the moduli space of the vector multiplet sector of the 4d $\CN=2$ EFT obtained upon dimensional reduction, whose Lagrangian reads  \cite{Ferrara:1988ff,Andrianopoli:1996cm,Lauria:2020rhc}
\be
S_{\rm 4d}^{\rm VM} =  \frac{1}{2\kappa_{4}^2} \int_{\R^{1,3}} R * \mathbbm{1} - 2 g_{ab} dT^a \wedge * d\bar{T}^{\bar{b}} +  \oh \int_{\R^{1,3}}  I_{AB} F^A \wedge *_4 F^B + R_{AB} F^A \wedge F^B ,
\label{SVM}
\ee
with $A = (0,a)$. Here $F^A$ are integrally-quantised two-form field strengths associated to the U(1) gauge bosons, with $F^0$ the graviphoton, see \cite{Marchesano:2022axe,Marchesano:2023thx}  for their precise definition. The moduli space metric in the large-volume regime is specified by
\be
g_{ab} = \frac{3}{2} \left( \frac{3}{2} \frac{\CK_a \CK_b}{\CK^2} - \frac{\CK_{ab}}{\CK} \right)  ,
\label{metric}
\ee
with $\CK_{abc} = \ell_s^{-6} \int \omega_a\wedge \omega_b \wedge \omega_c$ the triple intersection numbers of $X$, from where we build $\CK_{ab} = \CK_{abc} t^c$, $\CK_{a} = \CK_{abc} t^bt^c$ and $\CK =  \CK_{abc} t^at^bt^c = 6 {\cal V}_{X}$. The gauge couplings are  
\be
I\, =\, - \frac{\CK}{6}
\left(
\begin{array}{cc}
1 +   4 g_{ab} b^ab^b & 4 g_{ab} b^b \\ 
4 g_{ab} b^b & 4 g_{ab}
\end{array}
\right) , \qquad
R\, =\, -
\left(
\begin{array}{cc}
 \frac{1}{3} \CK_{abc}b^ab^bb^c  & \oh  \CK_{abc}b^bb^c 
 \\  \oh  \CK_{abc}b^bb^c 
 & \CK_{abc}b^c 
\end{array}
\right) . 
\label{IandR}
\ee
These expressions are a large-volume approximation that receive curvature and world-sheet instanton corrections. The corrections are encoded in the full prepotential 
\be
 {\cal F} = -\frac{1}{6} \cK_{abc}T^aT^bT^c + \oh K_{ab}^{(1)}T^aT^b + K_{a}^{(2)}T^a + \frac{i}{2} K^{(3)} + (2\pi i)^{-3} \sum_{\bm{k}} n_{\bm{k}}^{(0)} {\rm Li}_3 \left( e^{2\pi ik_aT^a} \right) ,
\label{fullF}
\ee
where the $K^{(i)}$ $i=1,2,3$ depend on topological data of $X$, namely
\begin{eqn}
K_{ab}^{(1)} = \frac{1}{2} {\cal K}_{aab} \, , \qquad K_{a}^{(2)} = \frac{1}{24\ell_s^6} \int_{X_6} c_2(X_6) \wedge \omega_a \, , \qquad
    K^{(3)} = \frac{\zeta(3)}{8\pi^3} \chi(X_6)\, .
    \label{Kcurv}
\end{eqn}
Moreover, Li$_3$ stands for the 3rd polylogarithmic function and $n_{\bm{k}}^{(0)}$ for the genus zero Gopakumar-Vafa (GV) invariant of the curve class $k_a {\cal C}^a$, with ${\cal C}^a$ such that $\ell_s^{-2} \int_{{\cal C}^a} \om_b = \delta^a_{b}$. In the large-volume regime the piece $\cF^{\rm cl} = -\frac{1}{6} \cK_{abc}T^aT^bT^c$ is the dominant one, and it is from this term that \eqref{metric} and \eqref{IandR} are derived. The corrections $K_{ab}^{(1)}$ and $K_a^{(2)}$ do not modify the moduli space metric, while $K^{(3)}$ and world-sheet instanton corrections do.\footnote{See e.g. \cite[Appendix B]{Marchesano:2022axe} for explicit expressions on how $K^{(3)}$ modifies the moduli space metric.} 
  We analyse their effect on the asymptotic moduli space  curvature in section \ref{s:Ftheory}.

Within this setup,  \cite{Marchesano:2022axe,Marchesano:2023thx} consider   infinite distance trajectories of the form
\be     \label{limits}
t^a = t^a_0 + e^a \phi, \quad \text{with} \quad \phi \to \infty ,
\ee
where $t^a_0 \gg 1$ represents a point in the large-volume region of the K\"ahler cone, and $\bm{e} = \{e^a \in \mathbbm{N}\}$ a Nef divisor class ${\cal D}_{\bm{e}} = \ell_s^{-2} e^a [\om_a]$. As pointed out in \cite{Lanza:2020qmt,Lanza:2021udy,Lanza:2022zyg}, by wrapping an NS5-brane on ${\cal D}_{\bm{e}}$ and computing the backreaction of the resulting 4d string, one recovers the field space trajectory \eqref{limits} when approaching the string core. To describe a trajectory purely in vector multiplet moduli space, the 10d dilaton must be rescaled as
\begin{equation}
g_s (\phi) = g_{s, 0}  \frac{{\cal V}_{X}^{1/2} (\phi)}{{\cal V}_{X,0}^{1/2}} \to \infty\, ,
\label{codila}
\end{equation}
taking us to a region of 10d strong coupling. As a result, there is a lattice of D($2p$)-branes wrapping $2p$-cycles of $X$ that are perceived from the 4d viewpoint as infinite towers of particles that become asymptotically massless along the limit, as predicted by the Distance Conjecture \cite{Ooguri:2006in}. The lightest  tower is always the one made up of D0-branes, whose mass scale $m_*$ behaves as
\begin{equation}
    \frac{m_*^2}{M_{\rm P}^2} \sim \left(\frac{\cal T}{M_{\rm P}^2} \right)^w \sim \phi^{-w} , \qquad w=1,2,3 \, ,
    \label{scaling}
\end{equation}
as we proceed along the limit. Here  $m_* \simeq m_{\rm D0}$, $M_{\rm P} = \kappa_4^{-1}$ and ${\cal T}$ is the tension of the 4d string made up from wrapping and NS5-brane on ${\cal D}_{\bm{e}}$. The integer $w$ was dubbed scaling weight in \cite{Lanza:2020qmt,Lanza:2021udy,Lanza:2022zyg} and it plays a prominent role in the classification of these limits. In particular, in the language of \cite{Grimm:2018ohb,Grimm:2018cpv,Corvilain:2018lgw} it corresponds to the singularity type that describes the limit. Moreover, following \cite{Lee:2019wij} one can relate it to the physical nature of the descending tower of states.  

Indeed, while from the 10d viewpoint the above trajectories take us to a strong coupling regime, in 4d terms we are led to a weakly coupled region, which oftentimes has a simple description in terms of a dual frame. This is manifest when applying the classification made in \cite{Corvilain:2018lgw,Lee:2019wij} to the above set of limits, which gives the following result:

\begin{enumerate}

    \item $w=3$ limits correspond to the case where ${\bf k} \equiv \CK_{abc}e^ae^be^c\neq 0$. These were dubbed type IV$_d$ limits in \cite{Corvilain:2018lgw}, and signal the decompactification to a 5d theory \cite{Lee:2019wij}. Their dual description is M-theory compactified on $S^1 \times X$, where the $S^1$ radius tends to infinity. 
    
    \item $w=2$ limits occur when ${\bf k} = 0$ and ${\bf k}_a \equiv \CK_{abc} e^be^c \neq 0$ for some $a$. They were dubbed type III$_c$ limits in \cite{Corvilain:2018lgw} and J-Class A limits in  \cite{Lee:2019wij}, are decompactification limits to a 6d theory. Their dual description consists of F-theory compactified on ${\bf T}^2  \times X$, with ${\bf T}^2$ taken to a decompactification limit. 
 
    \item $w=1$ limits happen whenever ${\bf k}_a  = 0$, $\forall a$. They were dubbed type II$_b$  limits in \cite{Corvilain:2018lgw} and J-Class B limits in \cite{Lee:2019wij}. Their dual description features a weakly-coupled,  asymptotically tensionless critical string \cite{Lee:2019wij}. 
\end{enumerate}

The subindices in II$_b$, III$_c$, and IV$_d$ correspond to non-negative integers that classify subclasses of limits. More precisely, it follows from  \cite{Corvilain:2018lgw} that $b=c+2 =d =r$, where
\be
r= {\rm rank}\, \K , \qquad \K_{bc} \equiv e^a \CK_{abc} \, .
\label{rank}
\ee
The geometric and physical meaning of this subindex is crucial for the analysis performed in \cite{Marchesano:2023thx}, directed to detect asymptotic divergences in the moduli space curvature. Geometrically, when $r < h^{1,1}(X)$ there are non-Nef divisors in $X$ whose volume remains constant along the limit, and that are represented by the elements of $\ker {\bf K}$. If  such divisors are effective then one can wrap NS5-branes on them, yielding non-critical strings whose tensions ${\cal T}_{\rm nc}$ scale like ${\cal T}_{\rm nc}^{1/2} \sim m_*$ along the limit. The same applies to D4-branes wrapping such divisors or D2-branes wrapping curves within them, both perceived as 4d particles with a mass larger than $m_*$ but with the same scaling. As a result, all these objects will play a role in the dual frame describing this limit, either in terms of a higher-dimensional and/or emergent string theory. 

Physically, a non-trivial $\ker {\bf K}$ signals that some eigenvalues of the gauge kinetic matrix $\tilde{I}_{ab} = 4 {\cal V}_{X} g_{ab}$ remain asymptotically constant along the limit, while elements outside of this kernel correspond to couplings that vanish asymptotically. This observation reflects directly on the asymptotic behaviour of the 4d moduli space scalar curvature, whose expression in terms of the approximated metric \eqref{metric} reads
\be
R_{\rm IIA}^{\rm cl} =  - 2n_V(n_V + 1) + 2 \CV_X  \tilde{I}^{ab}\tilde{I}^{cd}\tilde{I}^{ef}  \CK_{ace} \CK_{bdf} \, ,
\label{scalarIIA}
\ee
where $n_V = h^{1,1}(X)$  stands for  the number of vector multiplets and $\tilde{I}^{ab}$ is the inverse of $\tilde{I}_{ab}$. Indeed, a set of asymptotically constant entries for $\tilde{I}^{ab}$ together with non-vanishing triple intersection numbers multiplying them implies a curvature that scales as ${\cal V}_X \sim \phi^w$ along \eqref{limits}. Alternatively, one can see $\tilde{I}_{ab}$ as the moduli space metric measured in units of $m_*$. The existence of a non-trivial kernel for ${\bf K}$ is then interpreted as a subset of vector multiplets that remain dynamical at this scale along the limit, while the rest decouple together with gravitational interactions, due to their diverging kinetic terms in  units of $m_*$. Below this scale one finds a rigid 4d $\cN =2$ field theory (RFT) with no charged matter and a field space curvature $R_{\rm rigid}^{\rm cl} = \frac{1}{2}  \tilde{I}^{\sig\rho}\tilde{I}^{\tau\eta}\tilde{I}^{\mu\nu}  \CK_{\sig\tau\mu} \CK_{\rho\eta\nu}$, with indices only running over elements of $\ker {\bf K}$. A curvature divergence can then be traced back to a parametric hierarchy between moduli space metric eigenvalues, and a non-vanishing  curvature $R_{\rm rigid}$ for the sector that correspond to the smallest eigenvalues. In terms of the latter, one can express the curvature divergence simply as follows:
\begin{equation}
    R_{\rm IIA}^{\rm div}  \sim \frac{M_{\rm P}^2}{m_*^2} R_{\rm rigid}\, .
    \label{Rasym}
\end{equation}

 In other words, there is an interacting field theory that decouples from gravity, which is the source of the divergence. It was proposed in \cite{Marchesano:2023thx} that having an interaction that is parametrically stronger than gravity, i.e. a  gravity-decoupling theory, is a necessary ingredient to generate an asymptotic divergence for the moduli space scalar curvature. In this sense, whenever the strength of gravity is comparable to all other interactions, one would expect to recover the proposal made in \cite{Ooguri:2006in} that the curvature should be asymptotically negative. 

The different elements of this picture  were analysed in detail in \cite{Marchesano:2023thx} for $w=3$ limits, which also illustrate the interplay between the geometric and physical interpretation of $\ker {\bf K}$. Indeed, from the perspective of  M-theory on $S^1 \times X$, $w=3$ limits correspond to a growing $S^1$ radius combined with a finite-distance trajectory in 5d vector multiplet moduli space. In this 5d frame the type IIA K\"ahler variables $t^a$ are mapped to  $M^a = t^a/{\cal V}_X^{1/3}$, so that  the Calabi--Yau volume remains constant along the limit, and the curves and divisors that in the type IIA frame had constant volume now shrink to zero size.  Additionally, the second piece in \eqref{scalarIIA} can be written as $2 \cI^{ab} \cI^{cd} \cI^{ef}  \CK_{ace} \CK_{bdf}$, where $\cI_{ab}$ is the 5d gauge kinetic function. It follows that a curvature divergence in a type IIA $w=3$ limit corresponds, from the M-theory perspective, to a strong coupling, finite-distance boundary in 5d vector multiplet moduli space. By the analysis of \cite{Witten:1996qb} this occurs whenever an effective, contractible divisor ${\cal D}$ shrinks to a point and hosts a 5d SCFT. Such a divisor is precisely an element of $\ker {\bf K}$, and an M5-brane wrapped on it yields a 5d non-critical string charged under the 5d SCFT. Upon circle compactification one recovers the set of NS5-branes, D4-branes and D2-branes wrapped on the divisor and its curves, which are charged objects under the 4d RFT but with a tension above the 4d cut-off $m_*$. Finally, the topology of the (generalised del Pezzo) shrinkable divisor  implies that its  triple intersection numbers are non-trivial, such that $R_{\rm rigid} \neq 0$ and there is a curvature divergence. To sum up, a curvature divergence signals the presence of a rigid 4d theory that descends from a 5d SCFT, linked to each other via the massive spectrum of particles and strings that lie above the 4d cut-off $m_*$, at a constant scale in units of $m_*$ set by the initial point $\bm{t}_0$ of the limit.  

\begin{figure}[!t]
  \centering
\includegraphics[width=.75\linewidth]{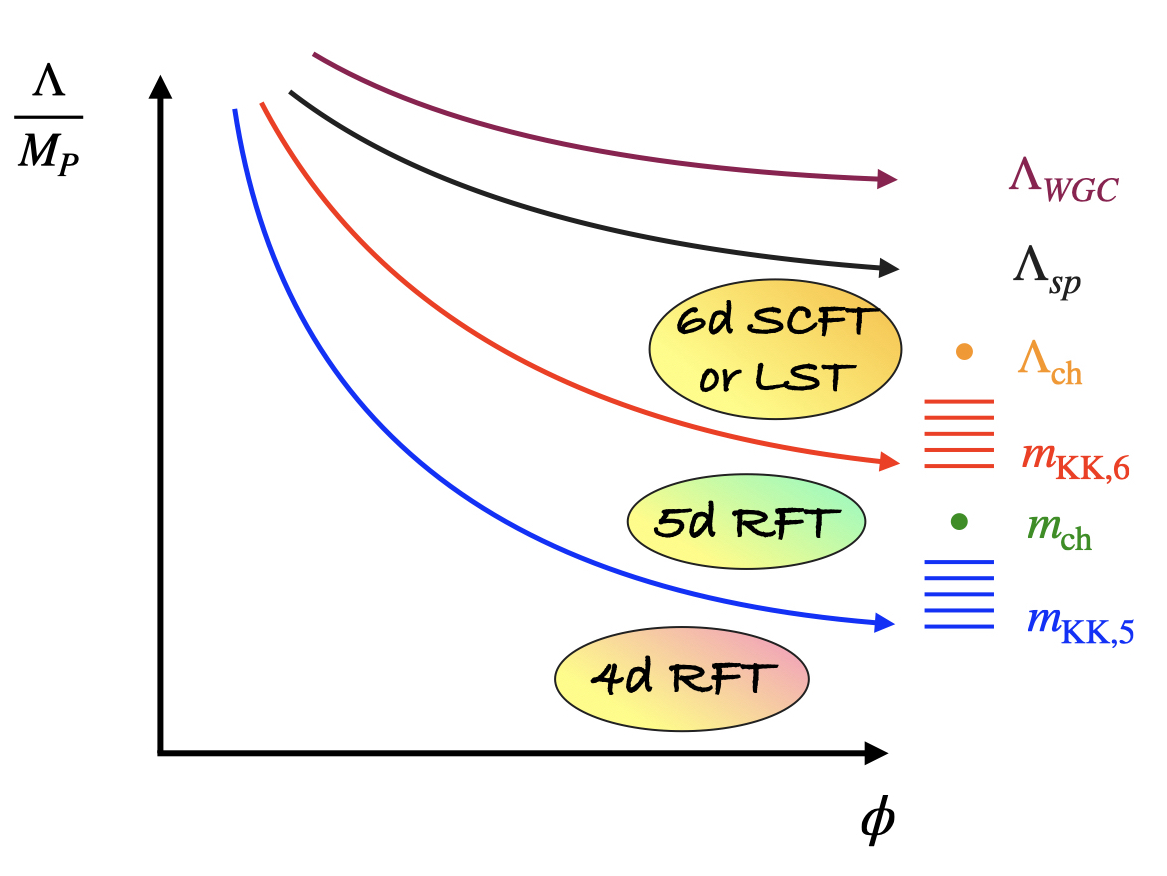}
\caption{Scales and rigid theories along $w=2$ and $w=1$ large volume limits. Here $m_{{\rm KK},5} = m_*$ stands for the 5d KK scale and $m_{{\rm KK},6}$ for the 6d KK scale, while $m_{\rm ch}$ represents an isolated state charged under the 4d RFT gauge group, and $\Lambda_{\rm ch}$ a tower of states with arbitrary charge under it. Depending on the limit $\Lambda_{\rm ch}$ can be above, at or below $m_{{\rm KK},6}$, and it is at this scale where the UV rigid theory is defined. }
\label{fig:genscales}
\end{figure}

In the following we extend this picture to $w=2$ and $w=1$ limits, by analysing the 4d RFTs that survive along these limits and the  theories that appear above $m_*$. Since we are considering type IIA large volume limits one always has a 5d M-theory description for them, with the difference that for $w = 2$ or $w = 1$ limits the trajectory in terms of 5d vector multiplet moduli space is also of infinite distance. As a result, one can again link the 4d RFT with a parent 5d theory that appears above $m_*$, and in most cases also with a  6d theory at energies just below the species scale $\Lambda_{\rm sp}$, see figure \ref{fig:genscales} for a schematic illustration. Just like the 4d rigid theory, we will find that such parent 6d theories decouple from the 6d gravity sector along the limit, just like 6d SCFTs  or 6d LSTs  do when geometrically engineered in F-theory compactifications \cite{Heckman:2013pva,Heckman:2015bfa,Heckman:2018jxk,Bhardwaj:2015oru,DelZotto:2023ahf}. 

Following the nomenclature in \cite{Lanza:2020qmt,Lanza:2021udy,Lanza:2022zyg} we dub the large-volume trajectories \eqref{limits} as EFT string limits. These are the simplest large-volume limits that one may look at to build moduli space trajectories where the scalar curvature diverges. However,  for our current purposes it will be useful to consider a more general set of limits based on {\em growth sectors}, in the sense of \cite{Grimm:2018cpv}, in which some K\"ahler moduli grow at a slower rate than others. To construct them, let us consider a set of vectors $\bm{e}_i$, $i =0 ,1,2, \dots$  with non-negative  entries and such that $\bm{e}_i \cdot \bm{e}_j = 0$, $\forall i \neq j$, where we are using the Cartesian product.\footnote{This is for instance the case for the so-called {\em elementary} EFT string limits \cite{Lanza:2021udy}, that generate the K\"ahler cone of $X$, and where only one component of $\bm{e}$ is non-vanishing and equal to one.\label{ft:elementary}} Then one can follow the trajectory
\be
\bm{t} = \bm{e}_0 \phi  +  \bm{e}_1 \phi^{\g_1} +  \bm{e}_2 \phi^{\g_2} + \dots \, ,
\label{growth}
\ee
with $1> \g_1 > \g_2 > \dots$ Here we define ${\bf K}_{ab} \equiv \cK_{abc} e_0^c$, and the scaling weight $w$  classification is based on the leading vector $\bm{e}_0$ contracted with the triple intersection numbers of $X$. 

The advantage of these more general limits is that along them the 5d KK scale $m_*$ can be decoupled parametrically from the 6d KK scale, as in figure \ref{fig:genscales}. Additionally, one can distinguish the scale $\Lambda_{\rm ch}$, which is where a tower of states with arbitrary charges under the 4d RFT appears, in the spirit of the tower/sublattice WGC \cite{Heidenreich:2016aqi,Montero:2016tif,Andriolo:2018lvp,Heidenreich:2019zkl}. In the large volume limits under consideration $\Lambda_{\rm ch}$ lies above $m_*$, but they scale differently along \eqref{growth}, see figure \ref{fig:genscales}. In particular, for $w=2$ and $w=1$ limits of the form \eqref{growth}, $\Lambda_{\rm ch}$ can lie at, above or below the 6d KK scale, indicating whether the 4d RFT should be understood microscopically in terms of a 5d or a 6d rigid theory. As we will see, both the dimension and the nature of the higher dimensional rigid theory can be detected by the degree of divergence of the moduli space scalar curvature. 

\section{Gravity decoupling and UV rigid theories}
\label{s:decoupling}

One of the main results obtained in \cite{Marchesano:2023thx} is that asymptotic divergences in the type IIA moduli space scalar curvature are sourced by field theory subsectors that decouple from gravity along the limit. This lesson, which is captured in the Curvature Criterion, was mainly tested along EFT string limits of the form \eqref{limits}, where the geometric picture is simpler. 
For our present purposes, it will be important to extend the relation between curvature divergences and gravity decoupling to the limits \eqref{growth}. In the following we would like to make this notion more precise, by using a specific criterion that captures when a field theory subsector decouples from gravity, and arguing that this is a necessary condition for the scalar curvature to diverge. 

Let us first notice that, in the context of large volume limits in type IIA vector moduli space, \eqref{scaling} can be written as
\be
\frac{m_*}{M_{\rm P}} \sim {\cal V}_X^{-1/2}\, ,
\label{m*}
\ee
and that this relation also applies to the more general set of limits \eqref{growth}. Additionally, the species scale behaves as
\be
\frac{\Lambda_{\rm sp}}{M_{\rm P}} \sim \phi^{-1/2}\, .
\label{Lamsp}
\ee
This behaviour was obtained explicitly in the case of EFT string limits in \cite{Marchesano:2022axe}. Along them, the bound proposed in \cite{Martucci:2024trp} is saturated, since  ${\cal T} \sim M_{\rm P}^2/\phi$. The result can be extended for more general set of  limits \eqref{growth} via the dependence of the genus-one topological free energy along the limit \cite{vandeHeisteeg:2022btw}, which in the type IIA large-volume regime provides the relation $M_{\rm P}^2 /\Lambda_{\rm sp}^2 \sim \int_{X} c_2 \wedge J$. 

Let us now focus on $w=3$ limits. In this case the rigid prepotential has a cubic term, which implies that 
\be
R_{\rm rigid} \sim g_{\rm rigid}^6\, .
\label{RiGid}
\ee
Here $g_{\rm rigid}$ is the gauge coupling of the rigid theory measured in the IR, or an average if $\dim \ker {\bf K} > 1$. Since in addition we have that ${\cal V}_X \sim \phi^3$ we find that the rhs of \eqref{Rasym} reads
\be
 R_{\rm IIA}^{\rm div}  \sim \frac{M_{\rm P}^2}{m_*^2} R_{\rm rigid} \sim \left(\frac{\Lambda_{\rm WGC}}{\Lambda_{\rm sp}} \right)^{2\nu}\, ,
  \label{Rasym3}
\ee
where $\nu =3 = w$ matches the value of the scaling weight, and we have defined
\be
\Lambda_{\rm WGC} \equiv g_{\rm rigid} M_{{\rm pl},d}^{\frac{d}{2}-1}\, ,
\label{LWGC} 
\ee
for $d=4$, with $M_{\rm P} \equiv M_{{\rm pl},4}$. That is, $\Lambda_{\rm WGC}$ is the cut-off scale predicted by the magnetic Weak Gravity Conjecture \cite{Arkani-Hamed:2006emk}, applied to the subsector of the 4d EFT that is the rigid theory. 

One can apply the same reasoning to $w=2$ limits based on smooth elliptic fibrations over a base $B_2$, that grows along the limit like $\phi^2$. There one finds  ${\cal V}_X \sim {\rm Area}(T^2) \cdot {\rm Vol}(B_2) \sim g_{\rm rigid}^{-2} \phi^2$, where we have used the asymptotic metric discussed in \cite[section 4.1]{Marchesano:2023thx}. In addition one finds that the rigid prepotential has no cubic terms so, if the rigid curvature is non-vanishing, it is of the form $R_{\rm rigid} \sim g_{\rm rigid}^6  \cO (e^{-2\pi t_{\rm rigid}})$, see next section for more accurate expressions. Taking  all this into account one finds a relation of the form
\be
 R_{\rm IIA}^{\rm div}  \sim \frac{M_{\rm P}^2}{m_*^2} R_{\rm rigid} \lesssim \left(\frac{\Lambda_{\rm WGC}}{\Lambda_{\rm sp}} \right)^{2\nu}\, ,
  \label{Rasym2}
\ee
where now $\nu=2=w$, again matching the value of the scaling weight. These relations indicate  that in order to have an asymptotically divergent curvature one must impose that
\be
\frac{\Lambda_{\rm WGC}}{\Lambda_{\rm sp}} \to \infty
\label{necond}
\ee
along the infinite distance limit. Because $\Lambda_{\rm WGC}$ is the natural cut-off of the 4d RFT and $\Lambda_{\rm sp}$ is the quantum gravity cut-off, one can interpret this quotient as measuring the strength of the 4d RFT gauge interactions versus the strength of the gravitational interactions, as done in \cite{FierroCota:2023bsp} to identify gauge theories that effectively decouple from gravity. In this sense, the condition \eqref{necond} for a divergent scalar curvature amounts to imposing that gravity is parametrically weaker than a subset of gravity-decoupling gauge interactions.  This gives a nice quantitative realisation of the Curvature Criterion, when applied to certain directions in vector multiplet moduli space. Just like in the CC, \eqref{necond} is a necessary condition but not a sufficient one for a curvature divergence. 

Notice that \eqref{necond} is always satisfied for EFT string limits of the form \eqref{limits} and with a non-trivial kernel, because this implies that $g_{\rm rigid}$ is constant along the limit. However, it can also be applied to limits of the kind \eqref{growth} where such gauge couplings are not necessarily constant. This will be our strategy in the following sections, where we analyse growth-sector limits where \eqref{necond} is satisfied. As mentioned above, this more general class of limits allows us to separate the energy scales at which a 5d and 6d KK towers appear, as well as the tower of arbitrary charges under the 4d RFT. The mass scale $\Lambda_{\rm ch}$ of the latter allows us to identify a higher-dimensional analogue of the 4d RFT, that we dub as UV rigid theory (UVRT). Just like the 4d RFT, this higher-dimensional theory will be decoupled from gravity along the limit.

Indeed, one interesting feature of the scales involved in \eqref{necond} is that they are invariant upon circle reductions. For the species scale this is obvious, since for decompactification limits it is simply the quantum gravity cutoff of the higher-dimensional theory. For $\Lambda_{\rm WGC}$ this follows from 
\begin{equation}
    \frac{1}{g_{d-1}^2}=\frac{2\pi R}{g_d^2}\,,\qquad M_{{\rm pl}, d-1}^{d-3} = (2\pi R)\, M_{{\rm pl}, d}^{d-2}\,,
\end{equation}
where $g_{d}$ and $M_{{\rm pl},d}$ are, respectively, the gauge coupling and Planck scale of the $d$-dimensional theory and $g_{d-1}$ and $M_{{\rm pl},d-1}$ their analogs after dimensional reduction on a circle of radius $R$. Using that $w=2$ and $w=1$ large-volume limits can be described in terms of F-theory on ${\bf T}^2 \times X$, it follows that the UV rigid theory of any 4d RFT will also satisfy \eqref{necond}, and so it decouples from gravity along the infinite distance trajectory. From this viewpoint it is no surprise that the UV rigid theories that we find are 5d SCFTs, 6d SCFTs and LSTs. The relation between a 4d RFT and a UVRT depends on the massive spectrum that one observes at $\Lambda_{\rm ch}$, as illustrated  in figure \ref{fig:genscales}.  Remarkably, we  find that the nature of the UV rigid theory along a given limit is encoded in the degree of divergence of the moduli space scalar curvature, more precisely in the exponent $2\nu$ that appears in \eqref{Rasym2}. Essentially, whenever $\Lambda_{\rm sp}$ can be made larger than $m_{{\rm KK},6}$, one should understand $\nu=3$ as indicating that the UV rigid theory  is a 5d SCFT, while $\nu=2$ signals a 6d SCFT and $\nu=1$ a 6d LST. Unlike in the simple examples discussed above, we will see that the matching $\nu =w$ not always holds, but more generally one finds the inequality $\nu \geq w$.


\section{Smooth fibrations and their F-theory uplift}
\label{s:Ftheory}

The simplest class of $w=2$ limits can be built from a Calabi--Yau $X$ that is a smooth elliptic fibration over a base $B_2$, with a single section. Such limits were already analysed in \cite{Marchesano:2023thx}, where it was shown that they do not give rise to divergences for the scalar curvature of the moduli space, except in those cases in which worldsheet instanton effects contribute to $R_{\rm rigid}$. In this section we sharpen this result by linking it to the massive spectrum that lies between $m_*$ and the species scale $\Lambda_{\rm sp}$. A similar strategy will be applied to more involved settings in the following sections. 

\subsection{Asymptotic curvature}
\label{ss:asymp_curv}

To describe this kind of type IIA limits, let us denote by $\om_a = \{\om_E, \om_\a\}$ the basis of K\"ahler cone generators of $X$, where $\om_E$ is dual to the elliptic fibre and $\om_\a = \pi^*(\om_\a^{\rm b})$ are the pulled-back K\"ahler cone generators of the base $B_2$. In this basis, the triple intersection numbers read
\be
\cK_{EEE} = \eta_{\a\b}c_1^\a c_1^\b\, , \qquad \CK_{EE\a} = \eta_{\a\b} c_1^\b\, , \qquad \CK_{E\a\b} = \eta_{\a\b}\, , \qquad \CK_{\a\b\g} = 0\, ,
\ee
where $c_1(B_2) = c_1^\a \om_\a^{\rm b}$ and $\eta_{\a\b}$ is a symmetric matrix with signature $(1, h^{1,1}(B_2) -1)$. To obtain a $w=2$ EFT string limit in this setup, one simply needs to consider a vector $e^a=(0,e^\a)$, such that $\eta_{\a\b}e^\a e^\b \neq 0$, and take the trajectory \eqref{limits}. However, similarly to \cite[section 4.2]{Marchesano:2023thx} it will be useful to consider the more involved trajectory
\be
t^\a = t^\a_0 +  e^\a \phi + \phi^\del, \qquad t^E = t^E_0 + \phi^\g, \qquad \g, \del \in [0,1) .
\label{w2limit}
\ee
Given that
\begin{equation}    \label{basisw2}
    {\bf K} =
    \begin{pmatrix}
   \eta_{\a\b} e^\a c_1^\b & \eta_{\a\b} e^{\b} \\  \eta_{\a\b} e^{\b} & 0
    \end{pmatrix} ,
\end{equation}
there is a non-trivial $\ker {\bf K}$ associated to this limit whenever $h^{1,1}(B_2) >1$. This kernel is given by  the vectors $(0, f^\b)$ such that $\eta_{\a\b}e^\a f^\b = 0$. Following \cite[section 4.2]{Marchesano:2023thx}, one can compute the rigid curvature associated to the kernel and see that, at the classical level, $R^{\rm cl}_{\rm rigid} = 0$. This is because, due to the structure of the triple intersection numbers,  at leading order the prepotential \eqref{fullF} is only quadratic in the K\"ahler moduli $T^\mu$ that correspond to $\ker {\bf K}$. More precisely one finds 
\be
{\cal F}_{\rm rigid}^{\rm cl} = - \oh T^E c_{\mu\nu} T^\mu T^\nu \, ,
\label{Frigidw2}
\ee
where $c_{\mu\nu} \equiv \eta_{\a\b} f^\a_\mu f^\b_\nu$ is the intersection matrix restricted to elements of $\ker {\bf K}$. Since $T^E$ is not a dynamical field of the rigid theory, at this level one finds a quadratic prepotential for the 4d RFT. As a result the gauge couplings are field-independent and the  rigid field theory metric is flat,  leading to a vanishing rigid curvature $R_{\rm rigid}^{\rm cl} \equiv  \frac{1}{2}  \tilde{I}^{\sig\rho}\tilde{I}^{\tau\eta}\tilde{I}^{\mu\nu}  \p_{\sig}\p_{\tau}\p_{\mu} {\cal F}_{\rm rigid}^{\rm cl} [\p_{\rho}\p_{\eta}\p_{\nu} {\cal F}_{\rm rigid}^{\rm cl}]^* = 0$.  The perturbative $\alpha'$ corrections will not change this result, and so only the exponential terms coming from world-sheet instanton corrections can generate a contribution to the rigid prepotential ${\cal F}_{\rm rigid}$ with a non-vanishing third derivative. For this there must be a contribution to the last term in \eqref{fullF} that remains constant along the limit \eqref{w2limit} and involves a K\"ahler modulus of the 4d RFT. If this occurs we have $R_{\rm rigid} \propto \cO(e^{-2\pi t^\mu})$ and an asymptotic curvature divergence of the form
\be
R_{\rm IIA} \sim  \cO(e^{-2\pi t^\mu})  \phi^{2(1-\g)}\, ,
\label{diverinst}
\ee
where we have used the asymptotic behaviour ${\cal V}_X \sim t^E {\rm Vol}(B_2) \sim \phi^{2+\g}$ and $\tilde{I}_{\sigma\rho} \sim t^E \sim \phi^\g$ \cite{Marchesano:2023thx}.

To find out whether exponential corrections to ${\cal F}_{\rm rigid}^{\rm cl}$ exist, the key question is whether or not a vector $(0,f^\b) \in \ker {\bf K}$ corresponds to an effective divisor of $X$. If it does, then it  necessarily corresponds to a vertical non-Nef divisor ${\cal D}_f$ that intersects with $B_2$ along a contractible curve ${\cal C}_f$ of the base. Such a curve will remain of constant area along the limit \eqref{w2limit}, provided that one chooses $\del =0$ in \eqref{w2limit}. It can thus host non-trivial worldsheet instanton corrections to the prepotential.  Let us then distinguish three different cases:

\begin{itemize}

\item[{\it i)}] $\ker {\bf K}$ does not contain any effective divisor.

\item[{\it ii)}] $\ker {\bf K}$ only contains one effective divisor ${\cal D}_f = (0, f^\b)$, such that $\eta_{\a\b} c_1^\a f^\b = 0$. 

\item[{\it iii)}] $\ker {\bf K}$ contains at least one effective divisor ${\cal D}_f = (0, f^\b)$ such that $\eta_{\a\b} c_1^\a f^\b \neq 0$. 
    
\end{itemize}

We now argue that only in case ${\it iii)}$ there is a curvature divergence, for the choice $\delta=0$. Indeed, in case ${\it i)}$ even for the choice $\del =0$ there cannot be any effective curves of the base that remain of constant area along the limit \eqref{w2limit}, because if such curves existed there would be an effective vertical divisor that belongs to $\ker {\bf K}$. Since the instanton charges $k_a$ run over classes of effective curves, and any curve that can contribute to $\cF_{\rm rigid}$ must have a component along the base, all possible world-sheet contributions to $R_{\rm rigid}$ vanish along the limit. In case ${\it ii)}$ there will be a single base curve ${\cal C}_f = {\cal D}_f \cdot B_2$ that is effective and remains of constant area along the limit if $\delta =0$. Because it is contractible, it has the topology of a $\mathbb{P}^1$, and moreover  the condition $\eta_{\a\b} c_1^\a f^\b = 0$ implies that the vertical divisor ${\cal D}_f$ is a product, of topology $\mathbb{P}^1 \times {\bf T}^2$. As such, it represents a sector of extended supersymmetry that does not give rise to world-sheet instanton corrections.\footnote{Using the identity $c_2(X) = 12 \om_E \wedge c_1(B_2) + c_2(B_2) - c_1(B_2)^2$\cite{Friedman:1997yq}, one can show that the condition $\eta_{\a\b} c_1^\a f^\b = 0$ is equivalent to  $c_2 \cdot {\cal D}_f = 0$, which implies the presence of 4d ${\cal N} =4$ supersymmetry in the sector hosted by ${\cal D}_f$.} In other words, the GV invariants associated to ${\cal C}_f$ vanish, and so $\cF_{\rm rigid}$ remains quadratic. Finally, in case  ${\it iii)}$ the curve ${\cal C}_f = {\cal D}_f \cdot B_2$ is a $\mathbb{P}^1$ in the base, that remains of constant area along the limit \eqref{w2limit} for the choice $\del=0$. Additionally, by construction its area will depend on a dynamical field of the 4d RFT. Because in this case there is no supersymmetry enhancement preventing a worldsheet instanton corrections to ${\cal F}_{\rm rigid}^{\rm cl}$, one expects a curvature divergence to occur. 

\subsection{Massive spectrum and UV rigid theory}

It is instructive to compare this result with the spectrum of states between $m_*$ and the species scale, which is given by $\Lambda_{\rm sp} \sim M_{{\rm pl},6} \sim \phi^{-1/2} M_{\rm P}$, with $M_{{\rm pl},6} = \ell_6^{-1}$ the six-dimensional Planck scale \cite{Marchesano:2022axe}. First, there are two Kaluza--Klein towers, signaling the decompactification to 5d and 6d, made up from D0-branes and from D2-branes wrapped on the fibre, respectively \cite{Corvilain:2018lgw,Lee:2019wij}. Using that ${\cal V}_X \sim \phi^{2+\g}$, one obtains the following mass scalings for these KK-towers
\be
m_{{\rm KK}, 5} \sim \phi^{-1-\frac{\g}{2}} M_{\rm P}\, , \qquad m_{{\rm KK}, 6} \sim  \phi^{-1+\frac{\g}{2}} M_{\rm P}\, .
\ee
 If ${\cal D}_f$ is not effective, all D2-branes wrapping a base curve have a mass that scales like $m_{\rm D2} \sim \phi^{ - \frac{\g}{2}} M_{\rm P}$, and so they lie parametrically above $m_{{\rm KK}, 6}$. Instead, if ${\cal D}_f$ is effective, we have a few candidates that can lie below such a tower. Let us consider a D2-brane wrapped on ${\cal C}_f$, a D4-brane wrapped on ${\cal D}_f$ and an NS5-brane on ${\cal D}_f$. For the first one, its mass always scales as
\be     \label{mD2,Cf}
m_{{\rm D2}, {\cal C}_f} \sim \phi^{-1-\frac{\g}{2} + \del} M_{\rm P}\, , 
\ee
and so it lies in between $m_{{\rm KK}, 5}$ and $m_{{\rm KK}, 6}$ whenever $\g \geq \del$. For the other two, we need to take into account that in case {\it ii)} Vol$({\cal D}_f) = t^E$Area$({\cal C}_f)$, while in case {\it iii)} Vol$({\cal D}_f) = t^E$Area$({\cal C}_f) + [\eta_{\a\b}c_1^\a f^\b] (t^E)^2$. From here one obtains:

\begin{subequations}
\label{w2cases}
\begin{eqnarray}
\label{w2case2}
    {\it ii)}  : & \quad &  m_{{\rm D4}, {\cal D}_f} \sim \phi^{-1+\frac{\g}{2} + \del} M_{\rm P}\,  , \qquad T_{{\rm NS5}, {\cal D}_f}^{1/2}  \sim \phi^{-1 + \frac{\del}{2}} M_{\rm P} \, ,  \\
    {\it iii)} \ \text{with} \ \g < \del : & \quad &  m_{{\rm D4}, {\cal D}_f} \sim \phi^{-1+\frac{\g}{2} + \del} M_{\rm P}\,  , \qquad T_{{\rm NS5}, {\cal D}_f}^{1/2}  \sim \phi^{-1 + \frac{\del}{2}} M_{\rm P} \, , 
     \label{w2case31} \\
    {\it iii)}\ \text{with} \  \g \geq \del :  & \quad &  m_{{\rm D4}, {\cal D}_f} \sim \phi^{-1+\frac{3\g}{2}} M_{\rm P}\,, \qquad T_{{\rm NS5}, {\cal D}_f}^{1/2}  \sim \phi^{-1 + \frac{\g}{2}} M_{\rm P}\,  .
\label{w2case32}
\end{eqnarray}
\end{subequations}
Notice that \eqref{w2case2} and \eqref{w2case31} have the similar scalings and satisfy the relations
\be
m_{{\rm D2}, {\cal C}_f} \simeq \frac{T_{{\rm NS5}, {\cal D}_f}}{m_{{\rm KK}, 6}}\, , \qquad 
m_{{\rm D4}, {\cal D}_f} \simeq \frac{T_{{\rm NS5}, {\cal D}_f}}{m_{{\rm KK}, 5}}\, .
\label{classt}
\ee
For \eqref{w2case32} only the last relation is satisfied, while the first one is corrected by $\phi^{\del-\g}$ on its rhs. The interpretation for this relation is the following: From the perspective of F-theory on ${\bf T}^2 \times X$ all these objects come from a single object in F-theory, namely a D3-brane wrapped on ${\cal C}_f$ that gives rise to a non-critical 6d string. When descending to 4d, one can wrap this string in different ways. If the string wraps the 6d circle of ${\bf T}^2$ one obtains a 4d particle dual to a type IIA D2-brane on ${\cal C}_f$, and if it wraps the 5d circle of ${\bf T}^2$ the particle in 4d is dual to a D4-brane on ${\cal D}_f$. Finally if the string does not wrap the ${\bf T}^2$ at all one obtains an NS5-brane on ${\cal D}_f$. The relations \eqref{classt} are then obtained from a classical dimensional reduction of a 6d string on a ${\bf T}^2$. The fact that they are not satisfied signals the presence of quantum effects that become more important as we approach a phase transition, as in \cite{Klemm:1996hh}. 

For $\g >0$ the 5d and 6d KK scales are asymptotically separated. Above $m_{{\rm KK}, 5}$  one obtains a 5d theory, while above $m_{{\rm KK}, 6}$ the theory is effectively six-dimensional. Just like the 4d RFT decouples from the 4d gravity sector, above $m_{{\rm KK},5}$ we expect a 5d sector that decouples from 5d gravity, and analogously above the 6d KK scale. In the following we will describe what these rigid theories are, focusing on the UV rigid theory that one  encounters at the scale $\Lambda_{\rm ch}$, where a tower of charges under the 4d rigid gauge group appears.

\begin{figure}[!t]
  \centering
\includegraphics[width=0.9\linewidth]{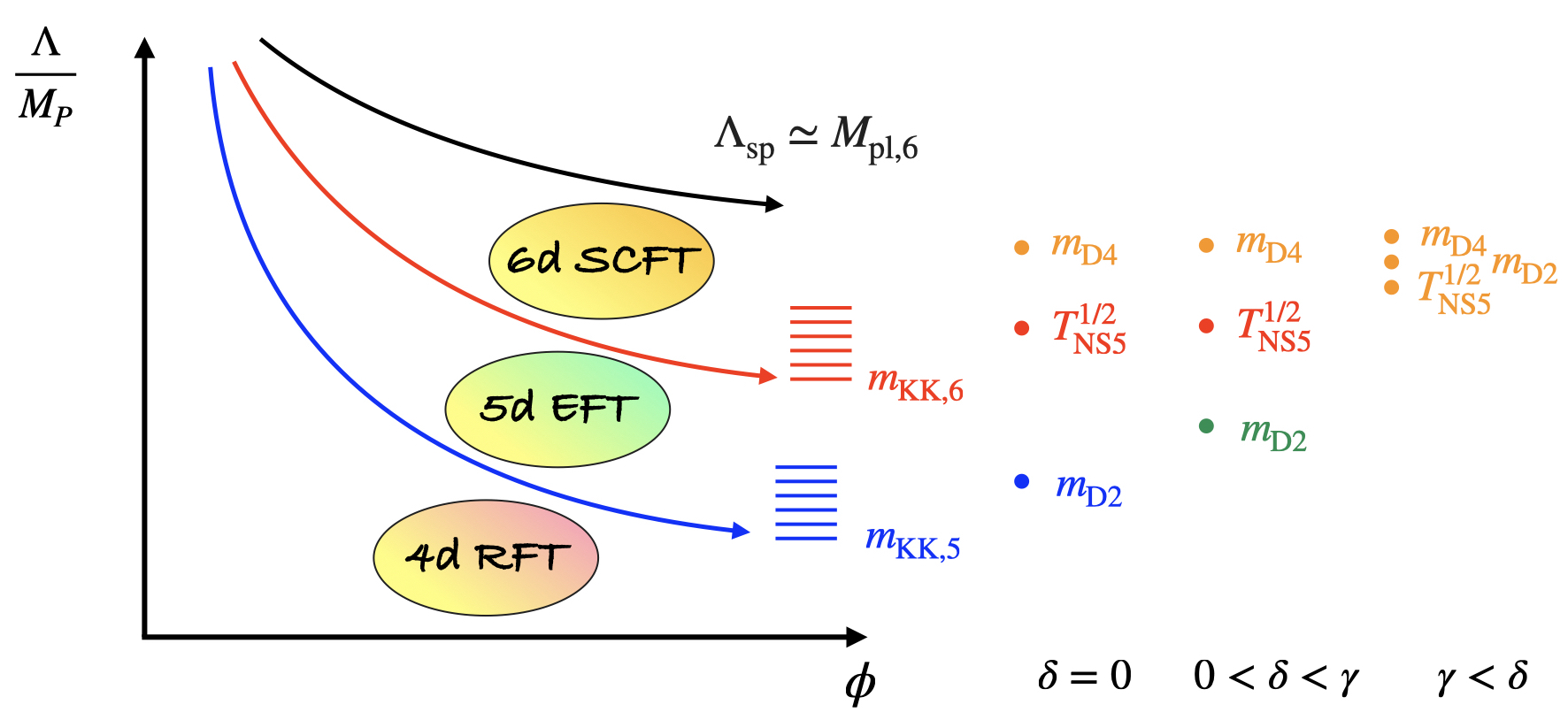}
\caption{Scales of the limit \eqref{w2limit}, case {\it iii)}. The UV rigid theory is defined at $\Lambda_{\rm ch}  \simeq {\rm max} \{m_{\rm D2}, m_{{\rm KK},6}\}$.} 
\label{w2scales}
\end{figure}

The most interesting class of limits corresponds to case ${\it iii)}$ and its different choices of parameters, as displayed in figure \ref{w2scales}. In this case we have  $\Lambda_{\rm ch} \simeq {\rm max} \{m_{{\rm D2},{\cal C}_f}, m_{{\rm KK},6}\}$, since at this scale bound states of D2-branes wrapping ${\cal C}_{f}$ and the elliptic fibre generate a cone of BPS states with arbitrary charges under the 4d rigid gauge group, see section \ref{ss:KMV} for an example.\footnote{An additional tower appears at $m_{{\rm D4}, {\cal D}_f}$ coming from the D4-brane on ${\cal D}_f$ with worldvolume fluxes. However, this always corresponds to a larger scale and so it does not play a role in our discussion.}

Let us first consider the choice $\g < \del$, which  corresponds to the standard F-theory limits in which the elliptic fibre is parametrically smaller than any curve in the base $B_2$. In M-theory variables $M^a = t^a/{\cal V}_X^{1/3}$ the fibre shrinks faster to zero size than any other curve, so one should be able to describe the  physics above $m_{{\rm KK},6}$ in terms of F-theory on $X$. Indeed, this choice gives a spectrum of the form \eqref{w2case31}, where all the states outside of the KK towers lie above $m_{{\rm KK},6}$ and so they can  be interpreted as a 6d non-critical string. Because there are only the two KK towers in between the 4d scale and the 6d scale, the 6d description of our system can be obtained by simply decompactifying the ${\bf T}^2$ and considering the 6d EFT of F-theory on $X$. As reviewed below, this is a simple 6d gravitational theory with a tensor branch, in terms of which the limit \eqref{w2limit} is mapped to a finite-distance trajectory. In the case at hand, the trajectory ends up in a 6d SCFT singular point  \cite{Heckman:2013pva,Heckman:2015bfa,Heckman:2018jxk}, where ${\cal C}_f$ shrinks to a point and the corresponding 6d non-critical string becomes tensionless. This 6d SCFT is precisely the subsector of the F-theory compactification that is selected by $\ker {\bf K}$, that  decouples from 6d gravity and displays a flat tensor branch metric $ds^2 = c_{\mu\nu} dj^\mu dj^\nu$. In the language of section \ref{s:decoupling}, this 6d SCFT corresponds to the UV rigid theory of the limit, defined at $\Lambda_{\rm ch} \simeq m_{{\rm D2},{\cal C}_f} > T_{{\rm NS5}, {\cal D}_f}^{1/2}$. Despite having a non-trivial UV rigid theory, we have seen that the 4d RFT does not yield a curvature divergence along this limit. The reason is that the rigid field theories below $T_{{\rm NS5}, {\cal D}_f}^{1/2}$ are too simple. Integrating out the 6d string modes one obtains a rigid field theory with the flat-metric tensor branch that describes the physics at $m_{{\rm KK},6}$. Upon dimensional reduction on a  ${\bf T}^2$ one recovers a 4d RFT with a flat metric, in agreement with the vanishing $R_{\rm rigid}$  obtained in our discussion above. 

A similar reasoning can be applied to the limits that correspond to case  {\it i)}, which also display the spectrum \eqref{w2case31}, but with the fixed value $\delta=1$. Here the difference is that $\Lambda_{\rm ch}$ lies above the species scale,  there is no tower of 4d RFT charges and the UV rigid theory is trivial. Correspondingly, the endpoint of the F-theory trajectory is at a nonzero distance from the 6d SCFT point, and all 6d non-critical strings are tensionfull. 

Let us now consider the range of parameters with $\del< \g$, that lead to the type IIA spectrum  \eqref{w2case32}. In this case we have that $m_{{\rm D2}, {\cal C}_f} < m_{{\rm KK},6}$ and so $\Lambda_{\rm ch} \simeq  m_{{\rm KK},6} \simeq T_{{\rm NS5}, {\cal D}_f}^{1/2}$. The 6d theory at this scale is described by F-theory on $X$, at the boundary of the K\"ahler cone where ${\cal C}_f$ is collapsed to a point. Our UV rigid theory is, as before, the 6d SCFT associated to the divisor ${\cal D}_f$, but now there is no trajectory in terms of F-theory on $X$: along the limit one simply sits at the origin of the 6d SCFT tensor branch, in agreement with the fact that $T_{{\rm NS5}, {\cal D}_f}^{1/2} \sim m_{{\rm KK}, 6}$ all along the trajectory. 

The 5d RFT that one recovers at lower energies is however more interesting than previously. In between the scales $m_{{\rm D2}, {\cal C}_f}$ and $ m_{{\rm KK},6}$ we have a hypermultiplet that must be integrated in the 5d theory, as  discussed in \cite{Witten:1996qb}. Then a cubic term appears in the prepotential of the 5d gravity-decoupled theory, that reads
\be
{\cal F}_{\rm rigid}^{\rm 5d} = -\oh M^E c_{\mu\nu} M^\mu M^\nu - \frac{n_{{\cal C}_f}}{2} (f_\mu M^\mu)^3\, .
\label{w25drigid}
\ee
where $f_\mu = \int_{{\cal C}_\mu} \om_a$ and $n_{{\cal C}_f}$ is the genus-zero GV invariant of the flop curve ${\cal C}_f$. Indeed, the 5d theory that one recovers above $m_{{\rm D2}, {\cal C}_f}$ is described by M-theory on $X$ at the locus where ${\cal C}_f$ is shrunk to zero size, which corresponds to a flop boundary in the K\"ahler moduli space of $X$.\footnote{Moving beyond the flop transition we obtain a different Calabi--Yau $\tilde{X}$ whose triple intersection numbers are
\be\nonumber
\cK_{abc}^{\tilde{X}} = \cK_{abc}^{X} - n_{{\cal C}_f} f_a f_b f_c
\ee
  One can understand this shift as the combined effect of integrating in the D2-brane on ${\cal C}_f$ and integrating out the D2-brane on $-{\cal C}_f$. The effect of just integrating-in the D2-brane on ${\cal C}_f$ on the couplings of the rigid theory therefore results in the 1/2 factor displayed in \eqref{w25drigid}. \label{ft:flop}} Such a cubic term however disappears when we descend to the 4d rigid theory, as we need to integrate out the D2-brane wrapped on ${\cal C}_f$. Particular attention deserves the case  $\del=0$, which implies  $m_{{\rm D2}, {\cal C}_f} \sim m_{{\rm KK}, 5}$. In this case,   when descending from 5d to 4d one will not only remove the cubic term from \eqref{w25drigid}, but because $m_{{\rm D2}, {\cal C}_f}/m_{{\rm KK}, 5}$ is a constant, integrating out the M2-brane will generate an instanton effect that can be computed  following \cite{Gopakumar:1998ii,Gopakumar:1998jq}. One then recovers a non-trivial exponential correction to the 4d prepotential
 \be
 \del {\cal F}_{\rm rigid}^{\rm 4d} \sim \cO \left(e^{-2\pi \frac{m_{{\rm D2}, {\cal C}_f}}{m_{{\rm KK}, 5}}}\right) \sim \cO \left(e^{-2\pi t^\mu}\right) \, ,
 \label{w2insta}
 \ee
 and from here an asymptotic curvature of the form \eqref{diverinst}, in agreement with the discussion above. That is, we see that \eqref{Rasintro} satisfied with $\nu = 2$ is linked to a UV rigid theory that is a 6d SCFT. 

Let us finally consider case {\it ii)}, which displays the spectrum \eqref{w2case2}. Here we again have that $\Lambda_{\rm ch} \simeq {\rm max} \{m_{{\rm D2},{\cal C}_f}, m_{{\rm KK},6}\} \geq T_{{\rm NS5}, {\cal D}_f}^{1/2}$, and so the UV rigid theory corresponds to a 6d SCFT associated to ${\cal D}_f$, which now has an extended ${\cal N}=2$ supersymmetry. At the level of the gravity-decoupled sector selected by $\ker {\bf K}$ one obtains a flat metric both for the 4d and 5d rigid theories. This is true even when  $\g > \del$ and $m_{{\rm D2},{\cal C}_f}$ and $T_{{\rm NS5}, {\cal D}_f}^{1/2}$ lie below the 6d KK scale. Indeed, from the 5d viewpoint they account for a charged hypermultiplet and its monopole string of the 5d rigid theory gauge group. Due to the extended supersymmetry of this configuration there is an additional vector multiplet charged under this group. Based on general results of rigid 5d $\cN =1$ field theories, one expects a change in the prepotential of the form
\be
\delta{\cal F}_{\rm rigid}^{\rm 5d} = \oh (n_V - n_H) (f_\mu M^\mu)^3
\label{rule5d}
\ee
where $n_H$ is the number of hypers and $n_V$ the number of vectors that one integrates in. Notice this indeed reproduces the second term in the rhs of \eqref{w25drigid}. In the present case, however, $n_V = n_H$ and there is no change in the 5d prepotential. Geometrically, this limit approaches an $su(2)$ boundary of the M-theory moduli space, such that above $T_{{\rm NS5}, {\cal D}_f}^{1/2}$ one expects an $su(2)$ enhancement in the 5d rigid theory. The enhanced supersymmetry of this sector also implies that, even for the choice $\del=0$, there will be no  prepotential term of the form \eqref{w2insta} when descending to 4d.

\subsection{The F-theory perspective}
\label{ss:F-th persp}

As mentioned above, the trajectories considered in this section can be interpreted as decompactification limits to a 6d theory, that can also be described by considering F-theory on ${\bf T}^2 \times X$. In this dual description the K\"ahler moduli of the base $B_2$ follow a path in moduli space that leaves ${\cal V}_{B_2} =$ Vol$(B_2)$ constant, while both radii of the ${\bf T}^2$ tend to infinity in 6d Planck units. The K\"ahler deformation of $B_2$ is seen as a finite-distance excursion in terms of the 6d EFT obtained from compactifying F-theory of $X$. This EFT only contains a tensor branch and has the following bosonic action 
\be
S_{\rm 6d}^{\rm TM} = \frac{2\pi}{\ell_6^4} \int_{\R^{1,5}} \left( R * \mathbbm{1} - g_{\a\b} \, dj^\a \wedge \ast dj^\b \right)\, .
\ee
Here $j^\a$, $\a=1,\dots , h^{1,1}(B_2)$, are real scalar fields and $g_{\a\b}$ is their moduli space metric given by
\be
g_{\a\b} = 2j_\a j_\b - \eta_{\a\b}\, ,
\ee
with $j_\a= \eta_{\a\b}j^\b$. The number of tensor multiplets is $n_T = h^{1,1}(B_2)-1$, as one  combination of the $j^\a$ is fixed by the constraint
${\cal V}_{B_2 }\equiv  \eta_{\a\b}j^\a j^\b=1$.

It is illustrative to express the trajectory \eqref{w2limit} in this 6d frame, by building a dictionary between type IIA and F-theory variables. Following \cite{Grimm:2013oga} let us first change the above K\"ahler cone basis to $\{\om_0, \om_\a\}$, where $\om_0 = \om_E -\frac{1}{2} c_1^\a \om_\a$. In this basis the  K\"ahler saxions are
\be     \label{shift_kahler}
t^0 = t^E\, , \qquad \tilde{t}^\a = t^\a + \frac{1}{2} c_1^\a t^E\, .
\ee
Using these, the F-theory fields can be expressed as
\be     \label{F-th_var}
2\pi R_5 = (t^E \CV_X)^\frac{1}{4} \, , \qquad 2\pi R_6 = [{\cal V}_X/(t^E)^{3}]^{1/4} \, , \qquad j^\a = \frac{1}{\sqrt{2}}  \tilde{t}^\a \sqrt{t^E/{\cal V}_X}\, ,
\ee
where $R_5$ and $R_6$ are the two radii of the ${\bf T}^2$ measured in units of $\ell_6 = M_{{\rm pl},6}^{-1}$. It is then easy to see that asymptotically the  trajectory \eqref{w2limit} translates into
\begin{align}     \label{6dlim}
R_5 \sim \phi^{\oh (1+\g)} & \, , \qquad  R_6 \sim \phi^{\oh (1-\g)}\, , \\\nonumber
j^\a = \frac{1}{\sqrt{\eta_{\bm{e}}}} e^\a + \frac{1}{2\sqrt{\eta_{\bm{e}}}} \left( c_1^\a - \frac{\eta_{\l \b}e^\l c_1^\b}{\eta_{\bm{e}}} e^\a \right) & \phi^{-(1-\g)} + \frac{1}{\sqrt{\eta_{\bm{e}}}} \left( 1 - \frac{ \sum_{\l} \eta_{\l \b} e^\b}{\eta_{\bm{e}}} e^\a \right) \phi^{-(1-\del)} + ...\, ,
\end{align}
where $\eta_{\bm{e}} = \eta_{\a\b}e^\a e^\b$. The tension of a D3-brane wrapping the curve ${\cal C}_f$ in the base then reads
\be
T_{{\rm D3},{\cal C}_f} = \ell_6^{-2} \eta_{\a\b} f^\b j^\a\,  M_{{\rm pl},6}^{2} =  \frac{\eta_{\a\b} f^\b}{\sqrt{\eta_{\bm{e}}}} \left( \oh c_1^\a \phi^{-(1-\g)} + \phi^{-(1-\del)} \right) M_{{\rm pl},6}^{2}\, .
\ee
Since, when wrapped on the 5d or 6d circles, the 6d string is dual to a D2 wrapping ${\cal C}_f$ and a D4 wrapping ${\cal D}_f$, respectively, we have $m_{{\rm D2},{\cal C}_f} = 2\pi R_6 \ell_6 T_{{\rm D3},{\cal C}_f}$ and $m_{{\rm D4},{\cal D}_f} = 2\pi R_5 \ell_6 T_{{\rm D3},{\cal C}_f}$. Using this we obtain the following scalings
\be
m_{{\rm D2}, {\cal C}_f} \sim \phi^{-\oh - \frac{\g}{2} + \del}  M_{{\rm pl},6} \, , 
\ee
and
\begin{subequations}
\begin{eqnarray}
    {\it ii)}  : & \quad &  m_{{\rm D4}, {\cal D}_f} \sim \phi^{-\oh+\frac{\g}{2} + \del}  M_{{\rm pl},6} \,  , \qquad T_{{\rm D3}, {\cal C}_f}^{1/2}  \sim \phi^{-\oh (1 - \del)} M_{{\rm pl},6} \, ,  \\
    {\it iii)} \ \text{with} \ \g < \del : & \quad &  m_{{\rm D4}, {\cal D}_f} \sim \phi^{-\oh+\frac{\g}{2} + \del}  M_{{\rm pl},6} \,  , \qquad T_{{\rm D3}, {\cal C}_f}^{1/2}  \sim \phi^{-\oh (1 - \del)} M_{{\rm pl},6} \, ,  \\
    {\it iii)}\ \text{with} \  \g \geq \del :  & \quad &  m_{{\rm D4}, {\cal D}_f} \sim \phi^{-\oh+\frac{3\g}{2}} M_{{\rm pl},6} \,, \qquad T_{{\rm D3}, {\cal C}_f}^{1/2}  \sim \phi^{-\oh (1 - \g)} M_{{\rm pl},6} \,  .
\end{eqnarray}
\end{subequations}
Finally, taking into account that $M_P^2 \sim  R_5 R_6 M_{{\rm pl},6}^2 \sim   \phi M_{{\rm pl},6}^2$, one can see that this result matches with \eqref{mD2,Cf} and \eqref{w2cases}. Let us stress that, in order to recover $m_{{\rm D2}, {\cal C}_f}$ in case $iii)$ with $\g \geq \del$, one needs to take into account the extra factor of $\phi^{\del-\g}$ due to quantum corrections.

From \eqref{6dlim} we see that we are taking a limit in 6d F-theory that corresponds to growing the curves of the base that intersect $e^\a$, while shrinking the rest. The curves that are shrinking correspond to the intersection of the divisors ${\cal D}_f \in \ker \bf{K}$ with the base and, since the volume of the base is kept constant, they are contractible curves in the base. These types of limits, in which some curves shrink with respect to the overall volume of the F-theory base, are gravity decoupling limits and have been widely studied to geometrically engineer 6d SCFTs, see \cite{Heckman:2018jxk} for a review. Each 6d SCFT is endowed with a negative definite intersection pairing matrix $A$, that defines the moduli space metric of the tensor branch of the SCFT. In this geometric engineering approach, this matrix corresponds to the intersection matrix of the contractible curves ${\cal C}_{f,\m}$ in the base, or equivalently the intersection matrix of the elements of the kernel ${\cal D}_{f,\m}$ with the base
\be
A_{\m\n} = {\cal C}_{f,\m} \cdot_{B_2} {\cal C}_{f,\n} = B_2 \cdot {\cal D}_{f,\m} \cdot {\cal D}_{f,\n} = c_{\m\n}\, .
\ee
From here and plain dimensional reduction on a ${\bf T}^2$, one then recovers the  prepotential \eqref{Frigidw2}.

\subsection{An example: the KMV conifold}
\label{ss:KMV}

We now illustrate the above discussion in a simple example, namely the KMV conifold, introduced in \cite{Klemm:1996hh} and recently  studied in the context of the Swampland Programme in \cite{Alim:2021vhs}. This geometry corresponds to a smooth elliptic fibration over an $\mathbb{F}_1$ base. The toric data and the GLSM matrix for this geometry are
\setlength{\arrayrulewidth}{0.2mm}
\renewcommand{\arraystretch}{0.8}
\begin{table}[H]
\begin{center}
\begin{tabular}{cccc|ccc}
& & & & ${\cal C}^1$ & ${\cal C}^2$ & ${\cal C}^3$ \\
\hline
1 & 0 & 0 & 0 & 0 & 0 & 1 \\
$-1$ & $-1$ & $-6$ & $-9$ & 0 & 0 & 1 \\
0 & 1 & 0 & 0 & 0 & 1 & 0 \\
0 & $-1$ & $-4$ & $-6$ & 0 & 1 & $-1$ \\
0 & 0 & 1 & 0 & 2 & 0 & 0 \\
0 & 0 & 0 & 1 & 3 & 0 & 0 \\
0 & 0 & $-2$ & $-3$ & 1 & $-2$ & $-1$
\end{tabular}
\end{center}
\label{tab:toric,GLSM,KMV}
\end{table}
\noindent Expanding the K\"ahler form in a basis of K\"ahler cone generators $J=t^a J_a$, $a=1,2,3$, the moduli $\{t^a\}$ are the volumes of the Mori cone generators $\{{\cal C}^a\}$ and we have the following triple intersection polynomial
\be
\cI(X) = 8 J_1^3 + J_1^2 \left( 3J_2 + 2J_3 \right) + J_1 \left( J_2^2 + J_2 J_3 \right) \, .
\ee
In this basis $t^1=t^E$ parametrises the volume of the elliptic fibre $\mathbf{T}^2$ and $\{t^\a\}=\{t^2,t^3\}$ parametrise the volumes of the fibre $\mathbb{P}^1_f$ and base $\mathbb{P}^1_b$ of $\mathbb{F}_1$, respectively. In this setup, the trajectory \eqref{w2limit} translates into
\be     \label{w=2lim_KMV}
t^1 \sim \phi^\g \, , \qquad t^2 \sim e^2 \phi + \phi^\del \, , \qquad t^3 \sim e^3 \phi + \phi^\del \, ,
\ee
with $e^2 >0$ and $e^3 \geq 0$, as required by the condition $\eta_{\a \b} e^\a e^\b \neq 0$. The matrix $\bf{K}$ for this limit reads
\be
\bf{K} =
\begin{pmatrix}
3e^2 +2e^3 & e^2 + e^3 & e^2\\
e^2 + e^3 & 0 & 0\\
e^2 & 0 & 0
\end{pmatrix} \, ,
\ee
and its kernel is generated by the vector $(0,f^\a) = \left( 0,e^2, -(e^2+e^3) \right)$. Since the effective divisors are given by positive integer combination of $\{J_1-2J_2-J_3, J_2-J_3, J_3\}$, one can check that the vector $(0,f^\a) \in \ker \bf{K}$ identifies an effective divisor ${\cal D}_f$ only in the case $e^3=0$, which we identify as the vertical divisor over the base curve ${\cal C}_f = \mathbb{P}^1_b$. In terms of the discussion in section \ref{ss:asymp_curv}, any  choice of the form $e^2,e^3 >0$ falls into case $i)$, while for $e^2>0, e^3=0$ we are in case $iii)$. Geometrically, this last case corresponds to growing the curve $\mathbb{P}^1_f$ in the base at the fastest rate, while the elliptic fibre and $\mathbb{P}^1_b$ are left constant or grow with slower rates (determined by $\g$ and $\del$, respectively). The objects that we are interested in, and whose spectrum is analysed and discussed in the previous sections, are a D2-brane wrapping the curve ${\cal C}_f$ and a D4-brane and NS5-brane wrapping ${\cal D}_f$, whose masses and tension are (here we set $e^3=0$ and $e^2=1$ for simplicity)
\be
\begin{split}
\frac{m_{{\rm D2},{\cal C}_f}}{M_{\rm P}} = \frac{\CV_{{\cal C}_f}}{\sqrt{\CV_X}} \simeq \frac{\sqrt{2} t^3}{\sqrt{t^E (t^2)^2}} \sim \phi^{\del-\gamma/2-1} \, ,\\
\frac{m_{{\rm D4},{\cal D}_f}}{M_{\rm P}} = \frac{\CV_{{\cal D}_f}}{\sqrt{\CV_X}} \simeq \frac{ (t^E)^2 + 2 t^E t^3}{\sqrt{2 t^E (t^2)^2}} \sim \phi^{-1+\frac{\g}{2}+\max\{\g,\del\}} \, ,\\
\frac{T_{{\rm NS5},{\cal D}_f}}{M_{\rm P}^2} = \frac{\CV_{{\cal D}_f}}{\CV_X} \simeq \frac{ (t^E)^2 + 2 t^E t^3}{t^E (t^2)^2} \sim \phi^{-2+\max\{\g,\del\}} \, .
\end{split}
\ee
We now show that, as argued above, only in case $iii)$ a divergence in the moduli space scalar curvature arises. As discussed, a divergent curvature signals the presence of a rigid field theory sector with $R_{\rm rigid}\neq 0$. This sector is identified by the components of the fields that correspond to $\ker \bf{K}$. Let us take a step back and consider some generic $e^\a \in \mathbb{Z}_{\geq 0}$. Then one can pick the following basis of divisors $\{J_E=J_1, J_e = e^\a J_\a, {\cal D}_f = f^\b J_\b\}$ and expand the K\"ahler form as
\be
J = t^E J_E + t^e J_e + t^f {\cal D}_f\, ,
\ee
Here $t^E$ controls the volume of the elliptic fibre, $t^e$ the volume of fastest growing base curve and $t^f$ the volume of the base curve that stays constant or grows slowest along the chosen limit. The modulus $t^f$ is the one that belongs to $\ker \bf{K}$ and that is dynamical in the rigid theory. One can then write down the rigid prepotential
\be
{\cal F}_{\rm rigid} = - \oh T^{E} c_{ff} (T^f)^2 + (2\pi i)^{-3} \sum_{\bf k} n_{\bf k} {\rm Li}_3 \left( e^{2\pi i k_a T^a} \right) \, ,
\ee
where $c_{ff} = \cK_{ffE} = \eta_{\a\b} f^\a f^\b$, $n_{\bm{k}}$ is the genus-zero GV invariant associated to the curve $k_a {\cal C}^a$ and the sum is restricted to curves with a non-vanishing component $k_f$ along the curve dual to ${\cal D}_f$. From here one can compute the rigid metric as
\be     \label{gff}
g_{f \bar{f}} = \text{Im} \left( \partial_{T^f}^2 {\cal F}_{\rm rigid} \right) = - c_{ff} t^E - \frac{1}{4\pi} \sum_{\bm{k}} n_{\bm{k}} k_f^2 \log \left[ 1- 2e^{-2\pi k_a t^a} \cos \left( 2\pi k_a b^a \right) + e^{-4\pi k_a t^a} \right]\, .
\ee
As already mentioned in section \ref{ss:asymp_curv}, the classical linear term in \eqref{gff} does not depend on the RFT moduli, implying that the classical rigid curvature is vanishing. However, the exponential corrections can generate a non-vanishing curvature. Notice that, in order for this to happen, there must exist a curve such that:
\begin{itemize}
\item $k_e = 0$, i.e., the curve does not have a component along the dual to $J_e$. Otherwise, the corresponding correction contains a term of the form $e^{-2\pi t^e}$ and vanishes asymptotically.
\item $k_f \neq 0$, i.e., the curve has a component along the dual to ${\cal D}_f$.
\item $n_{\bm{k}} \neq 0$, i.e., the GV invariant associated to the curve is non-vanishing.
\end{itemize}
Crucially, it turns out that these conditions can only be satisfied for the choice $e^3=0$ (and we set $e^2=1$ for simplicity), which corresponds precisely to case $iii)$, where ${\cal D}_f$ is an effective divisor. Given the GV invariants of the curves inside ${\cal D}_f$, e.g. in \cite{Alim:2021vhs}, one can check that there exists indeed a tower of states becoming light with arbitrary charge under the RFT sector. These states are given by D2-branes wrapping combinations of the curve ${\cal C}_f = \mathbb{P}^1_b$ and the elliptic fibre, such that  $\Lambda_{\rm ch} \simeq {\rm max} \{m_{{\rm D2},{\cal C}_f}, m_{{\rm KK},6}\}$. Thus, the sum in \eqref{gff} contains an infinite number of terms. However, at the level of instanton contributions to the prepotential a finite number of terms is sufficient to get a non-zero rigid curvature.  For instance, by keeping the only first contribution associated to the curve ${\cal C}^3 = \mathbb{P}^1_b$, with $-k_f = n_{\bm{k}} = 1$, one finds\footnote{For $\gamma>0$ this is, in fact, also the only contribution to the prepotential that does not vanish asymptotically.}
\be
R_{\rm rigid} \sim \frac{1}{2(t^E)^3} e^{-4\pi t^f}\, ,
\ee
where $t^f = t^3$ is the modulus of the RFT. Finally, notice that in the limit \eqref{w=2lim_KMV} this rigid curvature generates a divergence in the moduli space curvature of the gravitational theory only if $\del = 0$ and $t^f$ is constant. To see this explicitly, one can check that
\be
R_{\rm IIA}^{\rm div} \sim \CV_X R_{\rm rigid} \sim \frac{t^E (t^2)^2}{(t^E)^3} e^{-4\pi t^3} \sim \phi^{2(1-\g)} e^{-4\pi \phi^\del}\, ,
\ee
as anticipated in \eqref{diverinst}.

\section{Non-smooth fibrations and 6d SCFTs}
\label{s:nonsmooth}

Let us now consider the asymptotic behaviour of the curvature in a more general class of $w=2$ limits, namely in non-smooth elliptic fibrations over a base $B_2$.  Again, we consider a limit of the form \eqref{growth}, in which $\bm{e}_0$ satisfies the conditions of $w=2$ limits. This means that, if we interpret it as a divisor of $X$, it corresponds to a vertical divisor that projects to a Nef curve ${\cal C}_e^b$ of the base satisfying ${\cal C}_e^b\cdot_{B_2} {\cal C}_e^b\neq 0$. As in the smooth case, to have asymptotic divergences in the moduli space scalar curvature, one needs a non-trivial kernel for the matrix ${\bf K}_{ab} \equiv {\cal K}_{abc} e_0^c$. The question is then what are the vectors $f$ that belong to this kernel, and what is their geometric interpretation as non-Nef divisors of $X$. While in the case of smooth fibrations there are essentially two kinds of divisors, namely the shifted zero section ${\cal D}_E$ and the vertical divisors, for non-smooth fibrations we have a more involved divisor classification. These are:

\begin{itemize}

\item[-] Vertical divisors ${\cal D}_\a$. 

As in smooth fibrations, these are defined as pull-backs of divisors ${\cal C}_\a^{\rm b}$ of the base: ${\cal D}_\a = \pi^*({\cal C}_\a^{\rm b})$, $\a=1,...,h^{1,1}(B_2)$. That is, they can be thought of as fibrations of the generic elliptic fibre ${\cal E}$ over a base curve ${\cal C}_\a^{\rm b}$. If ${\cal C}_\a^{\rm b}$ is a non-Nef divisor that does not intersect ${\cal C}_e^b$, then ${\cal D}_\a$ will belong to $\ker {\bf K}$.

\item[-] Exceptional divisors ${\cal D}_I$. 

They arise from the resolution of singularities of the elliptic fibre along a divisor ${\cal C}_\a^{\rm b}$ in the base $B_2$. The generic fibre splits at these loci as ${\cal E} = {\cal E}_0 + \sum_I {\cal E}_I$  where, in the presence of a holomorphic zero section, ${\cal E}_0$ is the component of the fibre that is intersected by the zero section. A D2-brane wrapped on ${\cal E}_I$ corresponds to $W$-bosons of a non-Abelian gauge group with  algebra $\mathfrak{g}$. Since the ${\cal D}_I$ can be thought as fibering the component ${\cal E}_I$ over ${\cal C}_\a^{\rm b}$,  these divisors belong to $\ker {\bf K}$ if and only if the vertical divisor ${\cal D}_\a$ does as well.

\item[-] The shifted zero section ${\cal D}_0$. 

This is the divisor that we associate to the generic elliptic fibre. Following \cite{Grimm:2013oga,Grimm:2015wda},  one can relate this divisor to the zero section ${\cal D}_{\hat{0}}$ as ${\cal D}_0 = {\cal D}_{\hat{0}} - \oh \left( {\cal D}_{\hat{0}} \cdot {\cal D}_{\hat{0}} \cdot {\cal D}_\a \right) \eta^{\a\b} {\cal D}_\b$, with $\eta^{\a\b}$ the inverse of the intersection matrix $\eta_{\a\b}$ of the base $B_2$. Just like ${\cal D}_E$ in the  smooth case, this divisor never belongs to $\ker {\bf K}$ in $w=2$ limits.

\item[-] Additional rational sections $S_m$. 

Additional rational sections give rise to extra Abelian gauge theories of which a certain combination lifts to non-Cartan $U(1)$'s in 6d F-theory, see \cite{Weigand:2018rez,Cvetic:2018bni} for reviews. The correct linear combination of $U(1)$'s that does not have any admixtures from the KK-$U(1)$ or the Cartan $U(1)$'s is given by the Shioda map $\s(S_m)$. The gauge kinetic matrix for these $U(1)$ factors is 
\begin{equation}
    f_{mn} = - \int_{X} J \wedge \s(S_m) \wedge \s(S_n) + \frac{1}{4 \mathcal{V}_X} \left(\int_X J^2 \wedge \s(S_m) \right)\left(\int_X J^2 \wedge \s(S_n)\right) \,. 
\end{equation}
In the $w=2$ limits defined in \eqref{w2limit} the first term always scales like $\phi$ provided the height pairing $\pi_*[\s(S_m) \wedge \s(S_n)]$ is a non-contractible cycle in $B_2$, in which case the second term is sub-leading thanks to the properties of the Shioda map. It was shown in \cite{Lee:2018ihr} that, indeed, the height pairing for a rational section of an elliptically fibred Calabi-Yau threefold is not contractible. Therefore the divisor $\s(S_m)$ is never in the kernel of a $w=2$ limit which is consistent with the statement of \cite{Lee:2018ihr} that in F-theory Abelian gauge theories only survive as global symmetries in gravity decoupling limits. Therefore additional rational sections do not play a role in the following discussion. 

\item[-] Fibral divisors ${\cal D}_F$.

These arise when the fibration is non-flat, namely when the dimension of the fibre jumps over points of the base. As this happens one gets additional divisor classes, called fibral divisors, given by the fibres over such points. Non-flat fibrations are typically obtained upon flop transitions of curves of the base ${\cal C}_\a^{\rm b}$, with the fibral divisor being the image of the vertical divisor ${\cal D}_\a$ after the flop. An example of this transition is given by the flopped KMV conifold,  see section  \ref{ss:fibral}.

\end{itemize}

We refer the reader to \cite{Grimm:2013oga,Grimm:2015wda, Anderson:2016cdu} for a more detailed description of these divisors and their triple intersection numbers in flat fibrations. While in general we do not have an expression for the  intersection numbers in the Nef basis, we can still use our knowledge of these vertical divisors and the intuition developed in the last section to determine in which cases there is an asymptotic divergence of the scalar curvature. As in the case of smooth fibrations, this will be related to the kind of 4d RFT realised below $m_*$ and to its higher-dimensional UVRT. 

\subsection{Vertical divisors}

Let us first consider the case where only  vertical divisors belong to $\ker {\bf K}$.\footnote{The analysis of scaling weights made in  \cite{Marchesano:2023thx} for non-smooth elliptic fibrations corresponds to this case.} As in the smooth case, the 4d RFT only has the quadratic prepotential \eqref{Frigidw2} at the classical level, and one needs non-trivial worldsheet instanton corrections to generate a curvature divergence. For this, at least one divisor ${\cal D}_\a$ must be effective, so that the  intersection of ${\cal D}_\a$ with the base $B_2$ yields an effective curve ${\cal C}_\a^b$. The question is now whether or not ${\cal C}_\a^b$ hosts a non-trivial GV invariant. 

As we now argue, one essentially has the same possibilities as for a smooth fibration. Either ${\cal C}_\a^b$ corresponds to a base curve over which the elliptic fibration is trivial or it does not. In the first case ${\cal D}_\a \simeq \mathbb{P}^1 \times {\bf T}^2$, and we have a sector with extended supersymmetry, just like case {\it ii)} in section \ref{ss:asymp_curv}. The GV invariant for ${\cal C}_\a^b$ vanishes, and collapsing ${\cal C}_\a^b$ in the M-theory frame corresponds to reaching an $su(2)$ boundary.\footnote{Certain M-theory $su(2)$ boundaries are related to non-vanishing GV invarints, as recently analysed in \cite{Gendler:2022ztv}. However, the case at hand corresponds to $(g,N_F) = (1,0)$, displaying enhanced 4d ${\cal N} = 4$ supersymmetry.} Thus, no curvature divergence appears in this case.

The second case is analogous to case {\it iii)} in section \ref{ss:asymp_curv}:  Vol$({\cal D}_f)$ does not vanish when we collapse ${\cal C}_\a^b$ to a point, because 
\be
{\rm Vol} ({\cal D}_{\a})  =  {\rm Area} ({\cal C}_\a^b) \cdot {\rm Area} ({\cal E}) + n ({\rm Area} ({\cal E}))^2\, ,
\ee
for some $n \in \mathbb{Q}$. In the M-theory frame this correspond to a flop boundary, and hence to a non-vanishing GV invariant for ${\cal C}_\a^b$. The trajectory \eqref{w2limit} adapted to this case implies that
\be
{\rm Vol} (B_2) \sim \phi^2\, , \qquad {\rm Area} ({\cal E}) \sim \phi^\gamma\, , \qquad {\rm Area} ({\cal C}_\a^b) \sim \phi^\delta\,  \qquad \g, \del \in [0,1)\,.
\ee
As before, we have a curvature divergence as long as we choose $\delta =0$. The different choices for $\g$ and $\del$  will take us to the same scenarios as in figure \ref{w2scales}. For the choice $\g < \del$ we have $\Lambda_{\rm ch} = m_{{\rm D2}, {\cal C}_\a^b}$, while for $\del < \g$ we have $\Lambda_{\rm ch} = m_{\rm KK,6}$. In both cases, the 6d description is in terms of F-theory on $X$ and the UVRT is a 6d SCFT associated to the vertical divisor ${\cal D}_\a$, where the curve ${\cal C}_\a^b$ is shrunk to zero size. As we have explained at the end of subsection \ref{ss:F-th persp} for a smooth elliptic fibration, the parent 6d SCFT here contains as many tensor multiplets as vertical divisors in the kernel. One can extract the moduli space metric of the 6d tensor branch by computing the intersection matrix of such divisors with the base
\be
A_{\a\b} = B_2 \cdot {\cal D}_\a \cdot {\cal D}_\b = {\cal C}_\a^b \cdot_{B_2} {\cal C}_\b^b \, .
\ee

\subsection{Exceptional divisors}

Let us now consider the case for which an exceptional divisor ${\cal D}_{I}$ belongs to $\ker {\bf K}$. If this divisor projects down to the curve ${\cal C}_\m^{\rm b}$ in the base, it can only belong to $\ker {\bf K}$ if ${\cal C}_\m^{\rm b}$ does not intersect the  curve ${\cal C}_e^{\rm b}$ that determines the limit. This means that necessarily all the exceptional divisors associated to that curve and the vertical divisor ${\cal D}_\m$ also belong to $\ker {\bf K}$. Unlike in the previous case, this set of divisors has non-vanishing triple intersection numbers among themselves, which will generate cubic terms in the classical prepotential of the form
\be
{\cal F}_{\rm rigid}^{\rm cl} = - \oh T^{\cal E} c_{\mu\nu} T^\mu T^\nu - \oh \cK_{{\cal E}IJ} T^{\cal E} T^I T^J - \oh c_{\mu\mu} C_{IJ} T^\mu T^I T^J - \frac{1}{6} {\cal K}_{IJK} T^I T^J T^K  \, ,
\label{Frigidw2ex}
\ee
where $T^I$ is the complexified K\"ahler parameter dual to ${\cal D}_I$, $C_{IJ}$ is the coroot intersection matrix of the Lie algebra $\mathfrak{g}$ and $\cK_{IJK}$ are related to group theoretical information about the non-Abelian gauge group \cite{Bonetti:2011mw, Grimm:2011fx}. Here $T^{\cal E}$ is not a dynamical field of the rigid theory, but $T^I$ and $T^\m$ are dynamical. Note that we are using a divisor basis such that the gauge kinetic function and the moduli space metric are block diagonal at leading order, see \cite{Marchesano:2023thx} for details. The moduli $T^a$ are obtained by expanding the K\"ahler form in such a basis of divisors, for which the elements in general do not correspond to the K\"ahler cone generators. Suppose $\ker {\bf K}$ contains a vertical divisor ${\cal D}_\mu$ and an exceptional divisor ${\cal D}_I$ that are, respectively, elliptic and rational fibrations over a base curve ${\cal C}_\mu^{\rm b}$ with self-intersection $-1$. Thus, ${\cal D}_I$ implements an $\mathfrak{su}(2)$ enhancement along ${\cal C}_\mu^{\rm b}$. Then the classical rigid prepotential reads:
\be     \label{F_rig,su2}
{\cal F}_{\rm rigid}^{{\rm cl}, su(2)} = \oh T^{\cal E} (T^\mu)^2 - \oh \cK_{{\cal E}II} T^{\cal E} (T^I)^2 + T^\mu (T^I)^2 - \frac{1}{6} {\cal K}_{III} (T^I)^3 \, ,
\ee
and the corresponding rigid metric and its inverse in the saxionic coordinates $\{t^\m, t^I\}$ read
\be
\begin{split}
&g_{su(2),ab} =
\begin{pmatrix}
t^{\cal E} & 2 t^I \\
2 t^I & 2t^\m - \cK_{{\cal E}IJ} t^{\cal E} -\cK_{III} t^I
\end{pmatrix} \, ,\\
&g_{su(2)}^{ab} = \frac{1}{2 t^{\cal E} t^\m - \cK_{{\cal E}IJ} (t^{\cal E})^2 - \cK_{III} t^{\cal E} t^I - 4 (t^I)^2 }
\begin{pmatrix}
2 t^\m - \cK_{{\cal E}IJ} t^{\cal E} - \cK_{III} t^I & -2 t^I\\
-2 t^I & 
t^{\cal E}  
\end{pmatrix} \, .
\end{split}
\ee
From here it follows that the rigid curvature is
\be
R_{\rm rigid}^{{\rm cl}, su(2)} = \frac{t^{\cal E} \left( (\cK_{III}^2 -12 \cK_{{\cal E}II})(t^{\cal E})^2 + 12 \cK_{III} t^{\cal E} t^I + 96 (t^I)^2 + 24 t^{\cal E} t^\m \right)}{\left( 2 t^{\cal E} t^\m - \cK_{{\cal E}IJ} (t^{\cal E})^2 - \cK_{III} t^{\cal E} t^I - 4 (t^I)^2 \right)^3}\, ,
\ee
leading to a divergence via \eqref{Rasym} which is quadratic for the case of EFT string limits. To apply the philosophy of section \ref{s:decoupling} to this case, it is useful to consider more general limits based on growth sectors. Therefore consider the following scaling of the different curves:
\be\label{exceptionallimit}
{\rm Area} ({\cal C}_e^{\rm b}) \sim \phi\, , \quad {\rm Area} ({\cal C}_\mu^{\rm b}) \sim \phi^\delta\, , \quad {\rm Area} ({\cal E}) \sim \phi^\gamma\, , \quad
{\rm Area} ({\cal E}_I) \sim \phi^\beta \, , \quad   \b, \g, \del \in [0,1).
\ee
The scaling of ${\rm Area} ({\cal C}_e^{\rm b})$ is the same for any Nef curve of the base $B_2$, and it implies  ${\rm Vol} (B_2) \sim \phi^2$. The scaling of ${\rm Area} ({\cal C}_\mu^{\rm b})$ is that of any non-Nef curve of the base that does not intersect ${\cal C}_e^{\rm b}$, which in our $su(2)$ example amounts to one. Finally, Area(${\cal E}_I$) corresponds to the areas of the fibral curves ${\cal E}_I$, which constitute the exceptional divisors ${\cal D}_I$ when fibered over ${\cal C}_\m^b$. For consistency we must impose $\b \leq \g$. For simplicity we, however, restrict to the case in which all fibral curves scale in the same way, i.e., $\b = \g$. Later on we will illustrate the more general case with an example. Given the scalings \eqref{exceptionallimit} we expect to have
\be
t^{\cal E} \sim \phi^\g\, , \quad t^{\mu} \sim \phi^{\text{max}\{\g,\del\}}\, , \quad t^{I} \sim \phi^\g\, , 
\ee
which are found by inspection of explicit examples.\footnote{One should notice that the K\"ahler moduli $t^\m$ and $t^I$ in our basis are in general linear combinations of volumes of Mori cone generators and do not simply correspond to the areas of ${\cal C}_\m^b$ and ${\cal E}_I$.} From here it follows that 
\be
R_{\rm rigid}^{{\rm cl}, su(2)} \sim \phi^{-(\g + 2 \, \text{max}\{\g,\del\})} \implies R^{{\rm cl}, su(2)} \sim \phi^{2(1-\text{max}\{\g,\del\})}\, , 
\label{Rclasymex}
\ee
and the rigid theory gauge couplings scale like
\be
g_{{\rm rigid},\m} \sim \sqrt{g_{su(2)}^{\m\m}} \sim \phi^{-\g/2} \, , \qquad g_{{\rm rigid},I} \sim \sqrt{g_{su(2)}^{II}} \sim \phi^{-\oh \max\{\g,\del\}}\, .
\ee
This result saturates the inequality \eqref{Rasintro} with $\nu=w=2$. This suggests that the UVRT along this limit should correspond to a 6d SCFT with a gauge group $\mathfrak{g} = su(2)$.

To confirm this expectation, let us analyse the massive spectrum of the theory and the UVRT that one obtains above the 4d RFT scale. The asymptotic behaviour along \eqref{exceptionallimit} of the relevant mass scales reads as follows:
\be
\begin{split}
\frac{m_{\rm KK,5}}{M_{\rm P}} = \frac{1}{\sqrt{\CV_X}} \sim \phi^{-1-\g/2}\, , \qquad &\frac{m_{\rm KK,6}}{M_{\rm P}} = \frac{\CV_{\mathcal{E}}}{\sqrt{\CV_X}} \sim \phi^{-1+\g/2} \, ,\\
\frac{m_{{\rm D2}, {\cal C}_\mu^{\rm b}}}{M_{\rm P}} =\frac{\CV_{{\cal C}_\mu^{\rm b}}}{\sqrt{\CV_X}} \sim \phi^{\del-1-\g/2} \,, \qquad & \frac{m_{{\rm D2},{\mathcal{E}_I}}}{M_{\rm P}} = \frac{\CV_{\mathcal{E}_I}}{\sqrt{\CV_X}} \sim \phi^{\b-1-\g/2} \,, \\
\frac{T^{1/2}_{{\rm NS5},{\mathcal{D}_\mu}}}{M_{\rm P}} = \frac{\sqrt{\CV_{\mathcal{D}_\mu}}}{\sqrt{\CV_X}} \sim \phi^{\oh \max\{\g,\del\}-1}\, , \qquad &
\frac{T^{1/2}_{{\rm NS5},{\mathcal{D}_I}}}{M_{\rm P}} = \frac{\sqrt{\CV_{\mathcal{D}_I}}}{\sqrt{\CV_X}} \sim \phi^{\oh\b+\oh\max\{\b,\del\}-1-\g/2} \,, \\
\frac{m_{{\rm D4},{\mathcal{D}_\mu}}}{M_{\rm P}}  = \frac{{\CV_{\mathcal{D}_\mu}}}{\sqrt{\CV_X}}  \sim \phi^{\max\{\g,\del\}-1+\g/2}\,, \qquad 
&\frac{m_{{\rm D4}|_{\mathcal{D}_I}}}{M_{\rm P}}  = \frac{\CV_{\mathcal{D}_I}}{\sqrt{\CV_X}}  \sim \phi^{\b+\max\{\b,\del\}-1-\g/2} \,.
\end{split}
\ee
The main novelty with respect to previous cases is the presence of the scale  $m_{{\rm D2},{\mathcal{E}_I}}$, which is the mass of the $W$-boson of the $su(2)$ gauge theory. Above this scale, the $U(1)$ gauge group associated to the exceptional divisor $\mathcal{D}_I$ will be enhanced to $su(2)$. Notice that the restriction $\b \leq \g$ implies that $m_{{\rm D2},{\mathcal{E}_I}} \leq m_{\rm KK,6}$. Both in 5d and in 6d there will be a string associated to this non-Abelian sector, whose tension will be given by $T_{{\rm NS5},{\mathcal{D}_I}}$. Finally, $m_{{\rm D2}, C_\mu^{\rm b}}$ and $T_{{\rm NS5},{\mathcal{D}_\mu}}$ play the same role as in the case of a smooth elliptic fibration, with $m_{{\rm D2}, C_\mu^{\rm b}}$ signalling the scale where a flop curve shrinks to zero size.

Let us again consider the case $\b = \g$. Notice that this implies that all the components of the elliptic fibre parametrically scale in the same way. Futhermore, also the tensions of the NS5 branes wrapping ${\cal D}_\m$ and ${\cal D}_I$ have the same parametric scaling. As a consequence, this case is very similar to the smooth elliptic fibration and we have $\Lambda_{\rm ch} = \max \{ m_{{\rm D2},{\cal C}_\m^b}, m_{{\rm D2},{\cal E}_I} \}$, where $m_{{\rm D2},{\cal E}_I} \sim m_{\rm KK,6}$. The UVRT at this scale is still a 6d SCFT, associated to the kernel divisors ${\cal D}_\m$ and ${\cal D}_I$. But this time in addition to a tensor branch, we have a non-Abelian gauge group, realised by shrinking the curves ${\cal E}_I$ to zero size. When going down to 4d, integrating out all the massive objects, and restricting to the kernel one obtains a classical prepotential of the form \eqref{F_rig,su2}, where the cubic term in the modulus $T^I$ is responsible for the classical divergence of the moduli space curvature. In the case $\del=0$, one expects worldsheet instantons to correct the 4d prepotential that describes the 4d RFT below $m_{{\rm KK},5}$, along the lines of \eqref{w2insta}. However, in this case such corrections will be negligible for the asymptotic behaviour of the scalar curvature, compared to the classical divergence \eqref{Rclasymex}. As before, the moduli space metric of the 6d tensor branch is given by the intersection matrix of the kernel elements with the base
\be
A_{\m\n} = B_2 \cdot {\cal D}_\m \cdot {\cal D}_\n = {\cal C}_\m^b \cdot_{B_2} {\cal C}_\n^b \, .
\ee

\begin{figure}[!t]
\centering
\hspace{-1.2cm}
\includegraphics[width=1.\linewidth]{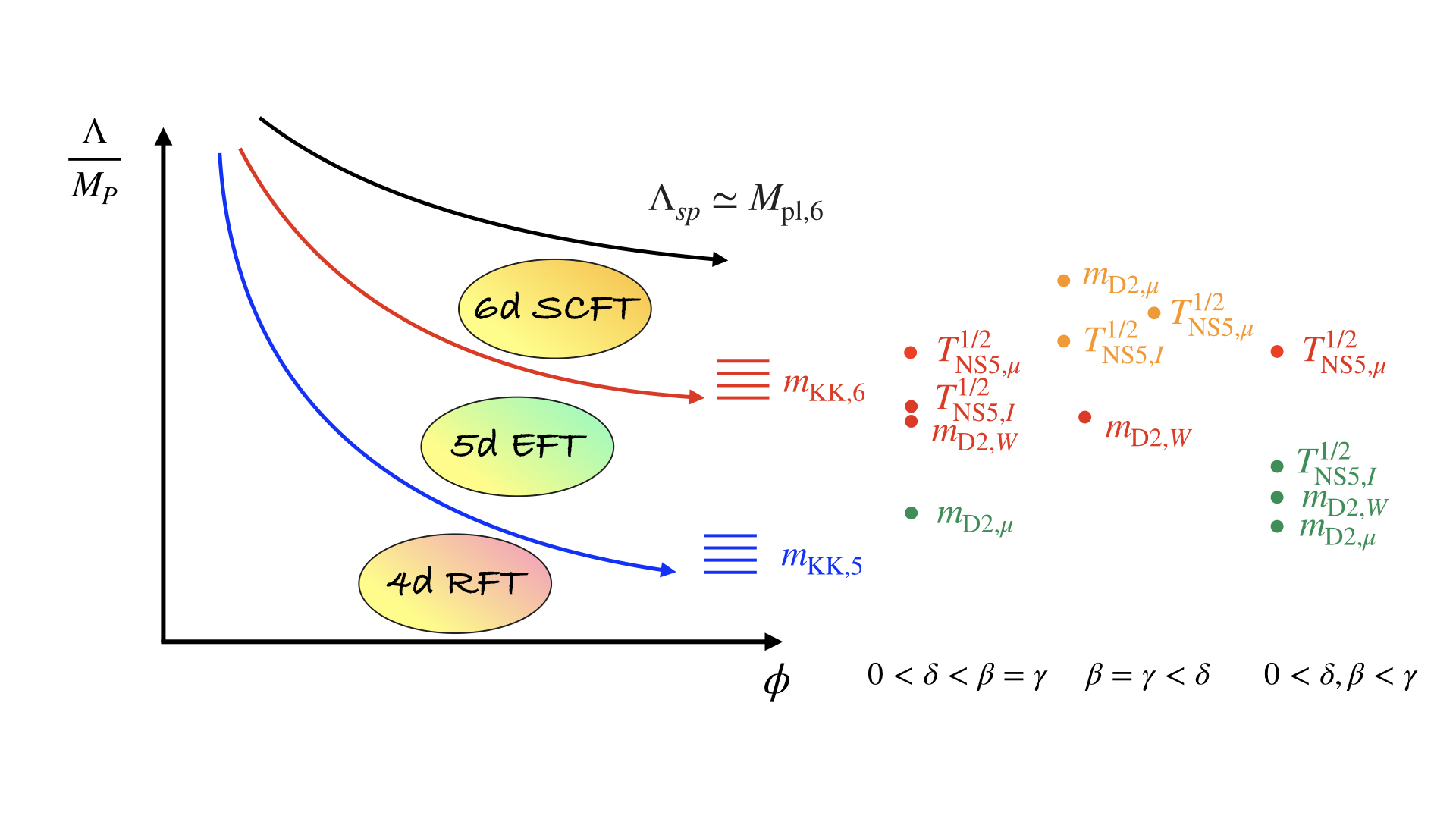}
\caption{Scales of the limit \eqref{exceptionallimit}. Here $m_{{\rm D2},W} = m_{{\rm D2},{\mathcal{E}_I}}$ is to the  $W$-boson mass that signals the enhancement to $su(2)$, $T_{{\rm NS5}, I}= T_{{\rm NS5},{\mathcal{D}_I}}$ the tension of the string of this $su(2)$ sector and $m_{{\rm D2},\mu} = m_{{\rm D2}, C_\mu^{\rm b}}$ the scale at which one reaches the flop boundary. For simplicity the scales associated to D4-branes have not been depicted.}
\label{fig:su2limit}
\end{figure}

The situation is different if we consider the case $\del, \b < \g$, which means that the fibral curve ${\cal E}_I$ is parametrically smaller than the generic fibre $\cal E$,  and hence the exceptional divisor ${\cal D}_I$ is smaller than the vertical one ${\cal D}_\m$. In this case, the D2's wrapping ${\cal E}_I$ and ${\cal C}_\m^b$ and the NS5 on ${\cal D}_I$ all lie parametrically below the 6d KK scale, see figure \ref{fig:su2limit}. As a consequence, $\Lambda_{\rm ch} = \max\{m_{{\rm D2}, C_\mu^{\rm b}}, m_{{\rm D2}, {\cal E}_I}, \} < m_{\rm KK,6}$, and one can decouple the SCFT sector associated to ${\cal D}_I$ below $m_{\rm KK,6}$. Therefore the UVRT of these limits is actually a 5d SCFT, rather than a 6d one. As we will illustrate in the following example, in this type of limits the relation \eqref{Rasintro} is realised with $\nu=3 \neq w$, which reflects the 5d nature of the UVRT.

\subsubsection*{Example: Genus-one fibration with two exceptional divisors}

To illustrate the above general discussion, we now consider a concrete Calabi--Yau threefold with $h^{(1,1)}=5$ studied in \cite{Lee:2018spm}. The threefold can be viewed as an elliptic fibration over $\mathbb{F}_1$ with two extra sections. Let us denote the basis of the Mori cone generators by $\{{\cal C}^a\}$, where ${\cal C}^2=\mathbb{P}^1_b$ and ${\cal C}^4=\mathbb{P}^1_f$ are the base and fibre of the $\mathbb{F}_1$ base, while ${\cal C}^1,{\cal C}^3,{\cal C}^5$ are fibral curves. The toric data and the GLSM matrix for this geometry are the following:
\setlength{\arrayrulewidth}{0.2mm}
\renewcommand{\arraystretch}{0.8}
\begin{table}[H]
\begin{center}
\begin{tabular}{cccc|ccccc}
  &  &  &  & ${\cal C}^1$ & ${\cal C}^2$ & ${\cal C}^3$ & ${\cal C}^4$ & ${\cal C}^5$ \\
\hline
$-1$ & 1 & 0 & 0 & 1 & $-1$ & $-1$ & 0 & 1 \\
0 & $-1$ & 0 & 0 & 0 & 0 & 0 & 0 & 1\\
1 & 0 & 0 & 0 & 1 & 0 & 0 & 0 & 0\\
$-1$ & 0 & 0 & 0 & 0 & 0 & 1 & $-2$ & $-1$\\
0 & 1 & 0 & 0 & $-1$ & 0 & 1 & 0 & 0\\
$-1$ & 1 & 1 & 0 & 0 & 1 & 0 & 0 & 0\\
0 & 0 & $-1$ & $-1$ & 0 & 1 & 0 & 0 & 0\\
$-2$ & 0 & 0 & 1 & 0 & 0 & 0 & 1 & 0 \\
0 & 0 & 0 & $-1$ & 0 & $-1$ & 0 & 1 & 0 \\
\end{tabular}
\end{center}
\label{tab:toric,GLSM,5mod}
\end{table}
\noindent In the basis of K\"ahler cone generators $\{J_a\}$ dual to $\{\mathcal{C}^a\}$, the intersection polynomial of $X$ reads
\be
\begin{split}
\mathcal{I} = & 16 J_1^3 + 4 J_1^2 J_2 + 30 J_1^2 J_3 + 6 J_1^2 J_4 + 18 J_1^2 J_5 + 8 J_1 J_2 J_3 + 2 J_1 J_2 J_4 + 4 J_1 J_2 J_5 +\\& 40 J_1 J_3^2 +
+ 11 J_1 J_3 J_4 + 18 J_1 J_3 J_5 + 2 J_1 J_4^2 + 7 J_1 J_4 J_5 + 4 J_1 J_5^2 +\\
&+ 10 J_2 J_3^2 + 3 J_2 J_3 J_4 + 4 J_2 J_3 J_5 + 2 J_2 J_4 J_5 +\\
&+ 44 J_3^3 + 13 J_3^2 J_4 + 18 J_3^2 J_5 + 3 J_3 J_4^2 + 7 J_3 J_4 J_5    + 4 J_3 J_5^2 +\\
&+ 2 J_4^2 J_5 + 2 J_4 J_5^2 \, .
\end{split}
\ee
In addition to the fibration strucutre described above, this model also admits an alternative, incompatible, genus-one fibration over $\mathbb{F}_1$ with two exceptional divisors and one bisection. Notice that the two fibrations are inequivalent since base and fibral curves are exchanged: in the second description, the base $\mathbb{F}_1$ contains the curves ${\cal C}^2=\mathbb{P}^1_b$, as before, and ${\cal C}^5=\mathbb{P}^1_f$, while ${\cal C}^1,{\cal C}^3,{\cal C}^4$ are the fibral curves constituting the genus-one fibre. The generic fibre corresponds to the class
\be
\mathcal{E} = 4{\cal C}^1 + 4{\cal C}^3 + 2{\cal C}^4 \, ,
\ee
and splits over codimension-one loci in the base that correspond to the two sections of the rationally fibred $\mathbb{F}_1$, given by ${\cal C}^2$ and ${\cal C}^2 + {\cal C}^5$. Fibering the blow-up curves over these loci we get two exceptional divisors. We consider the following $w=2$ limit
\be     \label{w=2_lim_exc}
t^5 \sim \phi \,, \qquad t^2 \sim \phi^\del \,, \qquad t^i \sim \phi^{\g_i} \,, \qquad i=1,3,4 \, .
\ee
In this limit the matrix
\be
\mathbf{K}_{ab} \equiv \mathcal{K}_{ab5} =
\begin{pmatrix}
18 & 4 & 18 & 7 & 4\\
4 & 0 & 4 & 2 & 0\\
18 & 4 & 18 & 7 & 4\\
7 & 2 & 7 & 2 & 2\\
4 & 0 & 4 & 2 & 0
\end{pmatrix}\,,
\ee
has rank 3 and a kernel generated by
\begin{equation}
    \ker \mathbf{K}_{ab} = \langle J_3-J_1, J_1-J_2-J_3+J_5 \rangle = \langle J_5-J_2, J_1-J_2-J_3+J_5 \rangle \,.
\end{equation}
One can check that the divisor $J_5-J_2$ is a vertical divisor, corresponding to a fibration of the generic elliptic fibre over the curve ${\cal C}^2$ in the base, while ${\cal D}_I = J_1-J_2-J_3+J_5$ is an exceptional divisor, corresponding to the curve ${\cal E}_I = 2{\cal C}^3 + {\cal C}^4$ fibered over ${\cal C}^2$. The volumes of the divisors in the kernel read
\be
\begin{split}
\CV_{J_5-J_2} &= t^2 \left( 4t^1 + 4t^3 + 2t^4\right) + 7 (t^1)^2 + 
5t^1 \left(2 t^3 + t^4\right) + \left(2 t^3 + t^4\right)^2 \, ,\\
\CV_{{\cal D}_I} &= t^2 \left(2t^3 + t^4\right) + \frac{1}{2}(2t^3+t^4)^2 \, .
\end{split}
\ee
To compare the limit \eqref{w=2_lim_exc} with the previous general discussion, one should notice that
\be
\b = \max \{\g_3, \g_4\}\, , \qquad \g = \max \{ \b, \g_1\}\, .
\ee
The scalar curvature in the limit scales as
\be
R_{\rm IIA} \sim f(t^1,t^2,t^3,t^4) \phi^2 \, ,
\ee
where $f(t^1,t^2,t^3,t^4)$ is a homogeneous function of the subleading moduli of degree $-2$. The rigid gauge coupling associated to the divisor ${\cal D}_I$ scales as
\be
1/g_{{\rm rigid},I}^2 \sim 2t^2 + 6t^3 + 3t^4 \, ,
\ee
such that
\be
\left( \frac{\Lambda_{\rm WGC}}{\Lambda_{\rm sp}} \right)^4 = \phi^2 g_{\rm rig}^4 \sim \frac{\phi^2}{(2t^2 + 6t^3 + 3t^4)^2} \, ,
\ee
where we used that the species scale is set by the 6d Planck mass. Comparing the above ratio to the scaling of the scalar curvature we have to distinguish two cases
\begin{itemize}
\item If at least one among $\{\del,\g_3, \g_4\}$ is larger than $\g_1$, which corresponds to either $\b, \g < \del$ or $\del < \b=\g$, then the asymptotic curvature scales as $$R_{\rm IIA} \sim \left( \frac{\Lambda_{\rm WGC}}{\Lambda_{\rm sp}} \right)^4$$ which hence satisfies \eqref{Rasintro} for $\nu=w=2$. In this case the D2-brane wrapping at least one curve in the exceptional divisor has a mass of the order of the 6d KK-scale, as shown in figure \ref{fig:su2limit}, such that $\Lambda_{\rm ch} \gtrsim m_{\rm KK,6}$ and the UVRT is identified as a 6d SCFT with a non-trivial gauge group. The fact that \eqref{Rasintro} is satisfied with $\nu=2$ confirms this expectation. 
\item If $\g_1 > \del, \g_3, \g_4$, or equivalently $\del, \b < \g=\g_1$, we find $$ R_{\rm IIA} \sim \frac{t^1}{\phi} \left( \frac{\Lambda_{\rm WGC}}{\Lambda_{\rm sp}} \right)^6 < \left( \frac{\Lambda_{\rm WGC}}{\Lambda_{\rm sp}} \right)^6 $$ satisfying \eqref{Rasintro} for $\nu=3 \neq w$. Since in this case all states making up the 5d SCFT associated to $\mathcal{D}_I$ have masses below the 6d KK-scale, we have $\Lambda_{\rm ch} < m_{\rm KK,6}$ and the UVRT is a 5d SCFT in accordance with our result that \eqref{Rasintro} is satisfied with $\nu=3$.
\end{itemize}

\subsection{Fibral divisors}
\label{ss:fibral}

Let us finally discuss the case where $\ker {\bf K}$ contains a fibral divisor ${\cal D}_F$. In general, we have that volume of this divisor and the area of its self intersection ${\cal C}_F = {\cal D}_{F} \cdot {\cal D}_{F}$ satisfy
\be
{\rm Vol} ({\cal D}_{F})  \leq  ({\rm Area} ({\cal E}))^2  \qquad \text{and} \qquad
 {\rm Area} ({\cal C}_F)  \leq  {\rm Area} ({\cal E})\, ,
\ee
where ${\cal E}$ represents the elliptic fibre. As a result, the NS5-brane wrapping ${\cal D}_F$ and the D2-brane wrapping ${\cal C}_F$ can lie below the 6d KK scale. If that happens, then we have $\Lambda_{\rm ch} \simeq m_{{\rm D2}, C_F} < m_{\rm KK,6}$ and the UVRT is a 5d SCFT. In that case, we expect \eqref{Rasintro} to be satisfied with $\nu = 3 \neq w$. If, instead, $\Lambda_{\rm ch}  \simeq  m_{\rm KK,6}$ we expect $\nu=w=2$. In the following, we illustrate how this case works in detail with an example.

\subsubsection*{Example: flopped phase of KMV conifold} \label{ss:KMVflop}

\begin{figure}[!t]
\centering
\hspace{-1cm}
\includegraphics[width=1\linewidth]{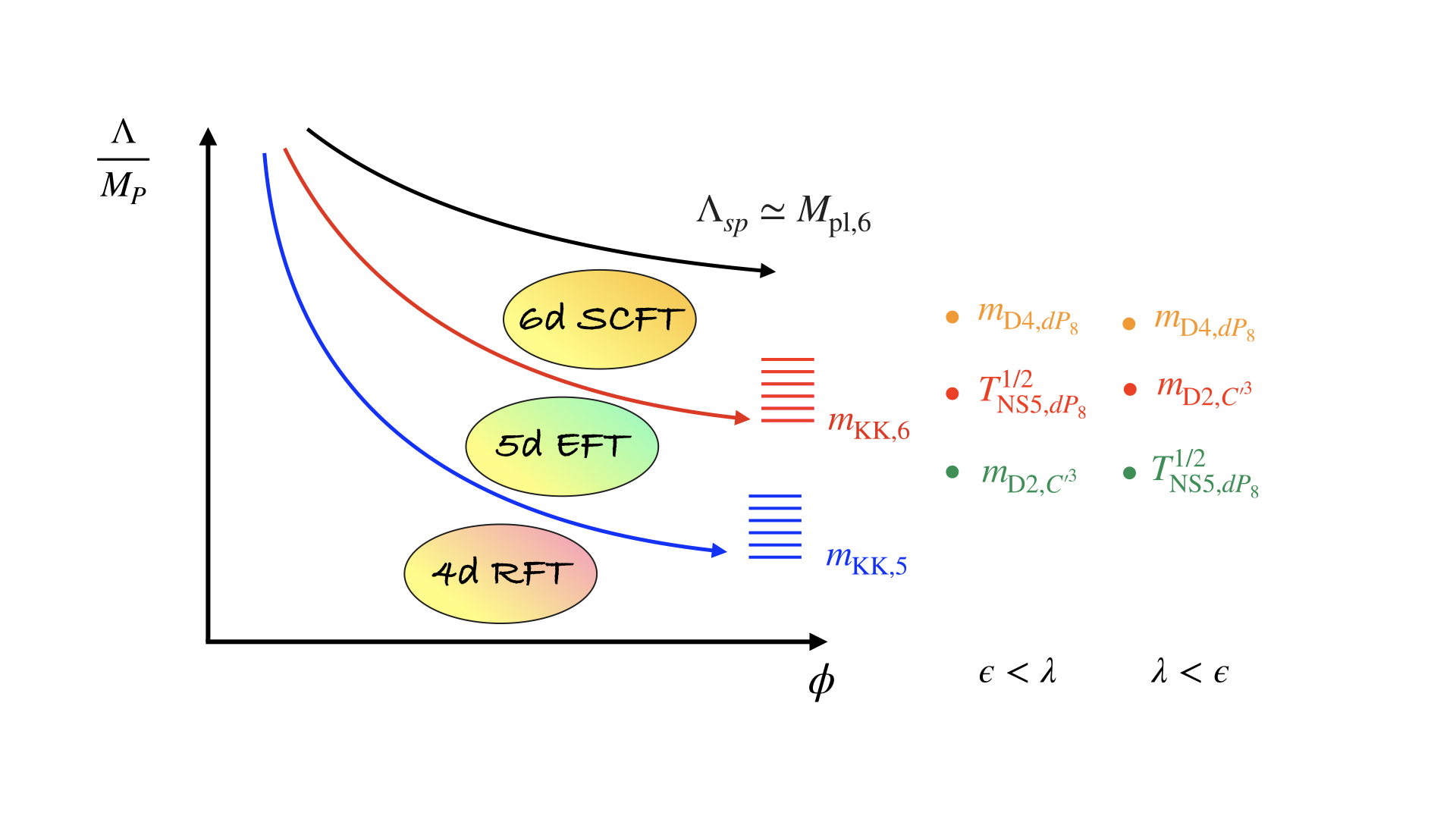}
\caption{Scales of the limit \eqref{w=2KMVflop}. In the case $\epsilon < \lambda$ one recovers an F-theory compactification, with a 6d SCFT that features a tensionless non-critical string. In the case $\lambda < \epsilon$ the UVRT is the 6d limit of a 5d SCFT, that appears at the scale $T^{1/2}_{{\rm NS5},dP_8}$.}
\label{KMVflop}
\end{figure}

The KMV conifold, discussed in section \ref{ss:KMV}, admits a flop transition, realised by shrinking the Mori cone generator ${\cal C}^3 = \mathbb{P}^1_b$ to zero size. The K\"ahler cone of the new phase is parametrised by the K\"ahler parameters $\{s^a\}_{a=1,2,3} = \{t^1+t^3, t^2+t^3, -t^3\}$, which control the volumes of the new Mori cone generators $\{{\cal C'}^a\}$. The triple intersection polynomial in the flopped phase reads
\be
\cI(X') = 8 J_1'^3 + J_1'^2 \left( 3J_2' + 9J_3' \right) + J_1' \left( J_2'^2 + 3J_2' J_3' + 9J_3'^2 \right) + J_2'^2 J_3' + 3 J_2' J_3'^2 + 9 J_3'^3 \, ,
\ee
where $\{J_a'\}=\{J_1,J_2,J_1+J_2-J_3\}$ are the new K\"ahler cone generators. In this phase the base $\mathbb{F}_1$ becomes a $\mathbb{P}^2$ surface, whose volume is controlled by $(s^2)^2$. Furthermore the elliptic fibration becomes non-flat, because the dimension of the fibre jumps over points of the base. The fibral fivisor over these points is a del Pezzo surface $dP_8$ whose volume scales as $(s^1)^2$ and which is to be identified with ${\cal D}_F$ in the discussion above. In this phase, we thus lose the smooth fibration structure, such that the analysis is different from a smooth elliptic fibration. Still, we can analyse the F-theory limits in the flopped phase by taking a growth sector of the form
\be     \label{w=2KMVflop}
s^1 \sim \phi^\l \, , \qquad s^2 \sim \phi \, , \qquad s^3 \sim \phi^\epsilon \, , \qquad \l,\epsilon \in (0,1) \, .
\ee
As in the unflopped phase, we are interested in the following mass scales
\be
\begin{split}
&\frac{m_{\rm KK,5}}{M_{\rm P}} = \frac{1}{\sqrt{\CV_X}}\, , \qquad \frac{m_{\rm KK,6}}{M_{\rm P}} = \frac{\CV_{T^2}}{\sqrt{\CV_X}}\, ,\\
\frac{m_{{\rm D2},{\cal C}'^3}}{M_{\rm P}} =& \frac{\CV_{{\cal C'}^3}}{\sqrt{\CV_X}} \, , \qquad \frac{m_{{\rm D4},dP_8}}{M_{\rm P}} = \frac{\CV_{dP_8}}{\sqrt{\CV_X}} \, , \qquad \frac{m_{{\rm D2},{\cal C}_F}}{M_{\rm P}} \sim \frac{\sqrt{T_{{\rm NS5},dP_8}}}{M_{\rm P}} =  \frac{\sqrt{\CV_{dP_8}}}{\sqrt{\CV_X}} \, ,
\end{split}
\ee
where we have added a D2-brane wrapping the curve ${\cal C}_F = {\cal C}'^1$. This curve can be identified with the self-intersection of $dP_8$. Multi-wrapped D2-branes on this curve are part of a tower of states dual to the tower of string excitations of the NS5-brane wrapped on ${\cal D}_F$. Using that $\CV_{\mathbf{T}^2} = s^1 + s^3$, $\CV_{{\cal C'}^3} = s^3$, $\CV_{{\cal C}_F} = s^1$ and $\CV_{dP_8} = \oh (s^1)^2$, we find in the limit \eqref{w=2KMVflop} 
\begin{align}\label{masses_floppedKMV}
\frac{m_{{\rm D2},{\cal C}'^3}}{m_{\rm KK,5}} \sim \phi^{\epsilon} \, , & \qquad \frac{m_{{\rm D4},dP_8}}{m_{\rm KK,5}} \sim \phi^{2\l}\, , \qquad \frac{m_{{\rm D2},{\cal C}_F}}{m_{\rm KK,5}} \sim \frac{\sqrt{T_{{\rm NS5},dP_8}}}{m_{\rm KK,5}} \sim \phi^{\l}\, ,\\
\frac{m_{{\rm D2},{\cal C}'^3}}{m_{\rm KK,6}} \sim \phi^{\epsilon-{\rm max}\{\l,\epsilon\}} \, , \quad & \frac{m_{{\rm D4},dP_8}}{m_{\rm KK,6}} \sim \phi^{2\l-{\rm max}\{\l,\epsilon\}} \, , \quad \frac{m_{{\rm D2},{\cal C}_F}}{m_{\rm KK,6}} \sim \frac{\sqrt{T_{{\rm NS5},dP_8}}}{m_{\rm KK,6}} \sim \phi^{\l-{\rm max}\{\l,\epsilon\}}\, .
\nonumber
\end{align}
The size of ${\cal C}'^3$ controls the mass of a single D2-particle charged under the rigid theory sector. Notice that, unlike in the original phase, this D2-brane is not part of any tower of states associated to some rigid divisor. On the other hand the area of ${\cal C}_F$ controls an infinite tower of states charged under the rigid theory, given by the NS5-brane wrapping the del Pezzo divisor $dP_8$ and D2-branes wrapping multiple times its self-intersection ${\cal C}_F$. These latter states constitute the SCFT sector that decouples in the limit \eqref{w=2KMVflop} and thus sets the scale $\Lambda_{\rm ch}= m_{{\rm D2},{\cal C}_F}$. We can distinguish two cases: $\epsilon<\l$ and $\l<\epsilon$, as illustrated in figure \ref{KMVflop}. For $\epsilon < \l$  we have $\Lambda_{\rm ch} \sim m_{\rm KK,6}$, the charged tower scales parameterically as the 6d KK scale and the UVRT is a 6d SCFT associated to the fibral divisor $dP_8$. In this case we expect the relation \eqref{Rasintro} to hold for $\nu=w=2$. For $\l < \epsilon$ instead, we have $\Lambda_{\rm ch} < m_{\rm KK,6}$ and the charged tower sits below the 6d KK scale. This means that effectively we have a 5d theory with tensionless strings and the UVRT is a 5d SCFT. Taking the limit $\phi\to \infty$ we thus take the 6d limit of a 5d SCFT in which the NS5-brane on $dP_8$ becomes a 6d tensionless string. The 6d theory then corresponds to an elliptic fibration over $\mathbb{P}^2$ with a non-minimal singularity over a point in $\mathbb{P}^2$ as befits a 6d SCFT (see for example \cite{Apruzzi:2018nre}). In this case we expect \eqref{Rasintro} to be satisfied with $\nu=3 \neq w$.

We can again study the behaviour of the 4d moduli space curvature in this limit, to check how it scales with respect to the gauge coupling of the rigid theory. Neglecting instanton corrections and only focusing on the classical part, we find that the leading behavior of the curvature in the limit \eqref{w=2KMVflop} is
\be\label{RIIA_KMVflop}
R_{\rm IIA} = \frac{s^1+s^3}{(s^1)^3} (s^2)^2 + \mathcal{O}(s^2) \sim \phi^{2 + \text{max}\{\l,\epsilon\} -3\l}\, ,
\ee
which diverges for any value of $\l$ and $\epsilon$. This is consistent with the fact that we are decoupling an SCFT sector associted to a del Pezzo divisor. Comparing the scaling of the curvature with
\be
\frac{\Lambda_{\rm WGC}}{\Lambda_{\rm sp}} \sim \phi^{\oh (1-\l)}\, ,
\ee
one can check that \eqref{Rasintro} is satisfied with an equality and $\nu=w=2$ for $\epsilon < \l$, while it is satisfied with an inequality and $\nu =3 \neq w$ for $\l < \epsilon$, confirming our expectations.


\section{Types of emergent string limits}
\label{s:typesw1}

We now turn our attention to $w=1$ limits. According to the general classification of section~\ref{s:typeIIA} these limits correspond to the case that the vector $\bm{e}$ appearing in 
\begin{equation}
    t^a = t^a_0 + e^a \phi \,,
\end{equation}
satisfies $\boldsymbol{k}_a = \mathcal{K}_{abc} e^b e^c =0$ $\forall a$, same for $\bm{e}_0$ in \eqref{growth}. In general, any Calabi--Yau threefold that has such a $w=1$ limit needs to exhibit a surface fibration over a $\mathbb{P}^1$ which becomes adiabatic in the limit $\phi\rightarrow \infty$ \cite{Lee:2019wij}. The surface fibre $S \rightarrow \mathbb{P}^1$ can either be a K3-surface or an Abelian surface and the asymptotic physics in $w=1$ limits is dominated by the NS5-brane wrapping the generic fibre. This wrapped brane yields a string in four dimensions which for the two fibre types is either dual to a critical heterotic or type II string. Since in the $w=1$ limits the string arising in the asymptotic limit is unique, so is the surface fibration corresponding to such a limit \cite{Lee:2019wij}. This means that there is a unique curve class in the Calabi--Yau threefold that can be identified with the base $\mathbb{P}^1_0$ of this fibration. Accordingly, the vector $\bm{e}$ has a single non-zero entry, $e^0\neq 0$, and we can split the basis of K\"ahler cone generators as $\omega_a=\{\omega_0,\omega_i\}$ such that $\cK_{00a}=0$ and the trajectory in moduli space is determined by 
\begin{equation}
    t^0 = t^0_0 + e^0 \phi \,,\qquad t^i = t^i_0 + \phi^{\gamma_i}\,,\qquad \gamma_i\in [0,1)\,. 
\end{equation}
The parameter $t^0$ is identified as the volume of the base $\mathbb{P}^1_0$ whereas the $t^i$ correspond to the volume of Mori cone generators $\mathcal{C}^i$ in the surface fibre. The matrix $\bf{K}$ defined in \eqref{rank} now takes the simple form 
\begin{equation}    \label{K_ij_w=1}
    {\bf K}_{ij} = \cK_{0ij}\,. 
\end{equation}
The rank $r$ of ${\bf K}$ can be computed following \cite{Cota:2022maf}. To that end we first notice that the surface fibre can degenerate over points $p$ in the base $\mathbb{P}^1_0$. Accordingly, the fibral Mori cone generators $\mathcal{C}^i$ can either be elements of the Picard lattice of the generic surface fibre or localised in the special fibres over the points $p$. Let us denote by $\mathcal{C}^\iota$, for $\iota\in \mathcal{I}_0$ some subset of indices, the Mori cone generators of $X$ corresponding to holomorphic curves in the generic fibre. We can further distinguish two broad classes of degenerations of the surface fibre depending on whether or not the entire surface splits into a union of multiple effective surfaces $S\rightarrow \cup_M  S_M$. We denote the Mori cone generators located in such multi-component surface fibres by $\mathcal{C}^\mu$, $\mu \in \mathcal{I}_\mu$  and those located in less severe degenerations by $\mathcal{C}^m$, $m\in \mathcal{I}_m$, with $\mathcal{I}_\mu$ and $\mathcal{I}_m$ disjoint set of indices. With this distinction the rank of ${\bf K}$ can be computed using the results of \cite{Cota:2022maf}, where it was argued that expressing \eqref{K_ij_w=1} in the dual basis of K\"ahler cone generators one has
\begin{equation}
    c_\mu \cK_{0 \nu a} = c_\nu \cK_{0 \mu a}\,,\qquad \forall \quad a \in \{0\}\cup \mathcal{I}_0 \cup \mathcal{I}_m \cup \mathcal{I}_\mu\, , 
\end{equation}
with $c_\mu$ a fixed vector specified by the degeneration details. It follows that a vector of the form $(0, d^\mu, 0, \dots, 0)$ with $d^\mu c_\mu =0$ belongs to $\ker {\bf K}$. Thus, whenever the generic surface degenerates into multiple components, the rank of ${\bf K}$ is reduced. Geometrically, the kernel of ${\bf K}$ corresponds to the divisor classes of the surface components localised at the degenerate surface \cite{Cota:2022maf}. 

To obtain a decoupled 4d RFT in a $w=1$ limit we hence require $X$ to be a surface fibration over $\mathbb{P}^1_0$ which degenerates over certain points in $\mathbb{P}^1_0$, in such a way as to produce additional divisors that are localised in the degenerate fibre. The 4d RFT is then associated to these localised divisors and its precise properties are determined by the type of degeneration of the surface. In the following we focus on the case that the generic surface fibre is a K3 (see section \ref{ss:Abelian} for the case of Abelian fibre) and that there is only one additional surface divisor giving a decoupled RFT. In other words, we consider a single degeneration of the K3-fibre over a point $p \in \mathbb{P}^1_0$ in which the generic surface $S$ splits into two components $S_0$ and $S_1$, with $S_1$ corresponding to the 4d RFT. To understand the physics of more general $w=1$ limits, one may consider these as building blocks of more general degenerations. 

On top of this assumption, one can distinguish two main classes of $w=1$ limits, in which:

\begin{itemize}
    \item[1)] The generic K3-fibre itself is elliptically fibred.  
    
    \item[2)]  The generic K3-fibre is not elliptically fibred.
\end{itemize}

In the first case, also the CY three-fold $X$ is an elliptic fibration over a base $B_2$, which is compatible with the K3 fibration over $\mathbb{P}^1_0$. Because of this, such a CY also gives rise to the $w=2$ large volume limits discussed in section \ref{s:nonsmooth}, which typically are deformations of the $w=1$ limits under consideration. In practice this implies that case 1)  corresponds to 4d RFTs with 6d UVRTs, and more precisely to 6d Little String Theories (LSTs) that can be engineered via F-theory on $X$, and deformed to the 6d SCFTs discussed in the previous section. It is important to notice that in case 1) one always has $\Lambda_{\rm sp} > m_{\rm KK,6}$, where the two quantities are controlled by $\sqrt{{\cal V}_{\rm K3}}$ and ${\cal V}_{\cal E}$, respectively. In some limits they scale in the same way asymptotically, but still can be separated by tuning the scaling of the subleading moduli, or the initial configuration $\{t^a_0\}$. In contrast, case 2) corresponds to 4d RFTs with 5d UVRTs, in the sense that one cannot decouple the 6d KK scale from the species scale, see below.

\subsubsection*{Refined Kulikov classification}

In general, the K3-degeneration type determines the properties of the 4d RFT that decouples from gravity in the $w=1$ limit. To study degenerations of K3 surfaces one considers a family $S_u$ of K3 surfaces for which the parameter $u$ takes value in a complex disk $D_u$. The degenerate fibre corresponds to the central surface $S_0$.  In the context of Calabi--Yau threefolds the disk can be viewed as the local neighborhood of the degenerate fibre in the base $\mathbb{P}^1$. A coarse classification of degenerations of K3 surfaces has been given by Kulikov \cite{Kulikov:1977,Person:1977} splitting the cases into Type I, II, and III, see \cite{Lee:2021qkx} for a recent review, \cite{Clingher:2003ui} for a more detailed discussion of the underlying mathematics, and \cite{Lee:2021usk} for applications to F-theory 8d infinite distance limits. The cases in which the K3 degenerates into a union of surfaces corresponds to Type II and III, which  are at infinite distance in the K3 moduli space. This classification assumes that the degeneration of the central fibre is semi-stable which can be achieved by birational transformation or base changes of the disk $D_u$. Whereas for the local model with base $D_u$ this is always possible, this is, however, not the case in the compact model which is a K3-fibration over $\mathbb{P}^1$. Two degenerate K3 surfaces that are related via base change lead to the same 4d RFT though their coupling to gravity is different. Since here we are interested in the gravity decoupling limit, we do not worry about this subtlety and focus on the original Kulikov classification. We are not going to delve into the details of the classification of Kulikov models and refer to \cite{Lee:2021qkx,Clingher:2003ui} for a detailed discussion. For our purposes here it is sufficient to think about the difference between Type II and III Kulikov models as follows: 

Calabi--Yau threefolds are expected to have a description in terms of a special Lagrangian fibration of a three-torus $T^3$ over $S^3$, with mirror symmetry corresponding to three T-dualities along such a $T^3$ \cite{Strominger:1996it}. For K3 surfaces we similarly expect them to possess a $T^2$ fibration over $S^2$. Notice that neither $S^2$ nor $T^2$ are holomorphic curves, and in particular $T^2$ should not be confused with the holomorphic fibre of an elliptic fibration. To avoid confusion we refer to the $T^2$-fibre that is the analogue of the $T^3$-fibre of Calabi--Yau threefold~\cite{Strominger:1996it} as $T^2_{\rm SYZ}$. The difference between Type II and III degenerations of K3 surfaces is now given by the number of legs of the $T^2_{\rm SYZ} \subset $ K3 that degenerate at the singular locus. For Type II Kulikov models only one leg of $T^2_{\rm SYZ}$ pinches whereas for Type III both legs of $T^2_{\rm SYZ}$ pinch. 

As mentioned above, when the K3 is elliptically fibred (case 1) above), F-theory compactified on $X$ provides the 6d UVRT of the 4d RFT. Due to this, the $w=1$ limit corresponds to the weak coupling limit for the heterotic theory compactified on a different K3 times a $T^2$. The Kulikov classification for the special case of elliptically fibred K3 surfaces has been refined in \cite{Lee:2021qkx, Clingher:2003ui}. For Type II degenerations one can distinguish two sub-cases depending on whether the leg of $T^2_{\rm SYZ}$ that pinches at the degeneration is located in the generic elliptic fibre or the base of the K3. The case with the pinching leg of $T^2_{\rm SYZ}$ being located in the base can be identified as the Type II.a Kulikov degeneration of \cite{Lee:2021qkx,Clingher:2003ui} whereas the Type II.b degeneration corresponds to the pinching leg of $T^2_{\rm SYZ}$ located in the elliptic fibre. In the Type II.a degeneration case the K3 surfaces split into the union of two rational elliptic fibrations that intersect over the generic elliptic fibre whereas in the Type II.b case the K3-surface splits into two rationally fibred surfaces over the base of the generic K3. Very roughly, one can now think of Type III.a degenerations as taking a Type II.a degeneration and further degenerating the second leg of $T^2_{\rm SYZ}$ over a base component. Type III.b degenerations are then obtained in a similar way from Type II.b degenerations. Hence, while also interesting, Type III degenerations take us away from the above assumption that the K3-fibre degenerates over $p \in \mathbb{P}^1$ into a union of just two surfaces. Let us stress that the actual classification of these models is more involved and has been carried out in great detail in \cite{Lee:2021qkx}, but that this very simplified picture is sufficient to understand the physics of $w=1$ limits. 

To sum up, in the case that the K3 is elliptically fibred, one can apply the classification  of \cite{Lee:2021qkx} to further distinguish between two subtypes of $w=1$, case 1) basic limits. Namely:

\begin{itemize}
    \item[1.a)] The generic K3-fibre  develops a degeneration of Type II.a over $p \in \mathbb{P}^1_0$.

     \item[1.b)] The generic K3-fibre  develops a degeneration of Type II.b over $p \in \mathbb{P}^1_0$.
    
\end{itemize}

Type III.a and Type III.b are more involved Kulikov degenerations and cannot be seen as building blocks in the sense above. 



\section{Emergent string limits and LSTs}
\label{s:LSTs}

Given the classification of $w=1$ limits in the last section we now focus on the 4d RFTs that decouple along the respective limits, and their associated UVRTs. In case the K3-fibre is  elliptically fibred one can separate $\Lambda_{\rm sp}$ from the 6d KK scale, and the UVRT is either a 6d LST or SCFT. If it is not, $\Lambda_{\rm sp} = m_{{\rm KK}, 6}$ and we instead recover a  5d LST or SCFT. 

\subsection{\texorpdfstring{$E_8\times E_8$}{}-type LSTs}
We start with the case that the K3-fibre over $p\in \mathbb{P}^1_0$ degenerates in such a way that the leg of $T^2_{\rm SYZ}$ that pinches is localised in the base of the elliptically fibred K3. The base $\mathbb{P}^1$ of the elliptically fibred K3 thus splits into a union of rational curves and the K3 surface splits into a union of rational elliptic surfaces as befits a Type II.a Kulikov degeneration \cite{Lee:2021usk}. 

The basis of Mori cone generators can now be split in the following way. First there is the curve $\mathcal{C}^0$ corresponding to the base of the K3-fibration. In addition we denote by $\mathcal{C}^1$ and $\mathcal{C}^2$ the two curves into which the base of the elliptic K3 splits over the degeneration point. Finally, there are the curves $\mathcal{C}^\iota$, $\iota \in \mathcal{I}_0$, in the generic K3-fibre that do not split at the degeneration point. In particular the elliptic fibre of the K3 surface $\mathcal{E}$ is of this latter type. Furthermore let $S_0$ be the class of the generic K3-fibre and the components in which this fibre splits over $p$ respectively by $\mathcal{E} \hookrightarrow S_1 \rightarrow \mathcal{C}^1$ and $\mathcal{E} \hookrightarrow S_2 \rightarrow \mathcal{C}^2$. As for $w=2$ limits it is instructive to consider the masses and tensions of the particles and strings of importance. First there is the 5d KK-scale that is again set by the mass of D0-branes and the 6d KK-scale set by the D2-brane mass on $\mathcal{E}$, i.e., 
\begin{equation}\label{eq:KKscalesw1}
    \frac{m_{\rm KK,5}}{M_{\rm P}} = \frac{1}{\sqrt{\mathcal{V}_X}}\,,\qquad \frac{m_{\rm KK,6}}{M_{\rm P}} = \frac{\mathcal{V}_{\mathcal{E}}}{\sqrt{\mathcal{V}_X}}\,.
\end{equation}
In addition the tension of the \emph{critical} string obtained from an NS5-brane wrapped on $S_0$ with tension 
\begin{equation}\label{eq:stringtension}
    \frac{T_{{\rm NS5}, S_0}}{M_{\rm P}^2} = \frac{\mathcal{V}_{S_0}}{\mathcal{V}_X}\,. 
\end{equation}
As argued in \cite{Lee:2019wij} the K3-fibre in type IIA string theory cannot shrink to zero size but at most has volume of order of the string scale. This implies $\sqrt{T_{{\rm NS5}, S_0}} \gtrsim m_{\rm KK,5}$. Similarly the tension of the critical NS5-brane string is always bounded by the 6d KK-scale such that we have the hierarchy of quantum gravity  scales $m_{\rm KK,5}\lesssim m_{\rm KK,6} \lesssim \sqrt{T_{{\rm NS5}, S_0}}$. In addition we have the scales of the 4d RFT associated to the degeneration 
\begin{equation}
    \frac{m_{{\rm D2}, {\mathcal{C}^i}}}{M_{\rm P}} = \frac{\mathcal{V}_{\mathcal{C}^i}}{\sqrt{\mathcal{V}_X}}\,,\qquad \frac{T_{{\rm NS5}, S_i}}{M_{\rm P}^2} = \frac{\mathcal{V}_{S_i}}{\mathcal{V}_X}\,,
\end{equation}
for $i=1,2$. With respect to the quantum gravity scales we have the hierarchy $m_{\rm KK,6} \lesssim \sqrt{T_{{\rm NS5}, S_i}}\lesssim \sqrt{T_{{\rm NS5}, S_0}}$ and $m_{\rm KK,5} \lesssim m_{{\rm D2}, {\mathcal{C}^i}}$. The spectrum of states is reminiscent of the $w=2$ limits for smooth fibrations discussed in section~\ref{s:Ftheory}. However, the physics of those limits differs from the $w=1$ limits discussed here. The main difference is the existence of a critical string whose tension is below the naive species scale associated to the 5d and 6d KK towers. As a consequence the 4d RFT does not correspond to a 6d SCFT compactified on a torus but instead to a 6d Little String Theory compactified on $T^2$. The origin of the 6d LST is most easily understood by considering the 6d theory corresponding to F-theory compactified on $X$ viewed as an elliptic fibration over some base $B_2$. In this theory the splitting of the base of the elliptic K3 into two curves $\mathcal{C}^1$ and $\mathcal{C}^2$ signals the existence of an additional tensor multiplet. Viewed as curves in $B_2$ both $\mathcal{C}^1$ and $\mathcal{C}^2$ have self-intersection $(-1)$ giving rise to E-strings when wrapped by D3-branes in F-theory. In other words the LST corresponds to curves intersecting as 
\begin{equation}\label{eq:E8E8}
    [\mathfrak{g}_1] \; 1 \;1\; [\mathfrak{g}_2] \,,
\end{equation}
where the two end components denote the flavor algebras of the LSTs corresponding to the perturbative heterotic gauge algebras, $g_1\times g_2 \in \mathfrak{e}_8\times \mathfrak{e}_8$.  Via duality to the heterotic string we can identify this LST as the worldvolume theory of an NS5-brane arising from a small instanton in the $E_8\times E_8$ heterotic string. We thus refer to these $w=1$ limits as LST limits of type $E_8\times E_8$. The moduli space of the 4d RFT descends from the tensor branch of the 6d LST. Therefore the classical rigid curvature of the 4d RFT vanishes and, similar to the $w=2$ limits for smooth elliptic fibrations, a curvature divergence is only generated for the $w=1$ limits of type $E_8\times E_8$ through instanton corrections in case the volume of either $\mathcal{C}^1$ or $\mathcal{C}^2$ is kept constant. 

As in the case of the $w=2$ limits we can identify the UVRT as the theory realised at the scale $\Lambda_{\rm ch}$ where a tower of charged states appear. In the simple case of the LST with local description in \eqref{eq:E8E8} there are two relevant charges of the LST corresponding to the $U(1)$'s associated to the $(-1)$-curves. From a 6d perspective, the tower of states charged under these $U(1)$'s correspond to multi-wrappings of the D3-brane strings with suitable amount of KK-charge arising at the scale $\Lambda_{{\rm ch},i}=\text{max}\,\{m_{\rm KK,6}, m_{D2|_{\mathcal{C}^i}}\}$. We can now distinguish two cases: 
\begin{equation}\label{casesE8E8}\begin{aligned}
    i)&\qquad \Lambda_{{\rm ch},1} \sim  \Lambda_{{\rm ch},2} \,,\\
    ii)&\qquad \Lambda_{{\rm ch},1} \succ  \Lambda_{{\rm ch},2} \,, \quad \textit{or vice versa.}
\end{aligned}
\end{equation}
In case $i)$ the UVRT is indeed the 6d LST. For $\Lambda_{{\rm ch},i}\succ m_{\rm KK,6}$ the moduli space of the 4d RFT is again the dimensional reduction of the (now two-dimensional) tensor branch of the 6d theory which is flat such that no curvature is generated. On the other hand, if $\Lambda_{{\rm ch},i}\sim m_{\rm KK,6}$ the UVRT is realised at the origin of the tensor branch and a curvature of the 4d RFT can be generated by instantons as in the case of $w=2$ limits for smooth elliptic fibrations. 

In case $ii)$, however, the UVRT only sees one of the towers associated to the two $(-1)$-curves in \eqref{eq:E8E8}. Instead of an LST the UVRT is therefore the 6d SCFT associated to this $(-1)$-curve in F-theory. Again, depending on the scaling relative to $m_{\rm KK,6}$ a curvature can only be generated by instanton effects in parallel to the discussion of the $w=2$ limits in section~\ref{s:Ftheory}.

\subsubsection*{Example: Smooth elliptic fibration over $dP_2$}

An LST limit of $E_8\times E_8$-type can be obtained for a smooth Weierstrass model over $B_2=dP_2$. This base can be viewed as a double blow-up of $\mathbb{P}^2$ in two points. The toric data and the GLSM matrix for this geometry are
\setlength{\arrayrulewidth}{0.2mm}
\renewcommand{\arraystretch}{0.8}
\begin{table}[H]
\begin{center}
\begin{tabular}{cccc|cccc}
 &  &  &  & ${\cal C}^0$ & ${\cal C}^1$ & ${\cal C}^2$ & ${\cal C}^3$ \\
\hline
0 & 0 & 0 & 1 & 0 & 0 & 0 & 3 \\
0 & 0 & 1 & 0 & 0 & 0 & 0 & 2 \\
0 & $-1$ & $-2$ & $-3$ & 0 & 0 & 1 & 0 \\
0 & $1$ & $-2$ & $-3$ & $-1$ & 1 & 0 & 0 \\
$1$ & $1$ & $-2$ & $-3$ & 1 & $-1$ & 1 & 0 \\
1 & 0 & $-2$ & $-3$ & 0 & 1 & $-1$ & 0 \\
$-1$ & 0 & $-2$ & $-3$ & 1 & 0 & 0 & 0 \\
0 & 0 & $-2$ & $-3$ & $-1$ & $-1$ & $-1$ & 1
\end{tabular}
\end{center}
\label{tab:toric,GLSM,dP2}
\end{table}
\noindent In a basis of K\"ahler cone generators $\{J_a\}$, the intersection numbers are given by
\be
\mathcal{I}(X) = J_3(J_1^2 + J_1J_0 +J_1J_2+ J_0J_2) + J_3^2(3J_1 +2 J_0 +2J_2) + 7 J_3^3 \, ,
\ee
where $J_{i} = \pi^*(j_i)$ for $i=0,1,2$ are the pull-back of the K\"ahler cone generators of $dP_2$ and $J_3=D_0 + \pi^* c_1(dP_2)$ is the shifted zero section. This geometry can be viewed as a K3-fibration over $\mathbb{P}^1_0$ in different ways. A $w=1$ limit is for example obtained if we make the K\"ahler parameter associated to $J_0$ very large. In this limit we can view $X$ as a K3-fibration over $\mathbb{P}^1_0$ with the class of the generic fibre given by $J_0$. We thus consider the scaling 
\begin{equation}\label{w1limitex1}
    t^0 \sim t^0_0\,\phi\,,\qquad t^i\sim \phi^{\delta_i}\,,\qquad i=1,2,3\,, 
\end{equation}
for $\delta_i\in[0,1)$. The generic K3-fibre is elliptically fibred with Picard lattice of rank 2 with the two holomorphic curves identified with the elliptic fibre of the K3 and its $\mathbb{P}^1$ base. The volume of the elliptic fibre is given by $t^3$, while the volume of the base of the generic K3-fibre by $t^1+t^2$. It is this base curve that splits into two components $\mathcal{C}^1$ and $\mathcal{C}^2$ over a point $p\in \mathbb{P}^1_0$ in line with our general discussion of LST limits of $E_8\times E_8$ above. The two curves $\mathcal{C}^{1,2}$ have self-intersection $(-1)$ such that the local LST can pictorially be described as in \eqref{eq:E8E8}. Since we are considering a smooth Weierstrass model the heterotic gauge group is broken entirely such that for this example the flavor groups of the LST are trivial. 

For this example the matrix \eqref{K_ij_w=1} takes the form
\begin{equation}
    \mathbf{K}_{ab} \equiv \mathcal{K}_{ab0} = \begin{pmatrix}
         0&0&0& 1\\ 0&0&0&0\\ 0&0&0& 1\\ 1&0&1&2
    \end{pmatrix}\,,
\end{equation}
which has rank 2 and a kernel generated by
\begin{equation}
    \ker \mathbf{K}_{ab} = \langle J_0, J_1-J_2 \rangle = \langle J_0+J_2-J_1, J_1-J_2 \rangle\,,
\end{equation}
where $S_1 = J_0+J_2-J_1$ and $S_2= J_1-J_2$ are the two components the K3 fibre splits into.  They both correspond to rational elliptic surfaces with base $\mathcal{C}^1$ and $\mathcal{C}^2$, respectively, and  intersect over the common elliptic fibre, hence this is a Type II.a Kulikov model. Notice that here the 6d KK scale is controlled by ${\cal V}_{\cal E} = t^3$ and the species scale by $\sqrt{{\cal V}_{\rm K3}} = \sqrt{(t^1 + t^2) t^3 + (t^3)^2}$, so that we have $\Lambda_{\rm sp} > m_{\rm KK,6}$, as anticipated above.

As expected from our general discussion the scalar curvature in the limit \eqref{w1limitex1} remains finite at the classical level but a divergence is generated by instanton corrections
\be
R \sim \phi\, t^3(t^1+t^2+t^3)\; \frac{1}{2(t^3)^3} n_{{\cal C}^i}^2 e^{-4\pi t^i} \, , \quad \text{with  }
\begin{cases}
i=1 \quad \text{if } \del_1< \del_2\\
i=2 \quad \text{if } \del_1> \del_2
\end{cases}\, ,
\ee
where we only kept the leading contribution that comes from the curve $k_a {\cal C}^a$ with smallest volume and non-vanishing $n_{\bf{k}}$, namely ${\cal C}^1$ if $\del^1 < \del^2$, or ${\cal C}^2$ if $\del^2 < \del^1$. The rigid gauge coupling associated to the divisors $S_1$ and $S_2$ scales as
\be
1/g_{\rm rig}^2 \sim t^3 \, ,
\ee
such that
\be
\left( \frac{\Lambda_{\rm WGC}}{\Lambda_{\rm sp}} \right)^2 = \phi g_{\rm rig}^2 \sim \frac{\phi}{t^3} \, ,
\ee
where we used that the species scale is set by the tension of the NS5-brane wrapping $J_0$. Again, we can compare the above ratio to the scaling of the scalar curvature. To that end we can distinguish the following two cases in which we obtain a diverging curvature corresponding to the two cases in \eqref{casesE8E8} (there are two additional cases obtained by exchanging $1\leftrightarrow 2$ which we do not discuss separately):
\begin{itemize}
\item[$i)$] If we choose, $\delta_1=0$ and $\delta_2 < \delta_3 < 1$ the asymptotic curvature scales as $$R_{\rm IIA} \sim \left( \frac{\Lambda_{\rm WGC}}{\Lambda_{\rm sp}} \right)^2 e^{-4\pi t^1} \lesssim \left( \frac{\Lambda_{\rm WGC}}{\Lambda_{\rm sp}} \right)^2 ,$$ which hence satisfies \eqref{Rasintro} for $\nu=w=1$. According to our general discussion in this case the UVRT is a 6d LST such that we indeed expect \eqref{Rasintro} to be satisfied with $w=1$.
\item[$ii)$] For $\delta_1=0$ and $\delta_3 < \delta_2<1$ we find $$R_{\rm IIA} \sim\phi^{\delta_2-\delta_3} \left( \frac{\Lambda_{\rm WGC}}{\Lambda_{\rm sp}} \right)^2 e^{-4\pi t^1} < \left( \frac{\Lambda_{\rm WGC}}{\Lambda_{\rm sp}} \right)^4,$$ which satisfies \eqref{Rasintro} for $\nu=2 \neq w$ signalling that our UVRT is a 6d SCFT in accordance with our general expectation.
\end{itemize}

\subsection{\texorpdfstring{$SO(32)$}{}-type LSTs}
\label{ss:so32}

We now turn to the case of a Calabi--Yau threefold that is fibred by an elliptic K3 over $\mathbb{P}^1_0$ which degenerates of Type II.b over a point $p\in \mathbb{P}^1_0$. As discussed in the previous section in this case the leg of $T^2_{\rm SYZ}$ located in the elliptic fibre of K3 degenerates over the entire base $\mathbb{P}^1$ of the K3 resulting in a singular K3. The resolved threefold thus has an exceptional divisor that is a rational fibration over the base $\mathbb{P}^1$ of the elliptic K3. As in the previous case let us denote by $\mathcal{C}^0$ the basis of the K3-fibration. Let us further denote by $[\mathcal{C}^1]$ the class of the base of the generic K3 and by $\mathcal{C}^1_p$ the corresponding curve of the degenerate K3 fibre. Over $\mathcal{C}^1_p$ the elliptic fibre $\mathcal{E}$ of the K3 splits into two components which we denote by $\mathcal{C}^i$, $i = 2,3$. In addition there are two divisors $S_2$ and $S_3$ identified with the components into which the degenerate K3 splits over $p$. The KK-scales and the tension of the critical string obtained from an NS5-brane wrapping the generic K3 fibre are again given by \eqref{eq:KKscalesw1} and \eqref{eq:stringtension}. The states characterising the 4d RFT are now given by the NS5-branes wrapping $S_2$ and $S_3$ as well as the D2-branes on $\mathcal{C}^{2,3}$. The latter can be viewed as W-bosons of an $\mathfrak{sp}_1$ algebra realised over $\mathcal{C}^1_p$. 

The physics of the RFT in this $w=1$ limit is very similar to the $w=2$ limits with exceptional divisors in the kernel of ${\bf K}$. However, in this case the 6d origin of the 4d RFT is again a Little String Theory. To see this we again consider F-theory on $X$ viewed as an elliptic fibration over a rationally fibred base $B_2$. In this theory a $\mathfrak{sp}_1$ gauge algebra is realised over the curve  $\mathcal{C}_p^1$ which is in the class of the generic rational fibre of $B_2$. In the gravity decoupling limit this theory is identified as an LST with local geometry 
\be\label{localgeometrySO(32)}
[\mathfrak{g}_{SO(32)}]\;\stackrel{\mathfrak{sp}_1}{0}\,,
\ee
corresponding to the worldvolume theory of a small instanton NS5-brane in the heterotic SO(32) string with gauge algebra $\mathfrak{g}\in \mathfrak{so}(32)$\,. We thus refer to the $w=1$ limits for this kind of geometries as LST limits of type SO(32).  In this case the moduli space of the 4d RFT is identified with the Coulomb branch of the $\mathfrak{sp}_1$ gauge theory upon compactification of the 6d LST on $T^2$. Due to the presence of charged states the Coulomb branch of the rigid theory has non-trivial curvature such that the asymptotic curvature diverges for the $w=1$ limits of type SO(32). 

In contrast to the previous case of $E_8\times E_8$-type LSTs, only one type of UVRT can be realised in this model, namely a 6d LST. The reason is the following: given the split of the generic fibre into two surfaces $\mathcal{S}_1$ and $\mathcal{S}_2$, we can still define two scales $\Lambda_{{\rm ch},i}$, $i=1,2$, that are determined by the mass of D2-branes inside the respective divisors. However, in this class of models it is not possible to parametrically separate these scales. To see this, we notice that in Kulikov Type II.b degenerations, the elliptic fibre of the generic K3 away from the degeneration splits into two components that are exchanged under monodromy around four branching points in the base $\mathbb{P}^1$ of K3~\cite{Clingher:2003ui,Lee:2021qkx}. Furthermore, the two surfaces $\mathcal{S}_1$ and $\mathcal{S}_2$ intersect over a curve that is the double-cover of the base $\mathbb{P}^1$ of $K3$ branched over the four points. For this reason, the two divisors $\mathcal{S}_1$ and $\mathcal{S}_2$ contain both curve classes in the elliptic fibre, such that $\Lambda_{{\rm ch},1}$ and $\Lambda_{{\rm ch},2}$ cannot be parametrically separated. The UVRT is, hence, always a 6d LST and, accordingly, \eqref{Rasintro} is always satisfied with $\nu=1$. Below we provide an explicit example confirming this. 

Let us notice that, unlike for the $E_8\times E_8$-type LSTs, the $SO(32)$-type LSTs cannot be viewed as two connected 6d SCFTs. This is clear from the local geometry in~\eqref{localgeometrySO(32)} from the single curve with vanishing self-intersection hosting the LST. Therefore, if there existed a limit in which \eqref{Rasintro} was satisfied with $\nu=2$, there would not be a candidate 6d SCFT that could serve as UVRT. It is thus consistent that the geometry of the Type II.b degeneration does not allow for this case. Similarly, by continuity, no regime can exist in which \eqref{Rasintro} is satisfied with $\nu =3$ since this would imply the existence of a regime in which $\nu=2$. The case of $SO(32)$-type LSTs hence gives very strong evidence for the relation between $\nu$ and the UVRT.

\subsubsection*{Example: Elliptic fibration over $\mathbb{F}_1$ with an exceptional divisor}

Let us consider the $SU(2)$ Little String Theory limit discussed in \cite{Hayashi:2023hqa}. We denote by $\{\mathcal{C}^a\}$, $a=0,\dots, 3$ the generators of the Mori cone of $X$. The manifold $X$ can be described as an elliptic fibration over $\mathbb{F}_1$, for which shrinking $\mathcal{C}^3$ leads to an $I_2$ singularity over curve in the class of the fibre $f=\mathcal{C}^1$ of $\mathbb{F}_1$. The base $h$ of $\mathbb{F}_1$ is identified with $\mathcal{C}^0$ and the generic elliptic fibre as ${\cal E} = \mathcal{C}^2+3\mathcal{C}^3$. The toric data and the GLSM matrix for this model read:
\setlength{\arrayrulewidth}{0.2mm}
\renewcommand{\arraystretch}{0.8}
\begin{table}[H]
\begin{center}
\begin{tabular}{cccc|cccc}
 &  &  &  & ${\cal C}^0$ & ${\cal C}^1$ & ${\cal C}^2$ & ${\cal C}^3$ \\
\hline
0 & 0 & 0 & -1 & 0 & 0 & 0 & 1 \\ 
0 & 0 & -1 & 0 & 0 & 0 & -1 & 1 \\ 
-1 & -1 & 2 & 3 & 1 & 0 & 0 & 0 \\ 
0 & 1 & 2 & 3 & 0 & 1 & 0 & 0 \\ 
0 & -1 & 2 & 3 & -1 & 1 & 0 & 0 \\ 
1 & 0 & 1 & 2 & 0 & 0 & 3 & -1 \\ 
1 & 0 & 2 & 3 & 1 & 0 & -3 & 1 \\ 
0 & 0 & 2 & 3 & -1 & -2 & 1 & 0 \\ 
\end{tabular}
\end{center}
\label{tab:toric,GLSM,SU(2)}
\end{table}
\noindent The intersection polynomial in a K\"ahler cone generators basis is given by 
\begin{equation}
\begin{aligned}
    I(X)=&\;188J_3^3 + 18J_3^2J_0 + 25J_3^2J_1 + 3J_3J_0J_1 + 3J_3J_1^2 + 68J_3^2J_2 + 6J_3J_0J_2\\ & + 9J_3J_1J_2 + J_0J_1J_2 + J_1^2J_2 + 24J_3J_2^2 + 2J_0J_2^2 + 3J_1J_2^2 + 8J_2^3\,.
\end{aligned}
\end{equation}
The above intersection polynomial makes apparent the K3-fibration of $X$ with base given by $\mathcal{C}^0$. We now take the $w=1$ limit corresponding to the scaling 
\be\label{limitw1ex2}
    t^0\sim t^0_0\,\phi\,,\qquad t^i\sim \phi^{\delta_i}\,,\qquad i=1,2,3\,.
\ee 
for $\delta_i\in [0,1)$. The curve in the class $[\mathcal{C}^1]$ over which we realize the gauge theory is identified as the self-intersection 0 curve hosting an LST. The 6d KK-scale is given by 
\be
 \frac{m_{\rm KK,6}}{M_{\rm P}}=\frac{t^2 + 3t^3}{\sqrt{\CV_{X}}}\,,
\ee 
and the matrix $\bf{K}$ for this example is given by
\begin{equation}
    {\bf K}=\cK_{0ab}= \begin{pmatrix}
        0&0&0&0 \\  0&0&1&3 \\ 0&1&2&6 \\ 0&3&6&18 
    \end{pmatrix}\,,
\end{equation}
which has rank 2 and a two-dimensional kernel 
\begin{equation}    \label{eq:kerSU2}
    \ker \cK_{0ab} = \langle J_0 , 3J_2-J_3 \rangle = \langle J_0+J_3 -3J_2 , 3J_2-J_3 \rangle \,,
\end{equation}
where $S_2 = 3J_2-J_3$ and $S_3=J_0+J_3 -3J_2$ have volumes 
\be\label{eq:volS2S3}
\CV_{S_2} = 2t^3(2t^2 +t^1)+8(t^3)^2\,,\qquad  \CV_{S_3} = t^1(t^2 +t^3)+(t^2+ t^3)^2\,,
\ee 
and intersect over the curve
\be
C= S_2 \cdot S_3 = 4\mathcal{C}^2 + 2\mathcal{C}^1 + 4\mathcal{C}^3\,.
\ee
Assuming that this curve is effective we can compute the genus of a generic representative as 
\begin{equation}
    2-2g = - \left(S_2+S_3\right)\cdot S_2 \cdot S_3 = 0\,,
\end{equation}
implying that the curve $C$ has genus one. We note that we can write $C$ as 
\begin{equation}
    C= 2\mathcal{C}^1 + 4 \left(\mathcal{C}^2 +\mathcal{C}^3\right)\,.
\end{equation}
Recalling that $\mathcal{C}^1$ denotes the class of the fibre of the Hirzebruch surface the curve $C$ can be identified as a bisection of the elliptic fibration of the degenerate K3 which is branched over four points each associated to the class $\mathcal{C}^2+\mathcal{C}^3$. This identifies the degeneration of the K3 surface as Type II.b in the refined Kulikov classification discussed in \cite{Lee:2021qkx}. Also in this example, one can check that $\Lambda_{\rm sp} > m_{\rm KK,6}$, since ${\cal V}_{\cal E} = t^2+3t^3$ and $\sqrt{{\cal V}_{\rm K3}} = \sqrt{t^1 (t^2 + 3 t^3) + (t^2 + 3 t^3)^2}$.

In the limit \eqref{limitw1ex2} the scalar curvature asymptotically scales as
\be
R_{\rm IIA} = \frac{16\left((t^2)^2 + t^2 t^1 + 6 t^2 t^3 + 3 t^1 t^3 + 9 (t^3)^2\right)}{(2 t^2 + t^1 + 2 t^3)^3}\,  \phi + \mathcal{O}(\text{const})\, ,
\ee
whereas the components of the gauge kinetic function associated to $S_2$ and $S_3$ along this limit scale like
\be
1/g_{\rm rig}^2 \sim 2t^1 + 4(t^2+t^3) \, .
\ee
Given that the species scale is again set by the tension of the NS5-brane wrapping the generic K3-fibre we find
\be
\left( \frac{\Lambda_{\rm WGC}}{\Lambda_{\rm sp}} \right)^2 = \phi g_{\rm rig}^2 \sim \frac{\phi}{2t^1 + 4(t^2+t^3)} \, .
\ee
Putting these relations together, one can check that $$R_{\rm IIA} \lesssim \left( \frac{\Lambda_{\rm WGC}}{\Lambda_{\rm sp}} \right)^2$$ which means that \eqref{Rasintro} is satisfied for $\nu=w=1$, as expected for an LST limit of $SO(32)$-type. Notice that, as we expected, in this example there cannot be a hierarchy between the scales $\Lambda_{{\rm ch},2}$ and $\Lambda_{{\rm ch},3}$ since the volumes cannot be decoupled as is clear from \eqref{eq:volS2S3}. 

\subsection{\texorpdfstring{$w=1$}{} limits without 6d origin} \label{ss:w=1_no6d}
So far we focused on K3-fibrations for which the generic K3-fibre is itself elliptically fibred. This had two main advantages: $i)$ the resulting 4d theory has a 6d origin, and $ii)$ the degenerations of the K3-fibre fit into a finer classification as in \cite{Lee:2021qkx}. In general, however, the generic K3-fibre does not have to be itself elliptically fibred for an emergent string limit to be taken. In this section we briefly discuss the corresponding $w=1$ limit and mostly focus on a simple realisation of a Calabi--Yau threefold that is K3-fibred but does not allow for an elliptic fibration compatible with this K3-fibration. 

Since these models do not have a direct 6d origin classifying the corresponding $w=1$ limits according to Little String Theories is not straight-forward. Still, even without an elliptic fibration the NS5-brane wrapped on the generic K3-fibre  gives rise to a string which is identified with the heterotic string compactified on a different K3 times a two-torus. However, the geometric moduli of this torus are fixed to the self-dual point such that there is no six-dimensional limit also on the heterotic side. To get a non-trivial RFT decoupling from gravity we again need the generic K3-fibre to degenerate over points in the base $\mathbb{P}^1_0$. For simplicity we again assume that the K3-fibre degenerates over a single point into two components $S_1\cup S_2$. As before, from the heterotic perspective this degeneration signals the presence of a small instanton NS5-brane. However, since there is no uplift to 6d, the NS5-brane worldvolume theory is compactified on a circle with self-dual radius in units of the heterotic string scale, which also sets the species scale. Notice that this implies that $m_{\rm KK,6}$ coincides with the species scale and the two cannot be separated, in contrast to the case of an elliptic K3 fiber.

To distinguish different limits we can again consider the scale at which we expect charged states under the rigid gauge theory. Towers of charged states arise from D2-branes wrapping curves inside $\mathcal{S}_1$ and $\mathcal{S}_2$ whose masses determine the scales $\Lambda_{{\rm ch},1}$ and $\Lambda_{{\rm ch},2}$ corresponding to the mass scale of charged towers. Since $m_{\rm KK,6} \sim \Lambda_{\rm sp}$ we need to have $\Lambda_{{\rm ch},i}\prec m_{\rm KK,6}$ in order to have a rigid theory decoupled from gravity also in the UV. If there is no hierarchy between the scales of the towers of states charged under the respective building blocks of the LST, i.e.,  $\Lambda_{{\rm ch},1}\sim \Lambda_{{\rm ch},2} $, the UVRT is indeed a 6d LST compactified on $S^1$. On the other hand, if $\Lambda_{{\rm ch},1}\prec \Lambda_{{\rm ch},2} $, or vice versa, the UVRT is a 5d SCFT given by the M5-brane wrapping the smaller $S_1$. In this case we thus expect \eqref{Rasintro} to be satisfied with $\nu =3$. Since we can continuously interpolate between the $\nu=1$ and $\nu=3$ case, there has to exist a parametric regime in which \eqref{Rasintro} is not satisfied for $\nu=1$ but is satisfied for $\nu=2$. This signals that we should view the 5d SCFT as a 6d SCFT compactified on an additional circle $S^1$ with radius fixed at the species scale (i.e. self-dual radius). Since there is no decompactification limit for this circle, we do not expect a major physical difference between the cases $\nu =2$ and $\nu=3$ in terms of the masses of charged states. 

\subsubsection*{Example: K3-fibration without compatible elliptic fibration}

We consider a non-smooth elliptic fibration over a $\mathbb{P}^2$ base with an extra section, which was studied for example in \cite{Morrison:2012ei, Hayashi:2023hqa, FierroCota:2023bsp}. This model has three K\"ahler moduli and a simplicial K\"ahler cone. In a basis of Nef divisors the triple intersection numbers are
\be
\mathcal{I}(X) = 18 J_0 J_1^2 + 6 J_0 J_1 J_2 + 2 J_0 J_2^2 + 9 J_1^3 + 3 J_1^2 J_2 + J_1 J_2^2 \, ,
\ee
This geometry can also be viewed as a K3-fibration which is incompatible with the original elliptic fibration and for which the base $\mathbb{P}^1$ is given by the curve $\mathcal{C}^0$ and the generic K3-fibre by $J_0$. The Picard lattice of the generic K3-fibre has rank 1 such that it does not allow for an elliptic fibration. Let us now consider the $w=1$ limit 
\begin{equation}\label{w1limitex3}
    t^0 = t^0_0\,\phi\,,\qquad t^i=t_0^i \phi^{\delta_i}\,, \quad i=1,2 \, ,
\end{equation}
for $\delta_i\in [0,1)$.  The matrix ${\bf K}$ for this family of limits is given by
\be\label{matrixKab}
{\bf K} \equiv \cK_{ab0} = 
\begin{pmatrix}
0 & 0 & 0 \\
0 & 18 & 6 \\
0 & 6 & 2 
\end{pmatrix},
\ee
has rank one and a kernel generated by
\be
    \ker \cK_{0ab} = \langle J_0 , J_1-3J_2 \rangle = \langle J_0-J_1+3J_2 , J_1-3J_2 \rangle \,,
\ee
where $S_1=J_0-J_1+3J_2$ and $S_2 = J_1-3J_2$ intersect over
\be
C= S_1 \cdot S_2 = 3\mathcal{C}^2 \,.
\ee
We identify $S_1$ and $S_2$ as the two components into which the generic fibre given by $J_0$ splits at the degeneration point. The respective volumes of these three divisors are given by 
\be 
\CV_{S_1} = 9(t^1)^2 + 6 t^1t^2+\frac12 (t^2)^2 \,,\qquad \CV_{S_2} = \frac12 (t^2)^2 \,,\qquad \CV_{J_0} = (3t^1 +t^2)^2\,. 
\ee
As mentioned before, in the heterotic dual of this geometry the 5d to 6d circle is frozen at the self-dual radius (the species scale). The associated KK tower is dual, in type IIA, to a tower of D2-branes multi-wrapping the self intersection of the generic K3
\be
J_0 \cdot J_0 = 3 {\cal C}^1 + {\cal C}^2\, .
\ee
This implies that in this class of limits we have $\Lambda_{\rm sp} = m_{\rm KK,6}$ which is controlled by $\sqrt{{\cal V}_{\rm K3}} = 3t^1 + t^2$. This is in contrast to what happens with an elliptic K3, where we always have $\Lambda_{\rm sp} > m_{\rm KK,6}$.

In the limit \eqref{w1limitex3}, the scalar curvature of the moduli space diverges as
\be
R_{\rm IIA}^{\rm cl} = \frac{6(3t^1+t^2)^2}{(t^2)^3} \phi + \mathcal{O}(\text{const})\, .
\ee
and the components of the gauge kinetic function associated to $S_1$ and $S_2$ along this limit scale as
\be
1/g_{\rm rig}^2 \sim 3t^2 \, ,
\ee
such that
\be
\left( \frac{\Lambda_{\rm WGC}}{\Lambda_{\rm sp}} \right)^2 = \phi g_{\rm rig}^2 \sim \frac{\phi}{3t^2} \, .
\ee
According to our general discussion we need to distinguish the limits in which there is a hierarchy between the volumes of $S_1$ and $S_2$ from those where no such hierarchy is realised. This translates into 
\begin{itemize}
\item For $\delta_2 \geq \delta_1$ we have $\CV_{S_1} \sim \CV_{S_2}$ such that $\Lambda_{{\rm ch},1}\sim \Lambda_{{\rm ch},2} $ such that the UVRT is identified as a 6d LST on a circle with self-dual radius. And indeed $$R_{\rm IIA} \sim \left( \frac{\Lambda_{\rm WGC}}{\Lambda_{\rm sp}} \right)^2$$ which satisfies \eqref{Rasintro} for $\nu=w=1$ as expected for a 4d RFT with an LST as UVRT.
\item In case $\delta_2 < \delta_1$ there is a hierarchy $\CV_{S_1}\succ \CV_{S_2}$ and we obtain a 5d SCFT as UVRT. We find $$R_{\rm IIA} \sim \left( \frac{t^1}{t^2} \right)^2 \left( \frac{\Lambda_{\rm WGC}}{\Lambda_{\rm sp}} \right)^2 < \left( \frac{\Lambda_{\rm WGC}}{\Lambda_{\rm sp}} \right)^6$$ which satisfies \eqref{Rasintro} for $\nu=2 \neq w$ for $\del_1-\del_2 \leq 1 - \del_1$ and $\nu=3\neq w$ otherwise consistent with our general expectation. 
\end{itemize}

\subsection{\texorpdfstring{$(2,0)$}{} LST limits}
\label{ss:Abelian}

So far we only considered Calabi--Yau threefolds with $w=1$ limits in which the emergent string is dual to a heterotic string. Still there can be limits in which the emergent string is a type II string. From the type IIA perspective this requires the Calabi--Yau threefold to have a fibration structure with generic fibre given by an Abelian fibre. The dual type II string theory is then compactified on a \emph{D-manifold} times a torus \cite{Lee:2019wij}. Also in these cases we can get a non-trivial 4d RFT that decouples from gravity in the $w=1$ limit. These arise if the generic Abelian fibre degenerates over points in the base of the Abelian surface fibration. From the dual type II perspective these degenerations correspond to 5-branes of type II string theory which give rise to $(2,0)$ LSTs compactified on a $T^2$. Since these LSTs have enhanced supersymmetry compared to their heterotic cousins, the rigid curvature of the 4d RFT vanishes in these cases such that there is also no curvature divergence in the respective asymptotic $w=1$ limits. Since these limits are not very interesting from the perspective of curvature, we do not discuss them any further.


\section{Summary}
\label{s:summary}

Let us summarise our findings so far, and in particular the casuistic that we have found when testing the inequality \eqref{Rasintro} across different kinds of limits. First, from the discussion of sections \ref{s:typeIIA} and \ref{s:decoupling}, it is easy to see that for EFT string limits \eqref{Rasintro} is saturated with $\nu = w$ matching the scaling weight. Indeed, in these cases by construction $g_{\rm rigid}$ is constant, and so is $R_{\rm rigid}$. Hence one simply has that
\be
R  \sim  \left(\frac{M_{\rm P}^2}{{\cal T}}\right)^{w} \sim \left(\frac{\Lambda_{\rm WGC}}{\Lambda_{\rm sp}} \right)^{2w} ,
\label{saturate}
\ee
where we have also used \eqref{scaling} and \eqref{Lamsp}, which implies that ${\cal T} \sim \Lambda_{\rm sp}^2$ along EFT string limits. In fact, following section \ref{s:decoupling} one can see that this result is also valid for $w=3$ limits \eqref{growth} based on growth sectors. Indeed, because these limits feature divisors with non-trivial triple self-intersection numbers, we have that $R \sim {\cal V}_X g_{\rm rigid}^6 \sim (\Lambda_{\rm WGC}/\Lambda_{\rm sp})^6$, saturating again the inequality with $\nu = w$, independently of the growth sector.

The casuistic is richer for $w=2$ and $w=1$ growth sector limits, as is manifest from tables \ref{tab:w2} and \ref{tab:w1}. However, we find that the inequality 
\be
w \leq \nu \leq 3
\label{range}
\ee
is always satisfied. In other words, the scaling weight $w$ is a lower bound for the index $\nu$. 

In tables \ref{tab:w2} and \ref{tab:w1}, we summarise the different origins of the curvature $R_{\rm rigid}$ of the four-dimensional rigid theory that leads to a curvature divergence in the different cases. If the rigid theory is trivial or has enhanced supersymmetry (16 supercharges) no rigid curvature is generated. If the UVRT is a six-dimensional SCFT or LST, a non-trivial $R_{\rm rigid}$ can be generated in two ways. If all states charged under the rigid gauge group arise as excitations of a 6d SCFT string, a rigid curvature can only be generated by instantons in the 4d theory. This is due to the fact that the tensor branch associated to the 6d string is flat and hence upon dimensional reduction no curvature is induced in the 4d theory at the classical level. On the other hand, if there are states charged under the gauge group that are not excitations of a 6d string, these do generate a rigid curvature. Examples of this are the W-bosons of non-Abelian gauge theories or isolated charged hypermultiplets that arise from M2-branes not contained in shrinking divisors. These give a classical contribution to $R_{\rm rigid}$ since they already induce a moduli dependence of the rigid gauge coupling in five dimensions. Finally, if the UVRT is a 5d SCFT charged states arise from excitations of the tensionless string in five dimensions. These give a contribution to $R_{\rm rigid}$ at the classical level as is already the case in the $w=3$ analysed in \cite{Marchesano:2023thx}.
 
\setlength{\arrayrulewidth}{0.2mm}
\renewcommand{\arraystretch}{1.1}
\begin{table}[!ht]
\begin{center}
\begin{tabular}{|c|c|c|c|c|c|}
\hline
$\ker {\bf K}$   & 4d RFT & UVRT & $R_{\rm IIA}^{\rm div}\lesssim \left(\frac{\Lambda_{\rm WGC}}{\Lambda_{\rm sp}}\right)^{2\nu}$ & $R_{\rm rigid}$ origin \\
\hline \hline 
No effective divisor & $U(1)$ & trivial & n/a & none \\ 
 \hline
Vertical Divisor  & \multirow{2}{*}{$U(1)$}& 6d SCFT & \multirow{2}{*}{n/a} & \multirow{2}{*}{none} \\
with $c_2 \cdot {\cal D} = 0$& &$\cN=2$, $\mathfrak{g}=\mathfrak{su}(2)$  & &\\
\hline
Vertical Divisor & \multirow{2}{*}{$U(1)$} & 6d  SCFT & \multirow{2}{*}{$\nu =2$} & \multirow{2}{*}{instantons in 4d} \\ 
with $c_2 \cdot {\cal D} \neq  0$ & & $\cN=1$, $\mathfrak{g}=0$& &\\
\hline 
 & \multirow{4}{*}{$U(1)^{{\rm rk}\, \mathfrak{g}}$} & 6d  SCFT & \multirow{2}{*}{$\nu=2$} & \multirow{2}{*}{W-bosons} \\ 
Vertical Divisor & & $\cN=1$, $\mathfrak{g}\neq 0$ &&\\ \cline{3-5} 
+ Exc. Divisors &  & 5d SCFT & \multirow{2}{*}{$\nu=3$}& W-bosons \\ 
 & &  $\cN=1$ , $\mathfrak{g}\neq 0$ &&+ 5d string\\ \hline 

\multirow{4}{*}{Fibral Divisor} & \multirow{4}{*}{$U(1)$} & 6d SCFT & \multirow{2}{*}{$\nu=2$}& \multirow{4}{*}{hyper in 5d }\\ 
&& $\cN=1$, $\mathfrak{g}=0$  &&\\\cline{3-4} 
& & 5d $\cN=1$ SCFT  & \multirow{2}{*}{$\nu=3$} &  \\ 
&&+ hyper &&\\\hline 
\end{tabular}
\end{center}
\caption{Casuistic of $w=2$ limits. \label{tab:w2}}
\end{table}

To understand the values for $\nu$ in table \ref{tab:w2}, let us consider the different $w=2$ limits in sections \ref{s:Ftheory} and \ref{s:nonsmooth}. In these cases we deal with an elliptic fibration of ${\cal E}$ over a two-fold base $B_2$, and
\be
{\cal V}_X \simeq {\rm Vol}(B_2) \cdot  {\rm Area}({\cal E}) \sim \phi^{2+\gamma} \, ,
\label{volw=2}
\ee
where we have assumed that ${\rm Area}({\cal E}) \sim \phi^\gamma$, as in  section \ref{s:Ftheory}. This in turn implies that 
\be
\frac{M_{\rm P}^2}{m_*^2} \sim \left(\frac{M_{\rm P}}{\Lambda_{\rm sp}} \right)^{4} \frac{m_{{\rm KK},6}}{m_{{\rm KK},5}} \implies R \sim g_{\rm rigid}^2  \frac{m_{{\rm KK},6}}{m_{{\rm KK},5}}\left(\frac{\Lambda_{\rm WGC}}{\Lambda_{\rm sp}} \right)^{2w}  .
\ee
That is, the departure from the relation $\nu = w$ is controlled by the quantity $g_{\rm rigid}^2 m_{{\rm KK},6}/m_{{\rm KK},5}$. In the case where the kernel only contains vertical divisors, we have that $g_{\rm rigid}^2 m_{{\rm KK},6} \simeq m_{{\rm KK},5}$, and so the relation $\nu = w$ is enforced, up to the caveat that the cubic derivative of the prepotential is generated by instantons and so in many limits the rigid curvature vanishes exponentially fast. In case this cubic term is present classically, which is the case for exceptional and fibral divisors, then one can use that $g_{\rm rigid}^{-2} \simeq t_{\rm rigid}$ to identify $g_{\rm rigid}^{-2} m_{{\rm KK},5}$ with the mass of a D2-brane wrapping the self-intersection of such divisors, which in turn coincides with the scale $\Lambda_{\rm ch}$. We then find that 
\be
R \sim \frac{m_{{\rm KK},6}}{\Lambda_{\rm ch}}\left(\frac{\Lambda_{\rm WGC}}{\Lambda_{\rm sp}} \right)^{2w}\, .
\label{Rw2}
\ee
Finally, as discussed in section \ref{s:nonsmooth}, for exceptional and fibral divisors $\Lambda_{\rm ch} \leq m_{{\rm KK},6}$. Precisely in case this inequality is not saturated we obtain $\nu = 3$ and the UVRT hosted by these divisors corresponds to a 5d SCFT.

\setlength{\arrayrulewidth}{0.2mm}
\renewcommand{\arraystretch}{1.1}
\begin{table}[!t]
\begin{center}
\begin{tabular}{|c|c|c|c|c|c|}
\hline
$\ker {\bf K}$   & 4d RFT & UVRT & $R_{\rm IIA}^{\rm div}\lesssim \left(\frac{\Lambda_{\rm WGC}}{\Lambda_{\rm sp}}\right)^{2\nu}$ & $R_{\rm rigid}$ origin \\
\hline \hline 
 & \multirow{4}{*}{$U(1)$} & 6d LST & \multirow{2}{*}{$\nu =1$}  & \multirow{4}{*}{instantons in 4d} \\ 
 Degenerate K3&&$\cN=1$, $\mathfrak{g}=0$& & \\\cline{3-4}
elliptic, Type II.a & & 6d SCFT & \multirow{2}{*}{$\nu =2$} & \\ 
&&$\cN=1$, $\mathfrak{g}=0$&&\\
\hline 
Degenerate K3 & \multirow{2}{*}{$U(1)$} &  6d LST & \multirow{2}{*}{$\nu =1$} & \multirow{2}{*}{W-bosons} \\ 
elliptic, Type II.b & & $\cN=1$, $\mathfrak{g}\neq 0$& & \\ 
\hline 
 & \multirow{5}{*}{$U(1)$} &  (6d LST)$\times S^1_{R=\sqrt{\alpha'}}$ & \multirow{2}{*}{$\nu =1$} & \multirow{2}{*}{W-bosons} \\ 
Degenerate K3 & & $\cN=1$, $\mathfrak{g}\neq 0$&&\\\cline{3-5}
non-elliptic & & 5d SCFT & \multirow{3}{*}{$\nu=2$ or $3$} &\multirow{3}{*}{5d string}\\ 
&&$\simeq$(6d SCFT)$\times S^1_{R=\sqrt{\alpha'}}$&&\\
&&$\cN=1$ && \\ \hline 
\end{tabular}
\end{center}
\caption{Casuistic of $w=1$ limits. \label{tab:w1}}
\end{table}

For $w=1$ limits the Calabi--Yau has the structure of the fibration of a surface $S_0$ over $\mathbb{P}^1_0$,  and we have that
\be
{\cal V}_X \simeq {\rm Vol}(S_0) \cdot {\rm Area}(\mathbb{P}^1_0) \sim {\rm Vol}(S_0)\, \phi \, .
\label{volw=1}
\ee
If moreover $S_0$ is an elliptically fibred K3, then the area of its generic fibre in string units corresponds to the quotient $m_{{\rm KK},6}/m_{{\rm KK},5}$, and we can write
\be
{\rm Vol}(S_0) \sim \frac{m_{{\rm KK},6}\, {\rm max}_i \{\Lambda_{{\rm ch},i}\} }{m_{{\rm KK},5}^2} \, ,
\ee
where $\Lambda_{{\rm ch},i} \leq \Lambda_{\rm sp}$ is defined for each of the divisors in the split  $S\to \cup_i  S_M$ as the scale of the tower of D2-branes wrapping its two-cycles. For Type II.a Kulikov degenerations we have that $g_{\rm rigid}^2 \simeq m_{{\rm KK},5}/m_{{\rm KK},6}$, from where we find 
\be
R \sim \frac{{\rm max}_i \{\Lambda_{{\rm ch},i}\} }{m_{{\rm KK},6}} \left(\frac{\Lambda_{\rm WGC}}{\Lambda_{\rm sp}} \right)^{2w} .
\label{Rw2a}
\ee
Moreover, we have $m_{{\rm KK},6} \leq \Lambda_{{\rm ch},i}$ and to have a rigid curvature generated by instanton effects, we must have at least one $\Lambda_{{\rm ch},i}$ saturating the inequality. Hence, for $\Lambda_{{\rm ch},i} \simeq m_{{\rm KK},6}$, $\forall i$ we recover the result $\nu = w =1$. On the other hand, in case some of the $\Lambda_{{\rm ch},i}$ are above this scale, the 6d KK scale and the species scales are separated and we instead have $\nu =2$. For the limits corresponding to $SO(32)$-type LSTs we can estimate $g_{\rm rigid}^2 \simeq m_{{\rm KK},5}/\min_i \{\Lambda_{{\rm ch},i}\}$. For the Type II.b limits discussed in section \ref{ss:so32} we moreover have that all the scales $\Lambda_{{\rm ch},i}$ are tied to each other and to $ m_{{\rm KK},6}$, hence the leading contribution to the curvature reads $R \sim \left(\Lambda_{\rm WGC}/\Lambda_{\rm sp}\right)^{2}$. That is, we obtain $\nu = w =1$, as in the explicit example of section \ref{ss:so32}.

In the case where $S_0$ is a K3 surface with no compatible elliptic fibration we have
\be
\text{Vol}(S_0) \sim \left ( \frac{\Lambda_{\rm sp}}{m_{\rm KK,5}} \right)^2\, ,
\ee
and we can estimate $g_{\rm rigid}^2 \simeq m_{{\rm KK},5}/\min_i \{\Lambda_{{\rm ch},i}\}$, so that
\be
R \sim \left ( \frac{\Lambda_{\rm sp}}{\min_i \{\Lambda_{{\rm ch},i}\}} \right)^2 \left ( \frac{\Lambda_{\rm WGC}}{\Lambda_{\rm sp}} \right)^{2w}\, .
\ee
From here it is clear that whenever there exists a $\Lambda_{\rm ch}$ that lies below $\Lambda_{\rm sp}$ we get $\nu > w=1$. As commented in section \ref{ss:w=1_no6d}, since the 5d to 6d circle is fixed at self-dual radius and $m_{\rm KK,6} = \Lambda_{\rm sp}$, there is no physical difference between $\nu =2$ and $\nu=3$.

The integers $w$ and $\nu$ can also be related with the singularity types and enchancements discussed in \cite{Grimm:2018cpv,Corvilain:2018lgw}. In the context of type IIA CY compactifications, there is a one-to-one correspondence between the scaling index $w$ of an EFT string limit of the form \eqref{limits} and the singularity type. In particular $w=1,2,3$ correspond to type II$_b$, III$_c$, IV$_d$, respectively, where $b, c+2, d$ correspond in each case to the rank of the matrix $\bf{K}$ in \eqref{rank}. In \cite{Marchesano:2023thx} it was shown that a necessary condition to have a divergent curvature is that the rank of $\bf K$ must be non-maximal, meaning that the subindex $b,c,d$ takes a non-maximal value. If one considers a growth sector for which the moduli scale with different rates, as in \eqref{growth}, then this translates into a chain of singularity enhancements as in \cite{Grimm:2018cpv,Corvilain:2018lgw}:
\be
{\bf A}_\a \to {\bf B}_\b \to {\bf C}_\g \to \dots
\label{singuchain}
\ee
Here ${\bf A}_\a$ is the type of singularity that we get if we set $\gamma_i =0$ in \eqref{growth}, ${\bf B}_\b$ the singularity obtained from setting $\gamma_1=1$ and $\gamma_{i>1} =0$, and so on. 

To illustrate the relation between the different enhancement chains and the degree of curvature divergence in \eqref{Rasintro}, let us for instance analyse the flopped phase of the KMV conifold, that we considered in section \ref{ss:KMVflop}. For the choice of parameters $\epsilon < \lambda$ we have the enhancement chain III$_0 \to$ IV$_3$ and $\nu = w = 2$, while for the choice $\lambda < \epsilon$ we have III$_0 \to$ IV$_2 \to$ IV$_3$ and $\nu > w$ instead. Given that $d=3$ is the maximal value for a singularity of the type IV$_d$ in this Calabi--Yau, we learn that the value of $\nu$ is nothing but the degree of the highest singularity with a non-maximal subindex along the chain. For $\epsilon < \lambda$ this is the initial singularity III$_0$, which leads to $\nu =2$, while for $\lambda < \epsilon$ it is IV$_2$, resulting into $\nu =3$. This rule, which  is verified in all other examples, explains why for EFT string limits (where there is no enhancement chain) we have $\nu = w$, as well as the range of values \eqref{range}.


\section{Conclusions}
\label{s:conclu}

In this work we have continued the study of the curvature of moduli spaces along asymptotic limits, initiated in \cite{Marchesano:2023thx}. To that end, we again focused on the vector multiplet sector of the 4d ${\cal N}=2$ supergravity EFT obtained by compactifying type IIA string theory on a Calabi--Yau three-fold. In this setup we studied the behaviour of the scalar curvature of the moduli space along a wide class of large volume limits, including not only EFT string limits \cite{Lanza:2020qmt,Lanza:2021udy,Lanza:2022zyg}, but also more generic limits defined in terms of growth sectors \cite{Corvilain:2018lgw,Grimm:2018cpv}. In \cite{Marchesano:2023thx} it was noted that whenever there is a positive divergence in the moduli space curvature, there is a 4d field theory sector, here dubbed 4d RFT, that stays dynamical while decoupling gravity. We have quantified this decoupling  through the ratio $\Lambda_{\rm WGC} / \Lambda_{\rm sp}$ (with $\Lambda_{\rm WGC} = g_{\rm rigid} M_{\rm P}$), which measures the relative strength of gauge and gravitational interactions. Following \cite{FierroCota:2023bsp}, one has a gauge subsector that decouples from gravity whenever $\Lambda_{\rm WGC} / \Lambda_{\rm sp} \to \infty$. In this context, it turned out to be useful to describe the scalar curvature divergences in powers of this quotient, more precisely as
\be
R^{\rm div}_{\rm IIA} \lesssim \left( \frac{\Lambda_{\rm WGC}}{\Lambda_{\rm sp}} \right)^{2 \nu}\, ,
\ee
where $\nu \in \{1,2,3\}$ measures the degree of divergence of the curvature. Using this parametrisation, one can build a dictionary that describes how  the different 4d RFTs uplift to higher dimensional non-gravitational theories (UVRTs) along each class of limits.   More precisely, we have found that there exists a relation between the parameter $\nu$ and the nature of the UVRT. In particular,  whenever the species scale is above the 6d KK scale, $\nu =3$ indicates that the UVRT is a 5d SCFT, $\nu =2$ that it is a 6d SCFT, while for $\nu =1$ we have a 6d Little String Theory. This dictionary is different is the species scale coincides with the would-be 6d KK scale, like in the case of non-elliptic K3's of section \ref{ss:w=1_no6d}, but then one simply needs to reduce the above UVRTs on a circle to obtain the new dictionary. 

An interesting observation is that the type of divergence in the moduli space curvature can give us information about the field content of the UVRT. Take, for instance, the limits for which the UVRT is an LST. If, in these limits, the curvature divergence arises at the classical level, the LST contains a non-trivial gauge group. Instead in case the divergence is instanton-generated, the LST can only contain a tensor branch. We also saw that the value of the index $\nu$, or equivalently the type of UVRT in a certain limit, is strictly related to the mass scale $\Lambda_{\rm ch}$. This mass scale corresponds to an infinite tower of states with arbitrary charges under the 4d RFT sector, describing the content of a strongly-coupled theory that the 4d RFT flows to in the UV. Of particular importance is the hierarchy between $\Lambda_{\rm ch}$, the 5d and 6d Kaluza--Klein scales and the species scale, as shown schematically in figure \ref{fig:genscales}. If $\Lambda_{\rm ch}$ lies between the 5d and 6d KK scales, the UVRT is a 5d SCFT. If, instead, it lies between $m_{\rm KK,6}$ and $\Lambda_{\rm sp}$ we have a 6d SCFT, whereas we recover a LST in case it coincides with the species scale. 

Our results can be extended in a number of directions. An obvious one is to test the picture that we have obtained them for more involved limits in Calabi--Yau manifolds with large number of K\"ahler moduli. Presumably, this should allow us to recover the plethora of UVRTs engineered via a local bottom-up approach from the top-down viewpoint of infinite-distance limits, see for instance \cite{Heckman:2013pva,Heckman:2015bfa,Heckman:2018jxk,Bhardwaj:2015oru,DelZotto:2023ahf}. In particular, one should be able to test our picture for LST limits corresponding to more involved degenerations of K3-fibers such as the Type III Kulikov degenerations and for further constructions in which the species scale coincides with the would-be 6d KK scale. Additionally, one could compare our results to the physics of infinite-distance limits beyond large-volume regimes. In particular one may consider field space trajectories that end up at conifold singularities and Seiberg-Witten points, a set of limits that is currently under investigation from the viewpoint of moduli space curvature divergences \cite{CMP}. Finally, by extending the analysis of \cite{Marchesano:2019ifh,Grimm:2019wtx,Baume:2019sry} it would be interesting to see how our results adapt to the hypermultiplet moduli space of type II Calabi--Yau compactifications, as well as how they generalise to 4d ${\cal N}=1$ setups. 

In perspective, it is quite remarkable that all the information about 4d field theories that decouple from gravity and their UV counterparts is encoded in something as simple as the field space scalar curvature. It suggests that, by looking at more detailed quantities such as the curvature tensor itself, one may learn further general lessons about those non-gravitational theories that can be coupled to gravity, and the precise way in which they do so.

\bigskip

\bigskip

\centerline{\bf  Acknowledgments}

\vspace*{.25cm}

We thank Alberto Castellano, Cesar Fierro Cota, Naomi Gendler, Damian van de Heisteeg, Alessandro Mininno, Jeroen Monnee, Lorenzo Paoloni and Timo Weigand for discussions.  F.M. and L.M are supported through the grants CEX2020-001007-S and PID2021-123017NB-I00, funded by MCIN/AEI/10.13039/501100011033 and by ERDF A way of making Europe. L.M. is supported by the fellowship LCF/BQ/DI21/11860035  from ``La Caixa" Foundation (ID 100010434). M.W. is supported by a grant from the Simons Foundation (602883,CV), the DellaPietra Foundation, and by the NSF grant PHY-2013858. F.M. and M.W. would like to thank the Erwin Schr\"odinger International Institute for Mathematics and Physics for hospitality during the completion of this work. L.M. would like to acknowledge the hospitality of the Department of Physics of Harvard University during the early stages of this work.





\bibliographystyle{JHEP2015}
\bibliography{papers}

\end{document}